\documentclass[aps,floats,showpacs,amsmath,epsf,twocolumn]{revtex4}
\usepackage{amsfonts, relsize, color}
\usepackage{graphics}
\usepackage{graphicx}
\usepackage{subfigure}
\usepackage{hyperref}

\begin{document}
\title{Quantum phases of interacting electrons in three-dimensional dirty Dirac semimetals}

\author{Bitan Roy}
\affiliation{Condensed Matter Theory Center and Joint Quantum Institute, University of Maryland, College Park, Maryland 20742-4111, USA}

\author{Sankar Das Sarma}
\affiliation{Condensed Matter Theory Center and Joint Quantum Institute, University of Maryland, College Park, Maryland 20742-4111, USA}

\date{\today}

\begin{abstract}
We theoretically study the stability of three dimensional Dirac semimetals against short-range electron-electron interaction and quenched time-reversal symmetric disorder (but excluding mass disorder). First we focus on the clean interacting and the noninteracting dirty Dirac semimetal separately, and show that they support two distinct quantum critical points. Using renormalization group techniques, we find that while interaction driven quantum critical points are \emph{Gaussian} (mean-field) in nature, describing quantum phase transitions into various broken symmetry phases, the ones controlled by disorder are \emph{non-Gaussian}, capturing the transition to a metallic phase. We classify such diffusive quantum critical points based on the transformation of disorder vertices under a \emph{continuous} chiral rotation. Our wek coupling renormalization group analysis suggests that two distinct quantum critical points are stable in an interacting dirty Dirac semimetal (with chiral symmetric randomness), and a multicritical point (at finite interaction and disorder) results from their interplay. By contrast, the chiral symmetry breaking disorder driven critical point is unstable against weak interactions. Effects of weak disorder on the ordering tendencies in Dirac semimetal are analyzed. The clean interacting critical points, however, satisfy the \emph{Harris criterion}, and are therefore expected to be unstable against bond disorder. Although our weak coupling analysis is inadequate to establish the ultimate stability of these fixed points in the strong coupling regime (when both interaction and disorder are strong), they can still govern crossover behaviors in Dirac semimetals over a large length scale, when either interaction or randomness is sufficiently weak. Scaling behavior of various physical quantities (e.g., spectral gap, specific heat, density of states, conductivity) and associated experimental signatures across various quantum phase transitions are discussed. 
\end{abstract}

\pacs{05.30.Rt, 64.60.ae, 71.30.+h, 11.10.Jj}

\maketitle

\section{Introduction}

Understanding the roles of electron-electron interactions and disorder in solid state systems are two separate problems of fundamental importance, and over the past few decades these questions have extensively been addressed in the context of Fermi liquids, although a definitive understanding of the ground state in such systems in the presence of both strong disorder and strong interaction is still a subject of debate~\cite{belitz, belitz-vojta, mott-davis, pollak, filkenstein, dashwang-1}. The Fermi liquid remains stable against weak electron-electron interactions in the renormalization group sense (i.e., the Fermi liquid phase is an infrared stable fixed point in dimensions two or higher), except for Kohn-Luttinger superconductivity at exponentially low temperatures. In general, weak interactions produce many-body renormalization effects, where various parameters of the noninteracting system (e.g., Fermi velocity), and physical observables (e.g., specific heat, compressibility, susceptibility) are renormalized by electron-electron interactions, which have been studied extensively in the literature~\cite{abrikosov, pines, shankar}. On the other hand, strong interactions could produce quantum phase transitions~\cite{sachdev} (often first order transitions) into symmetry-broken (e.g., ferromagnet or Wigner crystal) phases. Fermi liquids are also susceptible to disorder-induced randomness, and in particular sufficiently weak disorder destabilizes the Fermi liquid fixed point and gives rise to a stable diffusive metal in three dimensions. The diffusive metallicity is a manifestation of the Fermi system developing a finite relaxation time due to disorder scattering, leading to finite metallic conductivity. As the disorder gets stronger, such a metallic phase can encounter Anderson localization that has been a subject of intense research over the past fifty years~\cite{abrahams}. A question arises quite naturally regarding the fate of various phases in clean interacting and dirty noninteracting systems, when these two perturbations are present simultaneously, which remains an open problem.

Using field-theoretic renormalization group (RG) techniques, we here address this question for a disordered and interacting three dimensional Dirac semimetal (DSM), with the chemical potential being pinned exactly at the Dirac point~\cite{balatsky}. The simplicity of the noninteracting spectra, composed of linearly dispersing completely filled(empty) valence(conduction) band allows considerable progress, thus serving as a potential `toy' model for understanding the interplay of disorder and interaction in a specific situation (as long as neither is very strong).

In this work, we only consider repulsive local short-range electron-electron interaction (an extended Hubbard-type model) and short-range disorder. Thus, both interaction and disorder will be characterized only by a strength in our model, and no range parameter shows up explicitly. The key issue of interest in this work is the stability of the DSM to the varying strength of the interaction in the presence of disorder and vice versa. In particular, using RG techniques, we will study the possibility of various broken-symmetry phases (BSPs), which arise by destabilizing the DSM, as the interaction becomes stronger, and ask how the interaction-induced quantum phase transitions (QPTs) may depend on background disorder. In addition, we also shed light on the effects of interactions (when weak) on the disorder controlled phenomena in DSM.

Often two topologically distinct vacua are separated by a gapless phase. One celebrated example of such a gapless system is the three-dimensional DSM, describing the quantum critical point (QCP) between a trivial band insulator and strong $Z_2$ topological insulator, such as Bi$_2$Se$_3$, Bi$_2$Te$_3$, Bi$_{1-x}$Sb$_x$, each of which can be succinctly described by a single four-component massive Dirac fermion~\cite{fu-kane, model-TI, review-TI-1, review-TI-2}. Such a QCP is characterized by the dynamic critical exponent (DCE) $z=1$. In weakly correlated systems, such topological phase transition can be tuned by applying pressure or injecting impurities~\cite{hassan-cava, ando, hassan-neupane, armitage, TPT-BiT}. Recently discovered strongly correlated topological insulators, such as YbB$_6$, SmB$_6$~\cite{TKI-exp-1,TKI-exp-2,TKI-exp-3,TKI-exp-4,TKI-exp-5}, are  described in terms of three copies of inverted band massive Dirac fermions~\cite{roy-TKI, coleman-review}, and by applying pressure these systems can, in principle, be tuned through a gapless point~\cite{fisk-pressure}. In addition, various narrow gap semiconductors, such as, Pb$_{1-x}$Sn$_{x}$Te (hosting four copies of massless Dirac points at $L$ points of the Brillouin zone), Bi$_{1-x}$Sb$_x$, Hg$_{1-x}$Cd$_x$Te (both hosting single copy of Dirac point), become DSM for special values of $x$~\cite{dornhaus}. Furthermore, DSM can also be found as a stable phase in Cd$_2$As$_3$~\cite{cdas} and Na$_3$Bi~\cite{nabi} (referred as topological DSM) that support two copies of gapless Dirac cones. Therefore, understanding the stability of Dirac semimetals in the presence of interaction~\cite{goswami-chakravarty, hosur, isobe-nagaosa, gonzalez, nomura, kim-moon, throckmorton-3d, roy-goswami-sau} and disorder~\cite{fradkin, shindou, goswami-chakravarty, herbut-Imura, radziovsky, roy-dassarma, ominato, pixley}, and their interplay~\cite{goswami-chakravarty, kim-moon} are of definite fundamental importance, given the great deal of current experimental and theoretical interest in the subject, and may as well reveal some interesting interplays of band topology, electronic correlation and randomness. Due of the technical complexity of the analysis (presented in Sec.~\ref{dirac-intro}-\ref{interaction-disorder}), we will provide a synopsis of our main findings, emphasizing the various interaction driven broken symmetry phases and their robustness against quenched disorder. In Sec.~\ref{clean-interacting} and Sec.~\ref{dirty-DSM}, we respectively discuss the physics of clean interacting and dirty noninteracting DSM. These two sections provides a necessary background to appreciate the interplay of interaction and disorder in a three dimensional DSM. However, readers familiar with these two problems may wish to bypass these two sections and directly go to Sec.~\ref{interaction-disorder} where we address the competition between interaction and randomness in details.

\begin{figure*}[htbp]
\centering
\subfigure[]{
\includegraphics[width=7.5cm,height=5.0cm]{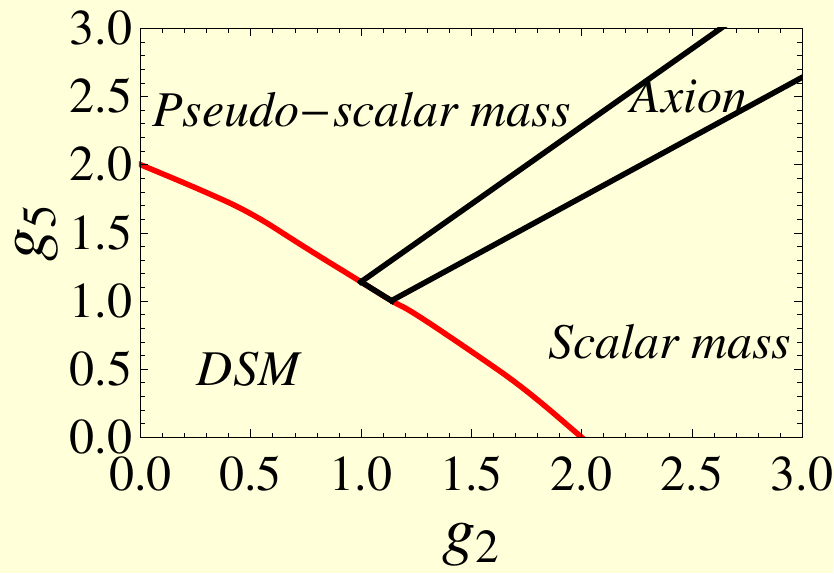}
\label{pd-DiracClean}  
}
\subfigure[]{
\includegraphics[width=7.5cm,height=5.0cm]{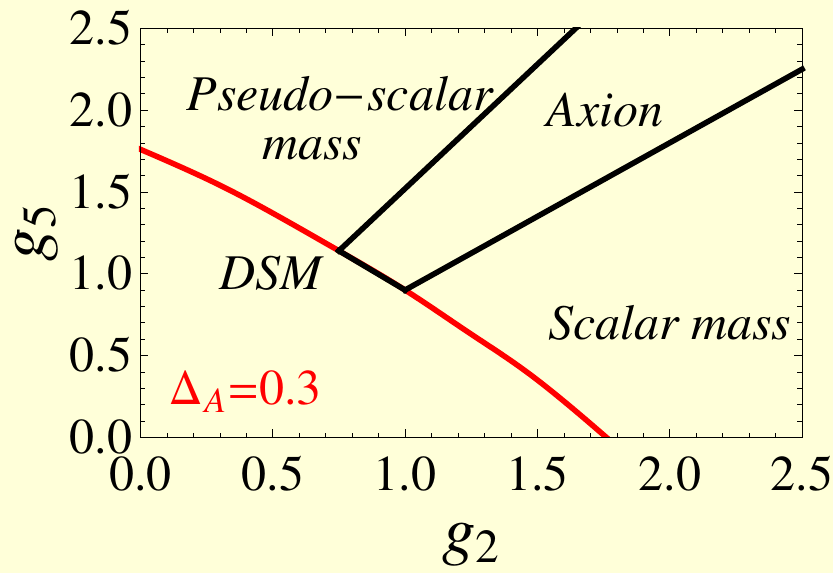}
\label{pd-Diracaxial}
}
\subfigure[]{
\includegraphics[width=7.5cm,height=5.0cm]{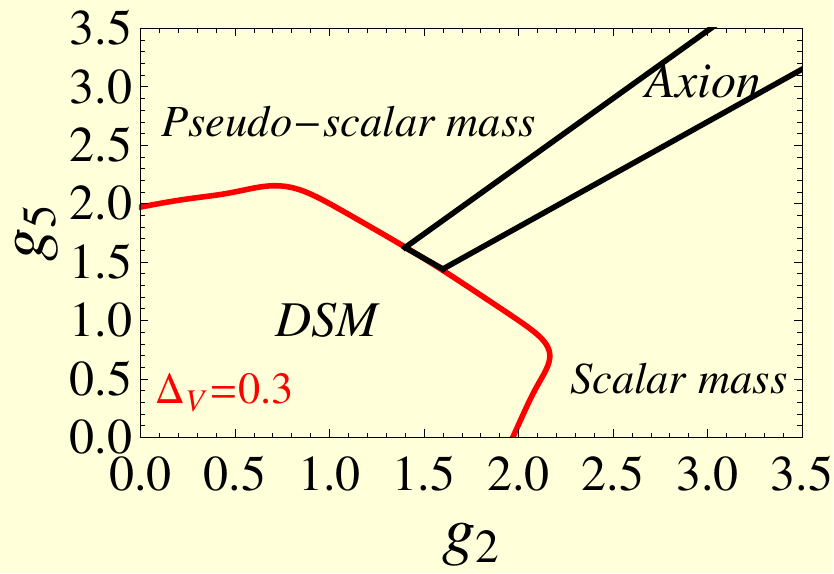}
\label{pd-Diracpotential}
}
\subfigure[]{
\includegraphics[width=7.5cm,height=5.0cm]{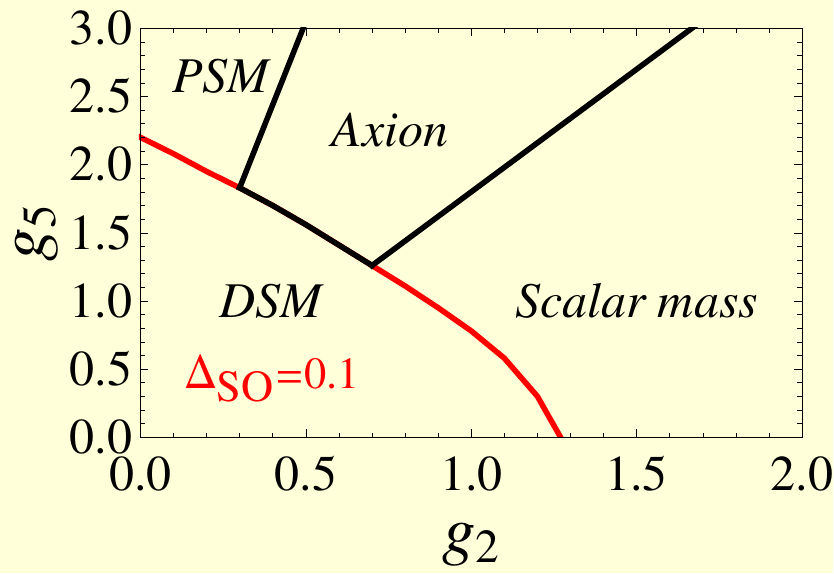}
\label{pd-Diracspinorbit}
}
\caption[] {(Color online) (a) Phase diagram of clean interacting three dimensional DSM. Here, $g_2$ and $g_5$ are two short-range four-fermion interactions, respectively supporting scalar and pseudo-scalar mass for Dirac fermions, when strong. When they acquire comparable strength, an axionic insulator (linear combination of scalar and pseudo scalar mass) is realized. In Sec.~\ref{Dirac-BSP} scalar and pseudo-scalar masses are denoted by $m_1$and $m_2$, and the axionic insulator is thus a linear superposition of $m_1$ and $m_2$. (b) Phase diagram of weakly disordered DSM in the presence of axial disorder ($\Delta_A=0.3<\Delta^\ast_A$). $\Delta^\ast_A=0.5$ is the critical strength (dimensionless) of axial disorder for DSM-CDM QPT. (c) Phase diagram of weakly disordered DSM in the presence of potential disorder ($\Delta_V=0.3<\Delta^\ast_V$). $\Delta^\ast_V=0.5$ is the critical strength (dimensionless) of potential disorder for DSM-CDM QPT. (d) Phase diagram of interacting DSM in the presence of weak spin-orbit disorder ($\Delta_{SO}=0.1<\Delta^\ast_{SO}$). $\Delta^\ast_{SO}=1.5$ is the threshold strength (dimensionless) of spin-orbit disorder for DSM-CDM QPT. Here PSM stands for pseudo-scalar mass. Even though, these results strictly hold for weak to intermediate interaction and disorder coupling strength, they may still control a large crossover regime to the eventual (unknown) strong coupling phase where interaction or/and disorder is/are strong~\cite{boundary}.}\label{Dirac-PD}
\end{figure*}

We begin with a discussion on the effects of interaction in clean DSM. Some valuable guidelines into the effects of electronic interactions can be obtained by combining the notions of scaling and renormalization group analysis. 

\begin{enumerate}

\item{ Notice that the density of states (DOS) in three dimensional DSMs vanishes as $\varrho(E) \sim E^2$ in the close vicinity of the band (Kramers degenerate) touching points (referred as Dirac points). Consequently, DSMs are extremely robust against sufficiently weak electron-electron interactions. In the language of RG, such stability stems from the fact that weak electron-electron interactions are \emph{irrelevant} perturbations near the non-interacting, infra-red stable Gaussian fixed point. }

\item{If, on the other hand, repulsive interactions are sufficiently strong, DSM can undergo QPTs and enter into a plethora of chiral symmetry breaking (CSB) BSPs (see Fig.~\ref{pd-DiracClean}, for example). When the net interaction acquires a strong attractive component, Dirac fermions can also condense into various superconducting ground states, a situation we ignore here and focus only on repulsive interaction. At $T=0$ fully gapped phases optimally lower the ground state energy at strong interactions. Otherwise, the QPTs in three dimensional DSMs are mean-field or Gaussian in nature, since the system lives at the \emph{upper critical dimension} ($d_u=3$). The interacting QCPs are characterized by DCE $z=1$ and the correlation length exponent (CLE) $\nu=\epsilon^{-1}_1$, where $\epsilon_1=d-z=2$ when $d=3$. Therefore, a pseudo Lorentz symmetry gets restored at the QCPs, but physical observable in the BSPs, such the fermionic mass gap, displays \emph{logarithmic} correction due to the violation of \emph{hyperscaling}~\cite{herbut-book}.}

\item{In this work, we address various instabilities in a interacting DSM, accounting for short-range [local in space and time (imaginary)] four-fermion interactions by using RG techniques. The CSB insulating states are (a)  parity (${\mathcal P}$) and time-reversal (${\mathcal T}$) odd insulator (pseudo-scalar mass), (b) only CSB, but ${\mathcal P}, {\mathcal T}$ symmetric insulator (scalar mass), and (c) ${\mathcal P}, {\mathcal T}$-odd axionic insulator. A representative phase diagram of clean interacting DSM is shown in Fig.~\ref{pd-DiracClean}~\cite{coulomb-ref}. Notice $d=z$ is the \emph{lower critical dimension} for DSM-BSP QPT and our RG study follows the spirit of an $\epsilon$-expansion around $d=1$.}

\item{The physical nature of the interaction driven BSPs, however, depends on the microscopic details of the system. Due to strong spin-orbit coupling in three dimensional DSM, the insulating phases can arise from underlying charge-density-wave (CDW) or spin-density-wave (SDW) orders. For example, if we subscribe to the standard representation of Dirac matrices (ones used in high energy physics)~\cite{peskin}, all insulating phases represent CDW orders. On the other hand, the representation of Dirac matrices is quite different for systems like Bi$_2$Se$_3$~\cite{model-TI}, and the scalar and pseudo scalar mass corresponds to CDW and SDW order, respectively, while these two density-wave orders coexist in the axionic insulating phase (see Sec.~\ref{Dirac-BSP} for details).}

\end{enumerate}

The fact that the three dimensional DSM is stable against sufficiently weak electron-electron interaction, and undergoes a QPT at finite coupling strength, is qualitatively similar to its two dimensional counterpart, such as single layer graphene. However, the QPT in two dimensional Dirac materials is non-mean field in nature~\cite{khveshchenko, herbut-original, HJR}, while that in three dimensions is Gaussian (mean-field). Our results are qualitatively similar to the ones found for Weyl semimtels, which also support linearly dispersing quasiparticle dispersion. But, in Weyl semimetals Kramers non-degenerate bands give rise to such conical dispersion and only the axionic insulator can be realized as a massive or insulating phase~\cite{aji, nandkishore, sczhang, nomura-Weyl}. Such contrasting outcomes in Weyl and Dirac semimetals stems from the fact that in the former system the chiral symmetry is intimately tied with the \emph{translational} symmetry, enforcing $g_2=g_5$, thus forbidding the realization of scalar and pseudoscalar masses separately.

A proper insight into the stability of ballistic quasiparticle excitations in a three dimensional DSM in the presence of random quenched disorder can also be gained from the scaling theory. The scaling dimension of disorder coupling $(\Delta)$ is $[\Delta]=2 z-d$. Therefore, in three dimensional DSMs, weak disorder is also an irrelevant perturbation, since $[\Delta]=-1$. However, beyond a critical strength of disorder, DSM can undergo a disorder driven QPT and becomes a compressible diffusive metal (CDM). In the CDM phase the DOS at zero energy, the quasiparticle lifetime, the mean free path, and the metallic conductivity [as $T \to 0$ (dc conductivity) or $\omega \to 0$ (ac conductivity)] are \emph{finite}. The nature of such disorder driven quantum criticality in the noninteracting dirty DSM and effects of disorder in transport phenomena has been a subject of intense analytical and numerical investigation in recent years~\cite{fradkin, shindou, goswami-chakravarty, herbut-Imura, radziovsky, roy-dassarma, ominato, pixley, dassarma-hwang-transport, fiete}. Our main results for noninteracting dirty DSM are announced below and also summarized in Table~\ref{table-intro}.

\begin{table}[h]
  \begin{tabular}{|c||c|c|c|c|c|c|c|}
     \hline 
		Disorder & CLE & DCE & $\varrho(E)$ & $C_v$ & $\sigma(\omega)$ & $\sigma(T)$ & Stability \\
		\hline \hline
		CSP & $\epsilon^{-1}_2$ & $1+\frac{\epsilon_2}{2}$ & $|E|$ & $T^2$ & $\omega^{\frac{2}{3}}$ & $T^{\frac{2}{3}}$ & Stable \\
		\hline
		CSB & $\epsilon^{-1}_2$ & $1+ \frac{9}{2} \epsilon_2$ & $|E|^{-\frac{5}{11} }$ & $T^{\frac{6}{11} }$ & $\omega^{\frac{2}{11}}$ & $T^{\frac{2}{11}}$ &  Unstable \\
		\hline
  \end{tabular}
\caption{ Critical exponents ($\nu$ and $z$) and scaling of DOS ($\varrho(E)$), specific heat ($C_v$), optical conductivity ($\sigma(\omega)$) and dc conductivity ($\sigma(T)$) near CSP and CSB disorder driven DSM-CDM QCPs. Scaling of physical quantities is quoted for $\epsilon_2=1$ or $d=3$. The last column shows that stability of various DSM-CDM QCPs against sufficiently weak short-range interaction, as one approaches from the DSM side.}\label{table-intro}
\end{table}

\begin{enumerate}

\item{We \emph{classify} various possible QPTs in the presence of generic, but time-reversal symmetric disorder in three dimensional DSM. We show that in the presence of chiral symmetry preserving (CSP) disorder (such as potential and axial disorder) the DSM-CDM QPT takes place through a line of QCPs, characterized by $\nu=\epsilon^{-1}_2$ and $z=1+\epsilon_2/2$, where $\epsilon_2=d-2$. On the other hand, when DSM hosts only CSB (such as spin-orbit) disorder, the DSM-CDM QPT is characterized by the exponents $\nu=\epsilon^{-1}_2$ ad $z=1+ 9 \epsilon_2/2$. Notice that $d=2$ is the lower critical dimension for DSM-CDM QPT, and our RG analysis can be considered as an $\epsilon$-expansion around the lower critical dimensions for the DSM-CDM QPT $d_l=2$. }

\item{Near the DSM-CDM QCP, the average DOS, conductivity (both dc and optical) can serve as bonafide order parameter. When the transition is driven by CSP (CSB) disorder, the DOS vanishes (diverges) as $\varrho (E) \sim |E| (|E|^{-5/11})$. Inside the quantum critical regime the optical conductivity scales as $\sigma(\omega) \sim \omega^{2/3}$ or $\sim \omega^{2/11}$, respectively, when DSM is subject to strong CSP or CSB disorder. Scaling of the dc conductivity follows that of optical conductivity upon replacing frequency ($\omega$) by temperature ($T$). The specific heat near these two types of DSM-CDM QCPs scales as $C_v \sim T^{2}, T^{6/11}$, respectively.  Otherwise, in DSM and CDM phases the specific heat scales as $C_v \sim T^3$ and $T$, respectively. }

\end{enumerate}        

As one keeps increasing the strength of disorder the CDM ultimately undergoes a second QPT at stronger disorder and becomes an \emph{Anderson insulator}~\cite{fradkin, pixley}. The Anderson transition in Dirac system is similar to the one in ordinary three dimensional metals, but goes beyond the scope of our weak coupling RG analysis. The notion of disorder driven QPT is also germane for Weyl semimetals that has also received ample attention in recent time~\cite{brouwer, altland, chen-Song, ohtsuki, roy-bera, hughes, nandkishore-disorder}. Notice that scalar mass disorder also breaks chiral symmetry. In the current work, we do not discuss the effect of strong mass disorder in a DSM. Nevertheless, we expect that for sufficiently weak mass disorder, sharp quasiparticle excitations in DSM remains stable~\cite{goswami-chakravarty}.

Stability of DSM against sufficiently weak interactions and disorder gives us an opportunity to study their interplay pursuing a weak coupling RG approach. However, the interaction and disorder controlled QCPs in three dimensional DSM can only be accessed from two different lower critical dimensions, respectively $d_l=1$ and $2$. Therefore, the competition between interaction and disorder cannot be addressed in terms of a unique $\epsilon$-expansion. To circumvent this challenge, we here invoke the notion of a \emph{double $\epsilon$-expansion}, where $\epsilon_1=d-1$ and $\epsilon_2=d-2$ respectively capture the strength of interaction and disorder couplings at various fixed points. Our central achievements are promoted below, and also reflect in Fig.~\ref{pd-Diracaxial}, \ref{pd-Diracpotential} and \ref{pd-Diracspinorbit}.

\begin{enumerate}

\item{Within the framework of a weak coupling RG calculation, we find that both interaction controlled Gaussian and CSP disorder controlled non-Gaussian QCPs are stable against sufficiently weak disorder and interaction, respectively. In other words, the Gaussian (non-Gaussian) interacting (disordered) critical point remains stable against turning on infinitesimal disorder (interaction). In addition, a multi-critical point (MCP) emerges from the competition between these two perturbations, where three different phases, namely the DSM, a BSP and the CDM meet. In contrast, CSB disorder controlled DSM-CDM QCP becomes unstable against sufficiently weak interactions. }

\item{The stability of disorder controlled QCPs against sufficiently weak local four-fermion interaction can be anticipated intuitively. Recall that the DCEs are $z=1+\epsilon_2/2$ and $1+9\epsilon_2/2$ at the CSP and CSB disorder driven DSM-CDM QCPs, respectively, and the bare scaling dimension of short-range interaction is $[g]=z-d$. Therefore, near these two dirty QCPs, $[g]=-2+\epsilon_2/2=-3/2$ and $-2+9 \epsilon_2/2=5/2$ for $\epsilon_2=1$, respectively. Hence, weak short-range interaction is an irrelevant perturbation near the CSP disorder driven QCP, but a relevant perturbation when the QPT from DSM to CDM is driven by the CSB disorder. Same results can be arrived at from a slightly different view point. Notice that the DOS at the DSM-CDM QCP driven by CSP (CSB) disorder vanishes (diverges) as one approaches the Dirac point ($E \to 0$), see Table~\ref{table-intro}. As a result, electronic interaction gets suppressed (enhanced) near the CSP (CSB) disorder controlled QCP.}

\item{The location of the MCP, together with the interacting QCP in the clean limit, determines the phase boundary between DSM and BSP for sufficiently weak disorder. We show that the axial disorder enhances all ordering tendencies in DSM, as shown in Fig.~\ref{pd-Diracaxial} (by suppressing the DSM regime, while enhancing the BSP regime). By contrast, the potential disorder favors the formation of the pseudo-scalar and scalar mass for Dirac fermion, but tends to defer the condensation of the axionic insulator and trivial $s$-wave superconductor, as shown in Fig.~\ref{pd-Diracpotential}. Phase diagram of interacting Dirac fermions in the presence of sufficiently weak spin-orbit disorder is shown in Fig.~\ref{pd-Diracspinorbit}. }

\item{Now we present a comparative discussion on the qualitative structure of the phase diagrams in Figs.~\ref{pd-DiracClean}, \ref{pd-Diracaxial}, \ref{pd-Diracpotential}, and \ref{pd-Diracspinorbit}. We note that the phase diagrams in clean interacting DSM, as well as the ones in the presence of axial and potential disorders, are symmetric under $g_2 \leftrightarrow g_5$. This outcome in the clean interacting system stems from the fact that the scalar mass, the pseudo-scalar mass and the axionic order can be \emph{chirally} rotated into each other (continuous chiral rotation by the Hermitian matrix $\gamma_5$, see Sec.~\ref{dirac-intro}). In the presence of axial and/or potential disorder, the underlying chiral symmetry of DSM remains unaffected, and the phase diagrams of an interacting, but weakly disordered DSM (subject to axial/potential disorders), as shown in Figs.~\ref{pd-Diracaxial} and \ref{pd-Diracpotential}, continues to enjoy the symmetric under $g_2 \leftrightarrow g_5$. By contrast, in the presence of spin-orbit disorder, the underlying chiral symmetry in a noninteracting DSM gets broken, and as a result the phase diagram in Fig.~\ref{pd-Diracspinorbit} lacks the symmetry under $g_2 \leftrightarrow g_5$.}

\end{enumerate}

We emphasize that our theory in the presence of both disorder and interaction (i.e. the upper right hand quadrant indicated by question marks in Fig.~\ref{OddP-int-b}) is necessarily approximate, and cannot access the most interesting regime of both strong interaction and disorder (since no RG expansion is possible in this regime). Our double $\epsilon$-expansion is specifically designed to study effects of weak interaction (disorder) on the dirty (interacting) system so that an expansion is possible around the respective critical dimensionality of one (two). Since no critical dimensionality exists for strong interaction and strong disorder (since they are strongly relevant perturbations), our method is not applicable in such a regime.

It should be noted that the ultimate stability of disorder and interaction controlled QCPs and MCPs in the strong coupling limit (when interaction and/or disorder are strong) cannot be established from a weak coupling RG analysis. Nevertheless, a few comments can be made on this issue. Notice that the clean interacting QCP satisfies the \emph{Harris criterion} $\nu <2/d$ in three dimensional DSM~\cite{harris}. Therefore, such a QCP in the presence of random mass or bond disorder, which is naturally generated inside a BSP in the disordered environment, should be unstable toward a new fixed point at finite interaction and disorder, where $\nu \geq 2/3$~\cite{chayes}. Effects of interactions inside the CDM phase \emph{cannot} be addressed within the weak coupling RG analysis, since ballistic Dirac fermion does not constitute the lowest energy excitations inside the CDM phase, and one must start with diffusive fermions to build the RG formalism, which flows immediately to strong-coupling allowing no simple answer. Nevertheless, our weak coupling analysis suggests that if disorder or interaction is sufficiently weak, the interaction and disorder (CSP) controlled QCPs respectively govern at least \emph{crossover} behavior of various physical quantities over a large length scale. Thus, we speculate that our obtained quantum phase diagrams shown in Fig.~\ref{Dirac-PD} could remain relevant over a large crossover regime before eventually giving away to the (unknown) strong coupling phase for large interaction/disorder. It is worth emphasizing that the ultimate fate of strongly interacting disordered fermions is still an unsolved problem in ordinary metals as well both in two and three dimensional systems.

The rest of the paper is organized as follows. In the next section, we present the effective low energy theory for clean noninteracting three dimensional DSMs. We derive the interacting model composed of local four-fermion interactions and classify various BSPs. Symmetry transformations of various time-reversal symmetric disorders and the notion of disorder averaging (replica formalism) are also introduced in Sec.~\ref{dirac-intro}. In Sec.~\ref{clean-interacting}, we analyze various interaction driven instabilities in a clean DSM, and address the emergent quantum critical phenomena, the scaling of various physical observables across the DSM-BSPs QCPs. Sec.~\ref{dirty-DSM} is devoted to study the effect of quenched disorder [both CSP (such as potential and axial disorders) and CSB (such as spin-orbit disorder)] in a noninteracting DSM. Here we classify the disorder driven DSM-CDM QCPs based on the notion of chiral symmetry and discuss the scaling properties of thermodynamic and transport quantities. Interplay of interaction and disorder in three dimensional DSMs is addressed in Sec.~\ref{interaction-disorder}. Our findings and future outlooks are summarized in Sec.~\ref{discussion}. Some additional technical details have been relegated to Appendices~\ref{coulomb-long-range}-\ref{double-epsilon}.

\section{Dirac semimetal}\label{dirac-intro}

\subsection{Non-interacting system}

The minimal model for a three-dimensional DSM is represented by the Hamiltonian
\begin{equation}\label{Dirac3D}
H_D= v_1 \alpha_1 k_1 + v_2 \alpha_2 k_2 + v_3 \alpha_3 k_3,
\end{equation}
where $v_{1,2,3}$ correspond to the Fermi velocities along $x,y,z$ directions, respectively, and $\alpha_{1,2,3}$ are three mutually anti-commuting four-dimensional Hermitian matrices, satisfying the Clifford algebra $\{ \alpha_j, \alpha_k \} = 2\delta_{jk}$ for $j,k=1,2,3$. The remaining two mutually anti-commuting matrices that together with $\alpha_j$'s close the Clifford algebra of five mutually anti-commuting matrices are $\beta$ and $\beta \gamma_5$, where $\gamma_5=i \alpha_1 \alpha_2 \alpha_3$. For the rest of the discussion we omit the velocity anisotropy and set $v_1=v_2=v_3=v$. The Dirac Hamiltonian then assumes a rotationally symmetric form $H_D=v \alpha_j k_j$, where summation over the repeated indices is assumed. Throughout this paper $\hbar$ is set to be \emph{unity}. The spinor basis is chosen as $\Psi^\top=\left( c_{+\uparrow}, c_{+\downarrow}, c_{-\uparrow}, c_{-\downarrow} \right)$, where $c_{\pm s}$ represent fermionic annihilation operators for even (+) and odd (-) parity states with spin projection $s=\uparrow, \downarrow$. Thus, a DSM can be realized from the mixing/hybridization between two orbitals with \emph{unit} angular momentum difference.

The imaginary-time ($\tau$) action associated with $H_D$ is
\begin{equation}
S_{0} = \int d^3x d \tau \: \bar{\Psi} \left(\gamma_0 \partial_\tau + v \gamma_j \partial_j \right) \Psi \equiv \int d^3x d \tau L_0,
\end{equation}
where $\bar{\Psi}=\Psi^\dagger \gamma_0$ is an independent Grassman spinor that transforms in conjugate representation and $\gamma_j =i \gamma_0 \alpha_j$ with $\gamma_0 \equiv \beta$. The $\gamma$ matrices satisfy the anti-commuting algebra $\left\{ \gamma_\mu, \gamma_\nu \right\}=2 \delta_{\mu \nu}$ for $\mu, \nu=0,1,2,3,5$. The action $S_0$ is invariant under a continuous global chiral $U_c(1)$ rotation: $\Psi \to e^{i \theta \gamma_5/2}\Psi$, $\bar{\Psi} \to \bar{\Psi} e^{i \theta \gamma_5/2}$. The action also exhibits invariance under various discrete symmetries: parity (${\mathcal P}$), charge-conjugation (${\mathcal C}$), and the reversal of time (${\mathcal T}$). Respectively under these discrete symmetry operations the spinor transforms according to ${\mathcal P} \Psi {\mathcal P}^{-1}=\gamma_0 \Psi$, ${\mathcal C} \Psi {\mathcal C}^{-1}=-\gamma_2 \Psi$, ${\mathcal T} \Psi {\mathcal T}^{-1}=-\gamma_1 \gamma_3 \Psi$~\cite{peskin}.

Such a non-interacting ground state can also be realized on a \emph{cubic lattice}. The tight-binding Hamiltonian that leads to the above Dirac Hamiltonian in the low-energy and long wavelength limit is given by~\cite{fu-kane, model-TI}
\begin{eqnarray}\label{latticeDirac}
H_{lat} =  t_1 \sum^{3}_{j=1} \alpha_j \sin (k_j a) + t_2 \; \beta \sum^{3}_{j=1} \left[ 1-\cos (k_j a ) \right],
\end{eqnarray}
where $a$ is the lattice spacing. The first term gives rise to massless Dirac fermionic excitations in the vicinity of \emph{eight} high symmetry points of the Brillouin zone. The second term, playing the role of a momentum dependent mass, is also known as \emph{Wilson mass}. With the above chosen form of the mass term, all massless Dirac points, except the one at the $\Gamma=\left(0,0,0 \right)$-point, are gapped out, and one realizes a single copy of the four-component massless Dirac fermion at the $\Gamma$ point (the Dirac point of the model). In the vicinity of the $\Gamma$ point, the tight-binding Hamiltonian assumes the form
\begin{eqnarray}
H^{\Gamma}_{lat} = v \sum^{3}_{j=1} \alpha_j k_j + b \beta \; k^2 + {\mathcal O} (k^3),
\end{eqnarray}
where $v=t_1 a$ and $b=t_2 a^2/2$ in the low-enegry and long wavelength limit. Therefore, the momentum dependent mass ($\sim b$) breaks the continuous chiral $U_c(1)$ symmetry of $H_D$ generated by $\gamma_5$, while leaving the discrete particle-hole symmetry unaffected, since $\left\{ H_{lat}, \beta \gamma_5 \right\} = 0$. The Wilson mass ($\sim b k^2$) does not break any bonafide discrete microscopic symmetry (${\mathcal P}$, ${\mathcal C}$, ${\mathcal T}$) of the system.

The momentum dependent mass is, however, an irrelevant parameter in the language of RG. Hence, there exists an infra-red stable fixed point with $b=0$, where the $U_c(1)$ chiral symmetry, generated by $\gamma_5$, gets restored and $H_{lat} \to H_D$. In this work we will address the stability of such infrared stable DSM in the presence of interaction and disorder, and for the rest of the discussion we set $b=0$. The chiral symmetric Dirac Hamiltonian can be embedded in a bigger system, which possesses a genuine $U(1)$ chiral symmetry at microscopic level, such as the topological DSM. Our goal is to demonstrate the role of interaction and disorder in the above four dimensional \emph{building block}. Although we are focusing on single-valley DSM here, if one completely neglects the inter-valley scattering processes (the exact large-N limit), our results can also be applicable for multi-valley DSMs.

\subsection{Electron-electron interaction and broken symmetry phases}\label{Dirac-BSP}

Next we focus on the collection of Dirac fermions interacting via short-ranged interactions. Interaction can be considered to be short-ranged if it vanishes for finite wave-vector and the least irrelevant interaction term is comprised of four-fermions that is local in space and time (imaginary). The interacting Lagrangian, compatible with various discrete symmetries (${\mathcal P}, {\mathcal C}, {\mathcal T}$) and preserves the rotational symmetry takes the form
\begin{eqnarray}\label{Lint3D}
L_{int} &=& g_1 \left( \bar{\Psi} \gamma_0 \Psi \right)^2 + g_2 \left( \bar{\Psi} \Psi \right)^2 + g_3 \left( \bar{\Psi} \gamma_0 \gamma_j \Psi \right)^2 \nonumber \\
&+& g_4 \left( \bar{\Psi} \gamma_0 \gamma_5 \Psi \right)^2 + g_5 \left( \bar{\Psi} i \gamma_5 \Psi \right)^2 +g_6 \left( \bar{\Psi} i \gamma_l \gamma_k \Psi \right)^2 \nonumber \\
&+& g_7 \left( \bar{\Psi} \gamma_5 \gamma_j \Psi \right)^2 + g_8 \left( \bar{\Psi} i \gamma_j \Psi \right)^2.
\end{eqnarray}
The total Lagrangian density is $L_t=L_0-L_{int}$, and therefore in this notation $g_j >0$ represents \emph{repulsive} interactions. Such a four-fermion interaction vertex is represented by the Feynman diagram ($i$) in Fig.~\ref{Feynman-Diag}. The strength of the four-fermion coupling constants depends on the short-ranged part of the Coulomb interaction which relies on various nonuniversal details (e.g., lattice structure, atomic orbitals constituting the DSM) of the system. Instead of delving into the microscopic details of coupling constants ($g_1-g_8$), we here study the low energy properties of the interacting model, defined in Eq.~(\ref{Lint3D}), using the RG method~\cite{coulomb-ref, long-range-explanation-1, long-range-explanation-2, semenoff, sankar-twoloop}. In Appendix~\ref{coulomb-long-range}, we account for the interplay between long-range and short-range Coulomb interaction in DSM.

The DOS scales as $\varrho(E) \sim E^{d/z-1}$, where $z$ is the DCE and $d$ is the dimensionality of the system. Therefore, in a three dimensional DSM ($d=3, z=1$) the DOS vanishes as $\varrho(E) \sim E^2$. Consequently, any weak short-range electron-electron interaction ($g_i$'s) is an \emph{irrelevant} perturbation near the non-interacting stable Gaussian fixed point, since its scaling dimension $[g]=z-d=-2$. Nevertheless, beyond a critical strength, when interaction can no longer be considered weak, Dirac fermions can be driven out of the semimetallic phase through a continuous QPT, leading to various BSP. At zero temperature, a BSP optimally lowers the free-energy by opening up a \emph{mass} gap at the Dirac point. Various possible order parameters or ``masses" that can develop in the strong coupling (repulsive) phase are the followings:

\begin{enumerate}
\item{ $\langle \Psi^\dagger \gamma_0 \Psi\rangle = m_1$, which breaks the continuous chiral $U_c(1)$ symmetry generated by $\gamma_5$, since $\{\gamma_0,\gamma_5\}=0$, but preserves all discrete symmetries (${\mathcal P},{\mathcal C}, {\mathcal T}$). The spectrum in the order phase is $\pm \sqrt{v^2 k^2 +m^2_1}$. In Fig.~\ref{pd-DiracClean} such an insulating phase is denoted by the \emph{scalar mass} phase.   
}

\item{ $\langle \Psi^\dagger i \gamma_0 \gamma_5 \Psi\rangle=m_2$ that breaks discrete ${\mathcal P}$, ${\mathcal T}$ as well as the continuous chiral $U_c(1)$ symmetry, but preserves ${\mathcal C}$ and ${\mathcal P} {\mathcal T}$ symmetries. Such an ordered phase is characterized by a constant axionic angle $\theta_{ax}= \mbox{sgn}(m_2) \frac{\pi}{2}$, and the spectrum of massive Dirac femrions is given by $\pm \sqrt{v^2 k^2 +m^2_2}$. In Fig.~\ref{pd-DiracClean} such an insulating phase is denoted by the \emph{pseudo-scalar mass} phase.
}

\item{$\langle \Psi^\dagger (\gamma_0 \cos \theta + i \gamma_0 \gamma_5 \sin \theta) \Psi\rangle= m_1 \cos \theta+ m_2 \sin \theta$, which also lacks the discrete ${\mathcal P}$, ${\mathcal T}$ and the continuous chiral $U_c(1)$ symmetries. In the ordered phase the quasiparticle spectrum assumes the form $\pm \sqrt{v^2 k^2 +m^2_1+ m^2_2}$. In the continuum description, $\theta$ is a continuous variable and the ordered phase is accompanied by a massless Goldstone mode. Such an insulator represents an \emph{axionic} phase of matter (proposed long ago in the context of high-energy physics~\cite{peccei-quinn, weinberg, wilczek}, and more recently in the context of magnetic topological insulators~\cite{vanderbilt-axion, zhang-axion}), and the Goldstone mode is dubbed as \emph{axion}. The axionic angle in this phase is a dynamic variable, and given by $\theta_{ax}=\tan^{-1}\left(m_2/m_1\right)$. In Fig.~\ref{pd-DiracClean} such an insulating phase is denoted by the \emph{axion} phase.  
}
\end{enumerate}

The physical nature of various insulating phases depends on the microscopic details. In the standard representation of Dirac matrices $\gamma_j = \tau_3 \otimes \sigma_j$ for $j=1,2,3$, $\gamma_0=\tau_2 \otimes \sigma_0$, and $\gamma_5=\tau_1 \otimes \sigma_0$~\cite{peskin}, where $\tau$ and $\sigma$ are two sets of Pauli matrices operating on parity and spin indices, respectively. Therefore, two masses for the Dirac fermion, represented by $\gamma_0=\tau_2 \otimes \sigma_0$ and $i \gamma_0 \gamma_5=\tau_3 \otimes \sigma_0$ correspond to CDW ordering. On the other hand, in Dirac systems like Bi$_2$Se$_3$, the representation of the $\gamma$ matrices is the following: $\gamma_1=\tau_2 \otimes \sigma_2$, $\gamma_2=\tau_2 \otimes \sigma_1$, $\gamma_3=\tau_1 \otimes \sigma_0$, $\gamma_0=\tau_3 \otimes \sigma_0$, $\gamma_5=\tau_2 \otimes \sigma_3$~\cite{model-TI}. Therefore, in this representation the scalar mass ($\gamma_0=\tau_3 \otimes \sigma_0$) represents a CDW, while the pseudo-scalar mass ($i \gamma_0 \gamma_5= \tau_1 \otimes \sigma_3$) corresponds to a SDW, and CDW and SDW orders coexist in the axionic insulating phase. Although the precise nature of the various interaction-driven BSPs can only be ascertained through microscopic calculations, all three BSPs in Fig.~\ref{Dirac-PD} represent some density-wave ordering, with spectral gaps opening up at the Dirac point.
 
In the continuum limit, $\theta_{ax}$ in the axionic insulating phase is a continuous variable, and thus the ordered phase can support \emph{line vortices}, which accommodate one dimensional gapless modes along its core. The one dimensional modes carry nondissipative current, which in turn is radially supplied from the bulk, according to the \emph{Callan-Harvey mechanism}~\cite{callan}.

Also, if the net electron-electron interaction acquires a strong attractive component, fermions can pair into various superconducting ground states. There are two condidates for fully gaped time-reversal symmetric superconducting ground states available for three dimensional Dirac fermions to condense into: (i) regular (topologically trivial) $s$-wave, and (ii) parity-odd topological superconductor~\cite{ohsaku, fuberg, GR-fieldtheory}. It is also conceivable to realize an \emph{axionic superconductor} with $p+is$ symmetry when the strengths of the pairing interactions in these two channels are comparable~\cite{axion-SC}. However, we do not address superconducting instabilities of DSM in this work, and restrict ourselves with repulsive interaction.

\subsection{Disorder}

\begin{table}[h]
  \begin{tabular}{|c||c|c|c|c|c|}
     \hline
\mbox{Bilinear} & ${\mathcal P}$ & ${\mathcal C}$ & ${\mathcal T}$ & $U_c$ & \mbox{Disorder average} \\
     \hline \hline
$\bar{\Psi} \gamma_0 \Psi$ & + & - & + & $\checkmark$ & $\langle \langle V_0 (\mathbf{x}) V_0 (\mathbf{x}') \rangle \rangle= \Delta_V \delta^{3}(\mathbf{x}-\mathbf{x}')$ \\
      \hline
$\bar{\Psi} \gamma_0 \gamma_5 \Psi$ & - & + & + & $\checkmark$ & $\langle \langle V_{A} (\mathbf{x}) V_{A} (\mathbf{x}') \rangle \rangle= \Delta_{A} \delta^{3}(\mathbf{x}-\mathbf{x}')$ \\
      \hline
$\bar{\Psi} \gamma_0 \mathbf{\gamma} \Psi$ & - & - & + & $\times$ & $\langle \langle V_{0i} (\mathbf{x}) V_{0j} (\mathbf{x}') \rangle \rangle= \Delta_{SO} \delta_{ij} \delta^{3}(\mathbf{x}-\mathbf{x}')$ \\
      \hline
  \end{tabular}
\caption{ First column represents various time-reversal symmetric disorder bilinears. Second, third, and forth column show their transformation properties under various discrete symmetries. $+,-$ correspond to even and odd. Transformation of disorder vertices under continuous chiral symmetry is shown in the fifth column. Double angular brackets in the right most column stand for disorder average with respect to Gaussian white noise distribution with zero mean~\cite{goswami-chakravarty}.}\label{table-1}
\end{table}

In this work, we also wish to address the stability of DSMs in the presence of quenched disorder. We here consider only the time-reversal symmetric disorders. All together, there are \emph{four} such candidates and the corresponding imaginary time (Euclidean) action is $S_D=\int d^3 x d \tau \; L_{d}$, where
\begin{eqnarray}
L_{d} = \bar{\Psi} \gamma_0 \big[ V_0(\mathbf{x}) + M(\mathbf{x}) \gamma_0 + V_{A}(\mathbf{x}) \gamma_5
+ V_{0j}(\mathbf{x}) \gamma_j \big] \Psi.
\end{eqnarray}
The physical meaning of various disorder vertices is representation dependent. In the chosen basis, $V_0(\mathbf{x})$ and $M (\mathbf{x})$ respectively represent random charge and mass scatterers. Strengths of random axial chemical potential and spin-orbit disorders are given by $V_{A}$ and $V_{0j}$, respectively. But, in this work we do not consider the mass disorder. Transformation properties of the disorder bilinears are summarized in Table~\ref{table-1}.

\begin{figure}[htb]
\includegraphics[width=8.2cm,height=7.0cm]{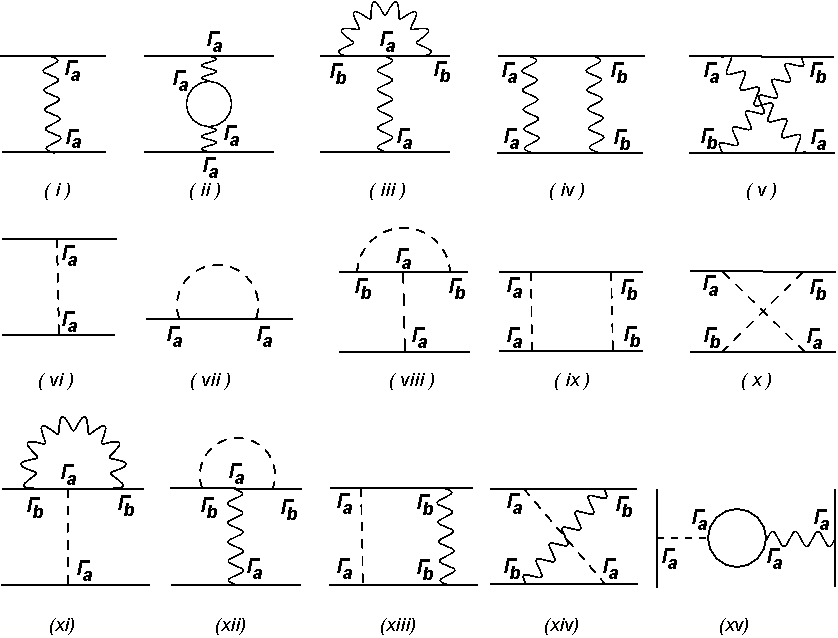}
\caption{First diagrams in first and second rows represent bare interaction and disorder vertices, respectively. Second diagram in central row accounts for the fermionic self-energy correction due to disorder. Rest of the diagrams in first and second row respectively give rise to renormalization of interaction and disorder couplings at one loop level. Diagrams shown in the third row capture the interplay of interaction and disorder at one-loop level. Here $\Gamma_a, \Gamma_b$ are four-dimensional matrices. All diagrams give rise to ultraviolet divergent contributions. }\label{Feynman-Diag}
\end{figure}

After performing the disorder average (assuming Gaussian white noise distributions with zero mean) for all disorder couplings, we obtain the replicated action
\begin{eqnarray}\label{disorder-action}
 \bar{S}_D = \int d^3x d\tau \: \bar{\Psi}_\alpha \left[\gamma_0 \partial_0 + v \gamma_j \partial_j \right] \Psi_\alpha
-\frac{1}{2} \int d^3x d\tau d\tau' \nonumber \\
\times \bigg[ \Delta_V (\bar{\Psi}_\alpha \gamma_0 \Psi_\alpha)_{x} (\bar{\Psi}_\beta \gamma_0 \Psi_\beta)_{x'} 
+ \Delta_{A} (\bar{\Psi}_\alpha \gamma_0 \gamma_5 \Psi_\alpha)_{x} \nonumber \\
\times (\bar{\Psi}_\beta \gamma_0 \gamma_5 \Psi_\beta)_{x'}
+\Delta_{SO} (\bar{\Psi}_\alpha \gamma_0 \gamma_j \Psi_\alpha)_{x}
 (\bar{\Psi}_\beta \gamma_0 \gamma_j \Psi_\beta)_{x'}  \bigg],
\end{eqnarray}
where $\alpha, \beta$ are the replica indices, $x \equiv (\mathbf{x},\tau)$ and $x' \equiv (\mathbf{x},\tau')$. The disorder vertex in the replicated theory is represented by the Feyman diagram $(vi)$ in Fig.~\ref{Feynman-Diag}.

Potential and axial disorders share an interesting symmetry. Both of them locally shift the chemical potential for left and right chiral fermions, while maintaining the overall charge-neutrality of the system. Potential disorder equally shifts the local chemical potential for left and right chiral fermions, while such shifts are of opposite sign for fermions with opposite chirality in the presence of axial disorder. Therefore, in the absence of any CSB perturbations, when left and right chiral worlds are decoupled, these two disorders are expected to behave identically, as we demonstrate explicitly in Sec.~\ref{dirty-DSM}.

The scaling dimension of any disorder couplings ($\Delta$) is $[\Delta]=2 z-d$, and hence with $z=1,d=3$, $[\Delta]=-1$. Therefore, weak quenched disorder is an \emph{irrelevant} perturbation in three dimensional DSMs. However, at stronger disorder DSM can undergo a QPT and enter into a CDM phase, where the DOS at zero energy becomes finite~\cite{fradkin, goswami-chakravarty, roy-dassarma, herbut-Imura, radziovsky, ominato, pixley}. Therefore, three dimensional DSMs offer a unique opportunity to study the interplay of interaction and disorder within the framework of an weak coupling RG analysis, at least when they are not too strong.

\section{Clean interacting system}\label{clean-interacting}

Let us first analyze the clean interacting system. As shown in Eq.~(\ref{Lint3D}) the interacting model, comprised of short-ranged electron-electron interactions in the three dimensional DSM, is described by eight coupling constants. However, not all of them are linearly independent. There exists a mathematical constraint, known as the \emph{Fierz indentity}~\cite{HJR} that restricts the number of linearly independent couplings to four and allows us to rewrite the remaining four-fermion terms as linear combinations of the independent ones (see Appendix~\ref{Append-fierz}). For convenience, we choose $g_1$, $g_2$, $g_4$ and $g_5$ as independent coupling constants. The interacting Lagrangian density then becomes
\begin{eqnarray}\label{int-reduced}
L_{int} &=& g_1 \left( \bar{\Psi} \gamma_0 \Psi \right)^2 + g_2 \left( \bar{\Psi} \Psi \right)^2+ g_4 \left( \bar{\Psi} \gamma_0 \gamma_5 \Psi \right)^2 \nonumber \\
&+& g_5 \left( \bar{\Psi} i \gamma_5 \Psi \right)^2.
\end{eqnarray}
Rest of the four quartic couplings, namely $g_3$, $g_6$, $g_7$, $g_8$, can be expressed as linear combinations of $g_1$, $g_2$, $g_4$ and $g_5$, as shown in Eq.~(\ref{constraintcouplings}) of Appendix~\ref{Append-fierz}. Notice that upon setting $g_1=g_4=0$ and $g_2=g_5$ we recover the celebrated Nambu-Jona-Lasinio model for the dynamic chiral symmetry breaking in particle physics~\cite{NJL}. However, to close the RG flow equations we need to account for two additional coupling constants $g_1$ and $g_4$. These two coupling constants, as we will show in a moment, only shift the locations of various QCPs, without altering the universality class of the transition or the nature of the BSPs at strong couplings.

Next we coarse-grain the interacting theory and compute the effective action to the quadratic order in terms of the four coupling constants ($g_{1,2,4,5}$). The relevant Feynman diagrams are shown in Fig.~\ref{Feynman-Diag}~$(ii)-(v)$. During this procedure, when we generate contact terms that are proportional to $g_{3,6,7,8}$, they are rewritten in terms of the original couplings by using Eq.~(\ref{constraintcouplings}). Therefore, the interacting theory [see Eq.~(\ref{int-reduced})] remains closed under the RG procedure to any order in perturbation theory. We integrating out the fast Fourier modes with $-\infty<\omega<\infty$ and $\Lambda e^{-l}<|\vec{k}|<\Lambda$ and rescale $\tau \to \tau e^{zl}$, $x \to x e^l$, and $\Psi \to \Psi e^{-dl/2}$ for casting the effective action into the original form. After defining the dimensionless couplings according to $2 g_j\Lambda^{\epsilon_1}S_d/[(2 \pi)^d v] \to g_j$, where $S_d$ is the surface area of $d$-dimensional unit sphere, we arrive at the following RG flow equations
\begin{eqnarray}\label{RGintcoup}
\frac{dg_1}{dl} &=& -\epsilon_1 g_1-\frac{1}{3} \left(g_1 g_2+ g_1 g_5+2 g_2 g_5 \right), \nonumber \\
\frac{dg_2}{dl} &=& -\epsilon_1 g_2 +g^2_2 -\frac{2}{3} \left( g_1 g_2 -g_2 g_5 +g_1 g_5 \right) \nonumber \\
& + & g_4 (g_2-g_5), \nonumber \\
\frac{dg_4}{dl} &=& -\epsilon_1 g_4 + \frac{1}{3} \left( g_1 g_2 + g_1 g_5 -4 g_2 g_5\right), \nonumber \\
\frac{dg_5}{dl} &=& -\epsilon_1 g_5 +g^2_5 -\frac{2}{3} \left(g_1 g_5-g_2 g_5+g_1 g_2 \right) \nonumber \\
&+& g_4 (g_5-g_2),
\end{eqnarray} 
where $\epsilon_1=d-z$. Any infinitesimally weak quartic coupling is an irrelevant perturbations above $d=z$, and the above coarse graining process should be understood as the $\epsilon$-expansion about the lower critical dimension ($d_{l}=z=1$). It is worth pointing out that the above set of flow equations display a symmetry under $g_2 \leftrightarrow g_5$, reflecting the underlying continuous chiral symmetry of massless Dirac fermions.

\subsection{Gross-Neveu model with $g_5$ or $g_2$}

Before analyzing the above set of coupled flow equations, we focus on a simpler model by setting $g_1=g_2=g_4=0$. The interacting model with only one coupling constant $g_5$ conforms to the Gross-Neveu model~\cite{gross-neveu}. The RG flow equation for $g_5$ is then given by 
\begin{equation}\label{g5RG}
\frac{d g_5}{dl} = -\epsilon_1 \;  g_5 + \; g^2_5,
\end{equation}
which exhibits a QCP at $g_5=g^\ast_5=\epsilon_1=d-z$, describing a continuous phase transition between the DSM and a ${\mathcal P}$, ${\mathcal T}$ symmetry breaking insulator. Inside the insulating phase the order parameter $\langle \bar{\Psi} i \gamma_5 \Psi \rangle =m_2 \neq 0$. The CLE for this transition is $\nu =(d-z)^{-1}$. For $d=d_u=z+2$, $\nu$ acquires the mean field value $1/2$, which demonstrates that $d_u=3$ is the upper critical dimension for such a transition in DSM. Due to the upper critical dimensionality, the \emph{hyperscaling} is violated through logarithmic corrections, which can be easily demonstrated by solving the corresponding gap equation (see Eq.~(\ref{gapeq}) below). The CLE $\nu=1/2$ is not an artifact of one loop calculation. From an $\epsilon$-expansion analysis of an appropriate order parameter field theory, known as the \emph{Gross-Neveu-Yukawa} formalism, around the upper critical dimension $d_u=3$, it can be shown that $\nu=1/2$ is an exact result (see Appendix~\ref{Append-GNY})~\cite{zinn-justin, moshe-moshe}.

The logarithmic correction to the mass gap can be obtained from the self-consistent gap equation~\cite{miransky}
\begin{equation}\label{gapeq}
\frac{1}{g_5}= \int \frac{d^3 \vec{k}}{(2 \pi)^3} \: \: \frac{1}{\sqrt{v^2 k^2 + m^2_2}},
\end{equation}
which yields a nonzero solution for $m_2$, when $g_5> g^c_5= \Lambda^2/(4 \pi^2 v^3)$, where $g^c_5$ is the critical strength of the interaction for insulation. In terms of a dimensionless quantity $\delta$, defined as
\begin{equation}\label{delta-def}
\delta=\frac{4 \pi^2 v^3}{\Lambda^2} \left( \frac{1}{g^c_5}-\frac{1}{g_5} \right)
\end{equation}
that measures the deviation from the critical point, the universal scaling of the dimensionless mass gap $\tilde{m}=m_2/(v\Lambda)$ is determined from the gap equation
\begin{equation}
\delta=1-\sqrt{1+\tilde{m}^2}+ \tilde{m}^2 \log \left( \frac{1+\sqrt{1+\tilde{m}^2}}{\tilde{m}} \right).
\end{equation}
The last term in the right hand side captures the logarithmic correction to the scaling of the mass. The above gap equation supports nontrivial solution of $\tilde{m}$ only for $\delta>0$ or $g_5> g^c_5$, i.e., there is no QPT with an emergent spectral gap unless $g_5$ is stronger than a critical strength.

In a similar spirit, we can set all the four-fermion interactions to zero except $g_2$. The flow equation of this model is given by Eq.~(\ref{g5RG}) after taking $g_5 \to g_2$. The QCP of this model is the placed at $g_2=g^\ast_2=\epsilon_1$, which describes a continuous transition out of DSM into a chiral $U_c(1)$ symmetry breaking fully gapped phase where $\langle \bar{\Psi} \Psi \rangle =m_1 \neq 0$. However, all the discrete symmetries $({\mathcal C}, {\mathcal P}, {\mathcal T})$ are preserved in the ordered phase. BSPs with finite $m_1$ and $m_2$ are respectively our scalar and pseudo-scalar gapped insulating quantum phases of Fig.~\ref{pd-DiracClean}.

\subsection{Generic interacting model}

We now proceed with the analysis of coupled flow equations in Eq.~(\ref{RGintcoup}). Besides the fully stable noninteracting Gaussian fixed point at $(g_1,g_2,g_4,g_5)=(0, 0, 0, 0)$, the above set of flow equations support \emph{four} QCPs, describing continuous transitions from DSM to various BSPs:

C2: $(g_1,g_2,g_4,g_5)=(0.185, -0.455, 0.405, 0.645) \epsilon_1$ dictates the transition to a ${\mathcal P}$, ${\mathcal T}$ symmetry breaking massive phase (pseudo scalar mass), since $g_5$ is the strongest coupling at this QCP. 

C3: $(g_1,g_2,g_4,g_5)=(0.185, 0.645, 0.405, -0.455) \epsilon_1$ describes the QPT into a CSB insulator (scalar mass). Notice, locations of two critical points C2 and C3 display a symmetry under $g_2 \leftrightarrow g_5$, stemming from the underlying continuous chiral symmetry of massless Dirac fermions.

C1: $(g_1,g_2,g_4,g_5)=(-0.125, 0.5, -0.375, 0.5) \epsilon_1$ corresponds to the transition into an insulating phase where $\langle \bar{\Psi} i \gamma_5 \Psi \rangle, \langle \bar{\Psi} \Psi \rangle \neq 0$, since $g_2=g_5$ at this QCP~\cite{multicritical-roy}. The ordered phase breaks ${\mathcal P}$, ${\mathcal T}$ and chiral $U_c(1)$ symmetry, and represents an axionic insulator. The order parameter in the axionic phase reads as $\langle \bar{\Psi} (\cos \theta + i \gamma_5 \sin \theta) \Psi\rangle$.

C4: $(g_1,g_2,g_4,g_5)=(-2,-1,0,-1)\epsilon_1$ is associated with the transition of massless Dirac fermions into the fully gapped $s$-wave superconductor, which can only be accessed when all the interactions are strongly attractive. 

The CLE at all QCPs is $\nu=\epsilon^{-1}_1$. The identification of an ordered phase in the vicinity of a particular QCP is substantiated from the computation of anomalous dimensions as well as the RG flow of susceptibilities of various fermion bilinears. The fermionic bilinear with the largest anomalous dimension develops a finite expectation value at a given QCP. Anomalous dimensions of various order parameters in the vicinity of each QCP and the flow of susceptibility are shown in Appendix~\ref{Append-susceptibility}.

Previously obtained QCPs at $(g_1,g_2,g_4,g_5)=(0,\epsilon_1,0,0)$ and $(0,0,0,\epsilon_1)$ appear as \emph{bicritical} points in the four dimensional coupling constant space. A bicritical point is characterized by two stable and two unstable directions. There exist two additional bicritical points at $(g_1,g_2,g_4,g_5)=$ $(-2.685, -2.52, 1.84, -0.165) \epsilon_1$ and $(-2.685, -0.165, 1.84, -2.52) \epsilon_1$. All bicritical points are symmetric under $g_2 \leftrightarrow g_5$ and they separate the basins of attraction of four QCPs.

\subsection{Summary}

To summarize this section, we argue that the interacting model for three dimensional Dirac fermions is constituted by four linearly independent local quartic interactions. Performing a weak coupling RG analysis, we show that strong enough interactions drive the DSM through continuous phase transitions into various BSPs, where the fermionic excitation spectrum is fully gapped or ``massive". A representative phase diagram of interacting DSM in the $g_2-g_5$ plane is shown in Fig.~\ref{pd-DiracClean}. Such a QPT is mean-field in nature, and various physical observable acquires logarithmic corrections, since the system lives at the upper critical dimension ($d_u=3$) and the hyperscaling hypothesis is violated. The critical exponents near such QCP are $\nu=1/2$ and $z=1$. Therefore, a \emph{pseudo Lorentz symmetry} emerges at each QCP~\cite{roy-lorentz}. These exponents govern the scaling behavior of various physical quantities. For example, the Fermi velocity scales as $v (\delta) \sim v_0 \; \delta^{\nu (z-1)}$, where $\delta$ measures the deviation from the QCP and $v_0$ is the \emph{bare} Fermi velocity. Therefore, with $z=1$, the Fermi velocity remains non-critical across the QPT. The residue of the quasi-particle pole remains finite in the entire semimetallic side of the transition, but vanishes smoothly at the QCP, beyond which gapless fermions cease to exist as sharp quasi-particle excitations, and a well defined energy gap opens up at the Dirac point, making the system an insulator. The critical temperature for DSM-BSP transition scales as $T_c \sim \delta^{\nu z}$.

One physical obstacle to observe such critical phenomena is how to tune the ratio of interaction to bandwidth that can drive the system from the DSM phase to various insulating phases. When the system is placed in a strong magnetic field the conical dispersion quenches into a set of Landau levels at energy $\pm \sqrt{2 n B+ v^2 k^2_z}$. In particular, the zeroth Landau level ($n=0$) is composed of two one-dimensional chiral modes with energies $\pm v k_z$. Hence, weak enough electron-electron interaction can drive the system into an insulating phase and develop a spectral gap ($m$) at the Dirac point due to the effective dimensional reduction of the system within the zeroth Landau level~\cite{catalysis-3D, roy-sau}. If the interaction strength is not too far from the semimetal-insulator QCP, the scaling of the mass gap assumes the form~\cite{roy-catalysis-scaling}
\begin{equation}
\frac{m}{v \Lambda} = \left( \frac{a}{l_B} \right)^z \: G \left(l_B \delta^{\nu} \right),
\end{equation}
where $l_B$ is the magnetic length and $\Lambda$ is the ultraviolet cut-off for Dirac dispersion, and $\delta$ measures the deviation from zero field QCP. Right at the semimetal-insulator quantum critical point the gap scales as~\cite{roy-sau}
\begin{equation}
m \approx v \sqrt{B} \left[ G(0) + c \log \left( \frac{B_0}{B} \right) \right],
\end{equation}
where $G(0)$ and $c$ are universal numbers and $B_0 \sim \Lambda^2$. The logarithmic correction in the scaling arises since the system lives at the upper critical dimension $d_u=3$.

\begin{figure}[htb]
\includegraphics[width=7.5cm,height=5.0cm]{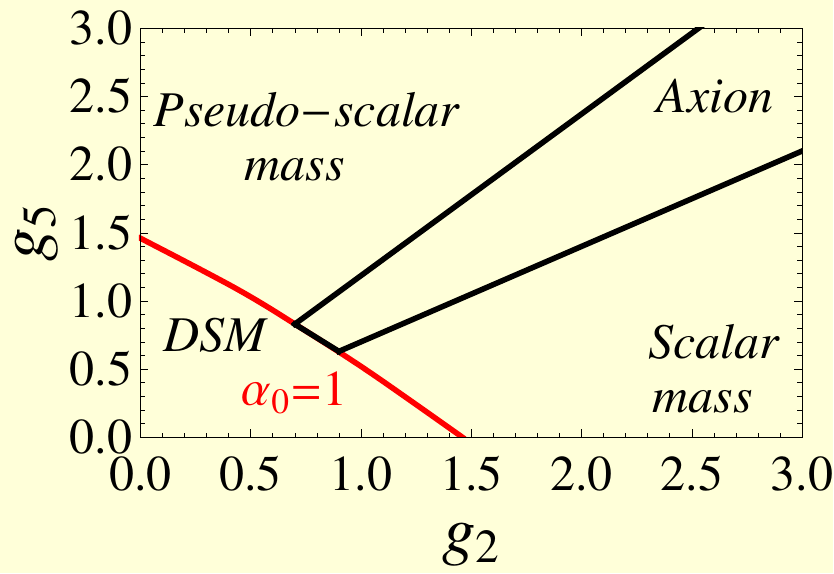}
\caption{Phase diagram of clean interacting DSM in $g_2-g_5$ plane, in the presence of long range Coulomb interaction. The strength of Coulomb interaction is set by the fine structure constant $\alpha=e^2/(4 \pi \varepsilon v)$, where $e$ is electronic charge, and $\varepsilon$ is the dielectric constant of the medium.  Here, the bare value of the fine structure constant is $\alpha_0=1$, see Appendix~\ref{coulomb-long-range}.}\label{pd-DiracCoulomb}
\end{figure}

So far we have considered only the short range components of the Coulomb interaction, and address the emergent quantum critical phenomena in three dimensional DSMs. The long range Coulomb interaction is a \emph{marginally irrelevant} perturbation in DSMs, and in its presence the fine structure constant of the medium decreases monotonically, but the Fermi velocity increases logarithmically~\cite{goswami-chakravarty, isobe-nagaosa, hosur, throckmorton-3d}. As shown in Appendix ~\ref{coulomb-long-range} that long range tail of the Coulomb interaction enhances the ordering tendency (insulation) in DSM, without altering the quantum critical behavior and the nature of BSPs, captured by the model composed of only short range interactions, and a representative phase diagram is shown in Fig.~\ref{pd-DiracCoulomb}. Comparing Figs.~\ref{pd-DiracClean} and \ref{pd-DiracCoulomb}, we find that the presence of long range Coulomb interaction shifts the phase boundaries between the DSM and various BSPs toward \emph{weaker} couplings. Thus by tuning the strength of the dielectric constant of the medium (since strength of bare Coulomb interaction is inversely proportional to the dielectric constant of the medium) one can drive DSM through QPTs and place it into various BSPs.

\section{Non-interacting dirty system}\label{dirty-DSM}

Next we focus on a noninteracting dirty DSM in the presence of various types of time-reversal-symmetric disorder. We take into account (a) CSP disorder, such as the regular potential disorder ($\Delta_V$) and axial disorder ($\Delta_{A}$), and (b) CSB disorder, such as random spin-orbit coupling ($\Delta_{SO}$) (but we do not consider mass disorder, which also breaks chiral symmetry). A Similar question was previously addressed in the context of the QPT between three dimensional topological and normal insulators, belonging to class AII~\cite{goswami-chakravarty}. Some of our results reconcile with the ones reported in Ref.~\cite{goswami-chakravarty}, after setting the band gap and $b$ (the momentum dependent chiral symmetry breaking Wilson mass) to zero. However, a classification (based on the chiral symmetry) of the disorder driven QPT in DSM is presently unavailable and constitutes the central theme of this section. In addition, this exercise provides a pedagogical introduction to the following section (see Sec.~\ref{interaction-disorder}), where we analyze the interplay of interaction and disorder.

\subsection{Chiral symmetric disorder}

\begin{figure}[htbp]
\centering
\subfigure[]{
\includegraphics[width=4.05cm,height=3.5cm]{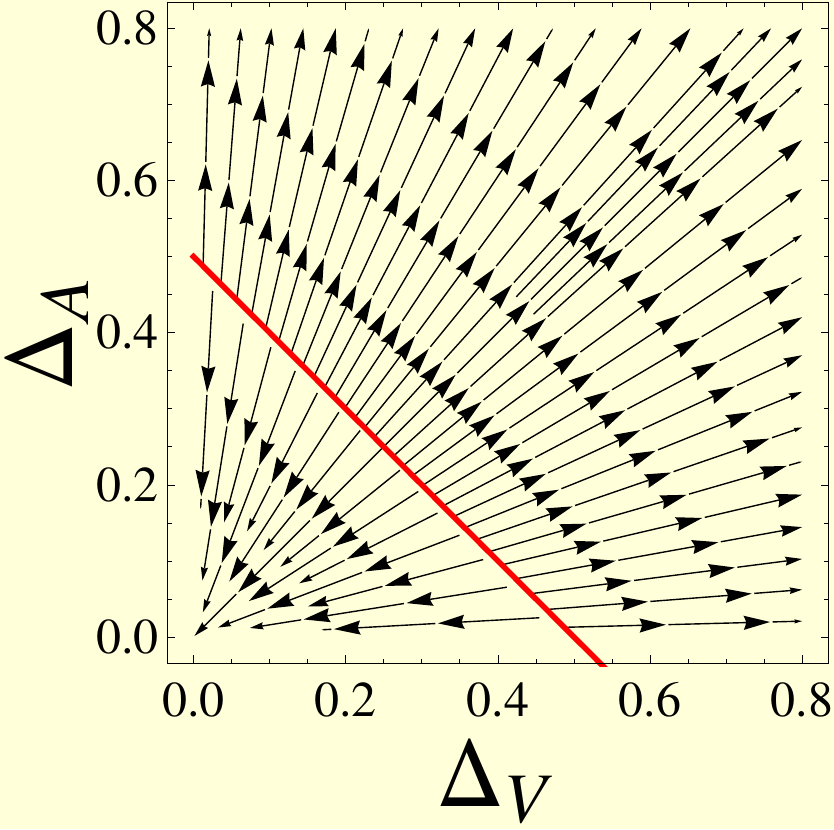}
\label{pot-oddP-dis-a}
}
\subfigure[]{
\includegraphics[width=4.05cm,height=3.5cm]{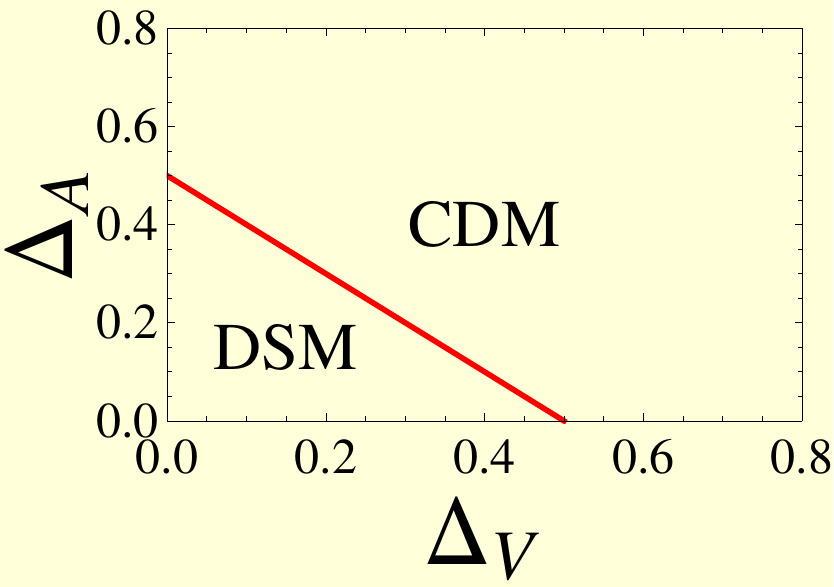}
\label{pot-oddP-dis-b}
}
\label{pot-oddP-dis}\caption[]{(a) RG flow and (b) phase diagram of disordered DSM in $\Delta_V-\Delta_{A}$ plane. The axes are measured in units of $\epsilon_2$. The red lines in figures (a) and (b) describe a line of QCPs and the phase boundary between DSM and CDM, respectively~\cite{goswami-chakravarty}. }
\end{figure}

First we focus on CSP disorder. The bare scaling dimension of disorder coupling $[\Delta_V]=[\Delta_A]=2z-d=-\epsilon_2$, where $z=1$ (for clean DSM) and $\epsilon_2=-1$ (setting $d=3$), dictates that DSM describes an infrared stable fixed point for sufficiently weak randomness. To capture a possible QPT beyond a threshold of disorder strength, we perform a perturbative RG calculation. The relevant Feynman diagrams to one loop order are shown in Fig.~\ref{Feynman-Diag} [$(vii)-(x)$]. The RG flow equations for the Fermi velocity and the disorder couplings are 
\begin{eqnarray} \label{dirtyflow}
\frac{d v}{d l} &=& v\left[z-1-\Delta_V-\Delta_{A} \right], \nonumber \\
\frac{d \Delta_V}{d l} &=&  -\epsilon_2 \Delta_V +2 \Delta^2_V +2\Delta_{A} \Delta_V, \nonumber \\
\frac{d \Delta_{A}}{d l} &=& -\epsilon_2 \Delta_{A} +2 \Delta_V \Delta_{A} +2\Delta^2_{A},
\end{eqnarray}
after integrating out the fast Fourier modes within the shell $\Lambda e^{-l}<|\vec{k}|<\Lambda$. We here define the dimensionless disorder couplings as $\Delta_V \Lambda^{\epsilon_2} S_d/[(2 \pi)^d v^2] \to \Delta_V$ and $\Delta_{A} \Lambda^{\epsilon_2}S_d/[(2 \pi)^d v^2] \to \Delta_{A}$. Keeping the Fermi velocity invariant under RG ($dv/dl=0$) we obtain a scale dependent DCE
\begin{equation}
z(l)=1+\Delta_V(l)+\Delta_{05}(l).
\end{equation}

If the system hosts only potential disorder, there exists a disorder controlled QCP (CV) at $\Delta_V=\Delta^\ast_V=\frac{\epsilon_2}{2}$, which describes a continuous QPT out of the DSM to a CDM. Within the one-loop RG calculation, the DCE and the CLE at this QCP (CV) are $z=1+\frac{\epsilon_2}{2}$ and $\nu=\epsilon^{-1}_2$, respectively~\cite{goswami-chakravarty, radziovsky, roy-dassarma}. Such a critical point can be seen in Fig.~\ref{pot-oddP-dis-a} on the $\Delta_{V}$-axis. If, on the other hand, DSM hosts only axial disorder ($\Delta_V=0$), a similar QPT takes place at $\Delta_{A}=\Delta^\ast_{A}=\frac{\epsilon_2}{2}$ (CA) [see $\Delta_A$-axis of Fig.~\ref{pot-oddP-dis-a}]. The critical exponents near this QCP (CA) are also $z=1+\frac{\epsilon_2}{2}$, and $\nu=\epsilon^{-1}_2$~\cite{goswami-chakravarty, roy-dassarma}.

In the presence of both potential and axial disorder, there exists a line of QCPs in the $\Delta_V-\Delta_{A}$ plane, determined by $\Delta^\ast_V+\Delta^\ast_{05}=\frac{\epsilon_2}{2}$ that also defines the phase boundary between clean the DSM and the CDM. The DCE along the entire line of critical points is $z=1+(\Delta^\ast_V+\Delta^\ast_{05})=1+\frac{\epsilon_2}{2}$ and the CLE $\nu=\epsilon^{-1}_2$. The RG flow and the corresponding phase diagram of dirty DSM in the $\Delta_V-\Delta_A$ plane are shown in Fig.~\ref{pot-oddP-dis-b}~\cite{goswami-chakravarty}.

The critical exponents ($\nu$ and $z$) along the entire line of critical points in the $\Delta_V-\Delta_A$ plane are identical. This intriguing outcome can be understood in the following way. Notice that the Clifford algebra (commuting or anticommuting) between the non-Hermitian elliptic Dirac Kernel ${\cal K}_D = \gamma_\mu \partial_\mu$ and the two matrices appearing at the CSP disorder vertices, namely $\gamma_0$ and $\gamma_0 \gamma_5$, are opposite to each other, but $(\gamma_0)^2=1$ and $(\gamma_0 \gamma_5)^2=-1$. In addition, no new disorder coupling gets generated through the loop corrections to any order in perturbation theory~\cite{roy-dassarma, comment-ladder-crossing}. Consequently, the diagramatic contributions from Fig.~\ref{Feynman-Diag} [$(vii)-(x)$] are identical for $\Gamma_a=\gamma_0$ and $\gamma_0 \gamma_5$, yielding a set of identical critical exponents along the entire line of QCPs. Such a result remains valid to all orders in the perturbation theory and a recent numerical analysis strongly supports this observation for two extreme limits, when DSM-CDM transition is tuned by (a) potential disorder, and $\Delta_A=0$, and (b) axial disorder, and $\Delta_V=0$~\cite{pixley}.

The above result can also be explained from a somewhat different perspective. In the absence of any CSB perturbation, the massless Dirac Hamiltonian can be decomposed into two isolated worlds of left and right chiral fermions. In each of these two disjoint sectors, $\Delta_{A}$ and $\Delta_V$ appear as regular potential disorder, but the former one carries a relative sign between them. Consequently, in the absence of any CSB perturbation, the set of critical exponents is identical along the entire line of QCPs in the $\Delta_V-\Delta_{A}$ plane.

Even though the critical exponents are identical at each point on the line of QCPs in $\Delta_V-\Delta_A$ plane, the anomalous dimension of a CSB fermionic bilinears changes continuously along the line of QCPs. For the purpose of demonstration, we here compute anomalous dimensions of the scalar mass ($m_1$) and the ${\mathcal P}, {\mathcal T}$-odd pseudo scalar mass ($m_2$), given by 
\begin{equation}\label{anomalousdim}
A_{m_1}= \left(\Delta_A-\Delta_V \right) \frac{\Lambda^{\epsilon_2} S_d}{(2 \pi)^d v^2}=A_{m_2}.
\end{equation} 
Therefore, $A_{m_1}/A_{m_2}$ varies continuously from $-1$ (on the $\Delta_V$-axis) to $+1$ (on the $\Delta_A$-axis) (in units of $\Lambda^{\epsilon_2} S_d/[(2 \pi)^d v^2]$) along the line of QCPs.

\subsection{Chiral symmetry breaking disorder}\label{CSB-disorder-dirty}

\begin{figure}[htbp]
\centering
\subfigure[]{
\includegraphics[width=4.05cm,height=3.5cm]{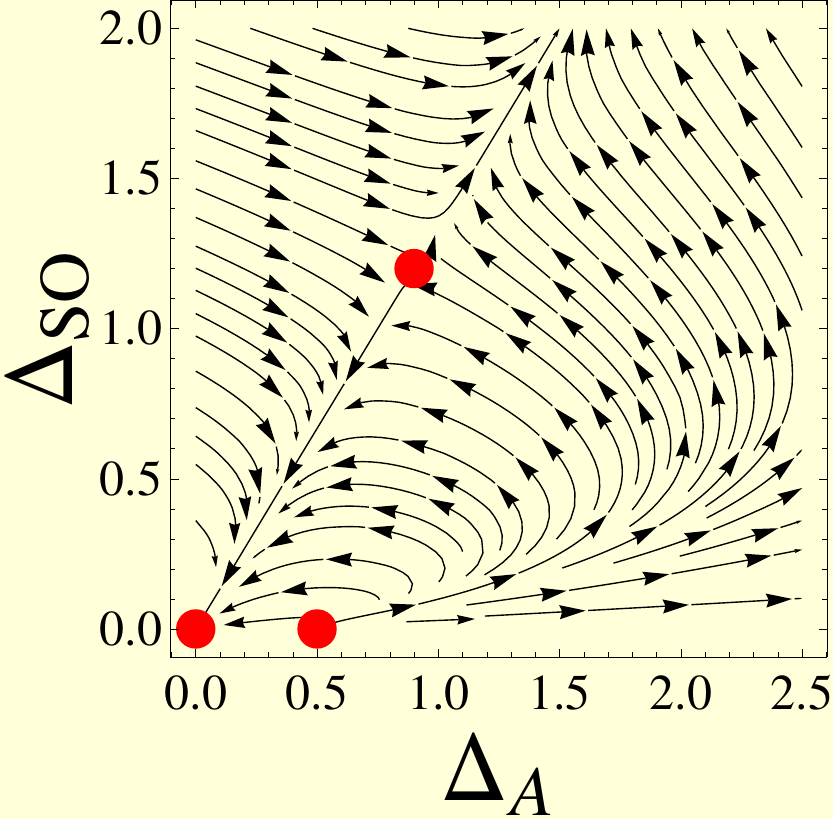}
\label{spinorbit-dis-a}
}
\subfigure[]{
\includegraphics[width=4.05cm,height=3.5cm]{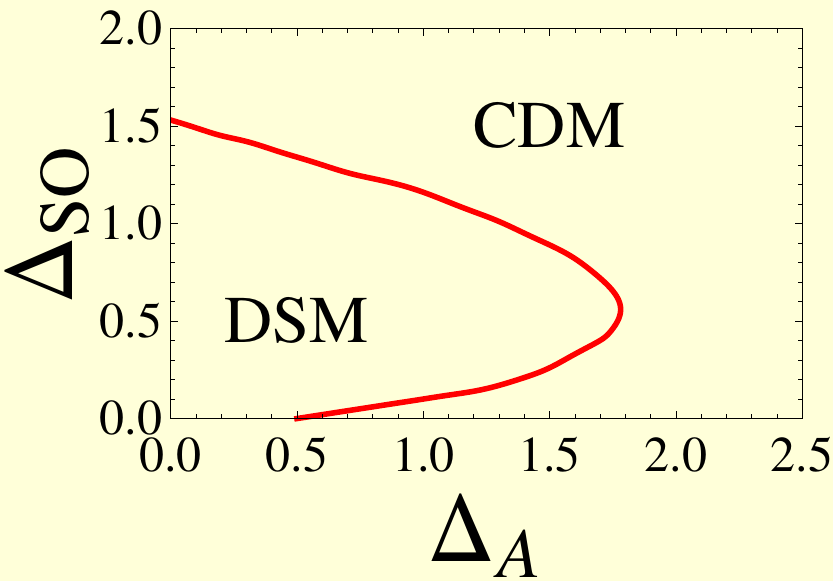}
\label{spinorbit-dis-b}
}
\label{spinorbit-dis}\caption[]{(a) RG flow and (b) phase diagram of disordered DSM in $\Delta_A-\Delta_{SO}$ plane. Notice that the phase boundary between DSM and CDM is determined by the irrelevant direction at the QCP~\cite{goswami-chakravarty}. Here, the axes are in units of $\epsilon_2$. }
\end{figure}

Next we seek to understand the effect of CSB disorders on DSM. Since we restrict ourselves to time-reversal symmetric disorder, the only candidate that breaks the chiral symmetry is spin-orbit disorder ($\Delta_{SO}$) (recall we here do not consider the mass disorder that also breaks chiral symmetry). The flow equations of various coupling constants to one-loop order are given by 
\begin{eqnarray}\label{CSB-disorder}
\frac{dv}{dl} &=& v(z-1-\Delta_{A}-3\Delta_{SO}), \nonumber \\
\frac{d \Delta_{SO}}{dl} &=& -\epsilon_2 \Delta_{SO}-\frac{2}{3} \Delta^2_{SO} + 2 \Delta_{SO} \Delta_{A}, \nonumber \\
\frac{d \Delta_{A}}{dl} &=& -\epsilon_2 \Delta_{A} +2 \Delta^2_{A}-6 \Delta_{A} \Delta_{SO} + 4 \Delta^2_{SO}.
\end{eqnarray} 
It is worth pointing out that the RG calculation is not closed with $\Delta_{SO}$. Through loop corrections (from diagrams $(ix)$ and $(x)$ in Fig.~\ref{Feynman-Diag}) the spin-orbit disorder generates axial disorder (notice that a term proportional to $\Delta^2_{SO}$ appears in $d \Delta_{A}/dl$). Hence, to close the RG equations we need to account for two disorder couplings $\Delta_{A}$ and $\Delta_{SO}$. Otherwise, keeping the Fermi velocity fixed under RG ($dv/dl=0$) we obtain a scale dependent DCE 
\begin{equation}
z(l)=1+\Delta_{05}(l) +3 \Delta_{SO}(l).
\end{equation}

The above set of flow equations supports three fixed points as shown in Fig.~\ref{spinorbit-dis-a} (left). (i) $(\Delta_{A},\Delta_{SO})=(0,0)$ describing the stable DSM, (ii) a fully unstable fixed point at $(\Delta_{A},\Delta_{SO})=(\frac{\epsilon_2}{2},0)$, (iii) a QCP at $(\Delta_{A},\Delta_{SO})=(\frac{9}{10},\frac{6}{5})\epsilon_2$. We note that the QCP residing on the chiral symmetric axis ($\Delta_{SO}=0$), becomes unstable in the presence of infinitesimal CSB disorder. A new critical point emerges from the competition between two types of disorder, characterized by DCE $z=1+ \frac{9}{2} \epsilon_2$ and CLE $\nu=\epsilon^{-1}_2$. The phase diagram of dirty DSM in the presence of CSB disorder is shown in Fig.~\ref{spinorbit-dis-b}.

The phase diagram in Fig.~\ref{spinorbit-dis-b} suggests an interesting possibility in the dirty noninteracting DSM subject to axial and spin-orbit disorder. When the strength of the axial disorder is such that $\Delta^\ast_A(=\frac{\epsilon_2}{2})<\Delta_A<1.75 \epsilon_2$, and one tunes the spin-orbit disorder, there is a very interesting re-entrant QPT with CDM-DSM-CDM phases showing up with increasing $\Delta_{SO}$ at fixed $\Delta_A$. Across the CDM-DSM and DSM-CDM phase transitions, the average DOS diverges with a unique power law dependence $\varrho(E) \sim |E|^{-5/11}$, as both of them are controlled by the QCP, located at $(\Delta_A, \Delta_{SO})=(\frac{9}{10}, \frac{6}{5}) \epsilon_2$, and the phase boundary between DSM and CDM in the entire plane is determined by the \emph{irrelevant} direction at the QCP.

\subsection{Scaling of physical observable}

Ballistic quasiparticles survive the onslaught of sufficiently weak but generic disorder (chiral symmetry preserving and breaking) in DSM. However, beyond a critical strength of disorder, DSM undergoes a continuous QPT and enters into a diffusive metallic phase. Subject to random impurities (time-reversal-symmetric) DSM can support two types QCPs, belonging to different universality classes (defined in terms of critical exponents $z$ and $\nu$): (i) In the presence of CSP disorder ($\Delta_V$ and $\Delta_A$) the QCPs and the line of QCPs are characterized by the exponents $z=1+\frac{\epsilon_2}{2}$ and $\nu=\epsilon^{-1}_2$; (ii) at the CSB disorder (such as $\Delta_{SO}$) driven DSM-CDM QPT the exponents are $z=1+\frac{9}{2} \epsilon_2$ and $\nu=\epsilon^{-1}_2$. These exponents control the scaling of various physical observables across the disorder driven DSM-CDM QPT.

In the metallic phase, the average DOS at zero energy becomes finite, which, therefore, serves the purpose of an order-parameter across the DSM-CDM QPT. The average density of states follows the scaling ansatz~\cite{herbut-Imura}
\begin{equation}
\varrho(E) =\delta^{\nu(d-z)} \: F(|E| \delta^{-\nu z}),
\end{equation} 
for energies much smaller than the bandwidth ($E \ll v \Lambda$). Here $F$ is an unknown, but universal scaling function. In the DSM phase $\varrho(E) \sim |E|^2$, whereas in the quantum critical regime $\varrho(E) \sim |E|^{d/z-1}$. In the metallic phase $\varrho(0)$ becomes finite. Therefore, near the CSP disorder driven QCP $\varrho(E)$ \emph{vanishes} according to $ |E|$, while in the close vicinity of CSB disorder driven QCP $\varrho(E)$ \emph{diverges} as $|E|^{-5/11}$ (setting $\epsilon_2=1$ for three dimensional DSM).

In addition, the quasiparticle lifetime, the mean-free path, and the metallic conductivity at $T=0$ are respectively finite and zero in metallic and semimetallic phases. Hence, at least in principle, these quantities may as well be considered as order parameters across the disorder driven QPT. As the dirty QCP is approached from the DSM side the residue of the quasiparticle pole vanishes smoothly, beyond which Dirac fermions cease to exist as sharp quasiparticles.

In the vicinity of DSM-CDM QCP, the specific heat ($C_v$) assumes the scaling form
\begin{equation}
C_v=T^{d/z} v^{-3} H \left( \frac{T}{\delta^{\nu z}}\right),
\end{equation}
when the temperature is much smaller than the bandwidth ($T \ll v \Lambda$), where $H$ is an unknown, but universal scaling function. In the Dirac semimetallic phase ($x \ll 1 $), $H(x) \sim x^{d(z-1)/z}$ and we recover $T^3$ dependence of the specific heat. In the quantum critical regime $H(x)$ is a universal function (independent of $\delta$) and $C_v \sim T^{d/z}$. On the other hand, in the metallic phase $H(x) \sim x^{1-d/z}$, yielding $C_v \sim T$. Within the one loop calculation, $C_v \sim T^2$ (setting $\epsilon_2=1$) within the critical regime of CSP disorder driven QCPs. Distinct power law behaviors and crossover between them have recently been established in a numerical work, which has further been exploited to estimate the extent of the quantum critical regime at finite temperatures~\cite{pixley}. On the other hand, $C_v \sim T^{6/11}$(setting $\epsilon_2=1$) in the proximity of CSB disorder driven DSM-CDM QCP, which, however, remains to be observed in numerics.

The (frequency dependent) optical conductivity at $T=0$ in a dirty DSM follows the universal scaling form~\cite{juricic}
\begin{equation}
\sigma(\omega)=\delta^{\nu(d-2)} \: \mathcal{F} \left( \omega \delta^{-\nu z}\right),
\end{equation}
where $\mathcal F$ is an unknwon but universal scaling function. Inside the DSM phase the optical conductivity scales as $\sigma(\omega) \sim \omega$, when the frequency is much smaller than the bandwidth ($\omega \ll v \Lambda$). In the quantum critical regime $\sigma(\omega) \sim \omega^{(d-2)/z}$. Finally, in the metallic phase the optical conductivity at zero frequency becomes finite, i. e., $\sigma(\omega \to 0)=$finite. Therefore, in the vicinity of chiral symmetry preserving and breaking disorder driven QCP, $\sigma(\omega) \sim \omega^{2/3}$ and $\omega^{2/11}$ (setting $\epsilon_2=1$), respectively. The dc conductivity (zero frequency) also possesses a similar scaling behavior when $T \ll v \Lambda$ (upon taking $\omega \to T$ in the scaling form of optical conductivity)~\cite{wegner}. Therefore, distinct power law behavior of DOS (measured through compressibility), specific heat, conductivity (both ac and dc) in the critical regime can serve as diagnostic tools to determine the nature of the disorder scatterer (CSP or CSB) driving the DSM-CDM QPTs in various Dirac materials.   

It should be noted that as one keeps increasing the strength of disorder in DSM, the CDM phase ultimately undergoes a second QPT, becoming an Anderson insulator~\cite{fradkin, pixley}. However, the weak coupling approach from the DSM side obviously cannot capture the CDM-insulator transition. The critical exponents for the Anderson transition are also quite different from those of the DSM-CDM transition, for example $z=d=3$ for Anderson transition in three dimensions~\cite{wegner}. In addition, across the Anderson transition the average DOS does not display any critical behavior, rather the \emph{typical DOS} serves the purpose of an order parameter~\cite{abrahams, pixley}. The Anderson localization transition in Dirac materials has recently been addressed in a numerical work~\cite{pixley}, but is obviously out of scope for present work, restricted to the weak to intermediate disorder strength.

\section{Interplay of interaction and disorder}\label{interaction-disorder}

\begin{figure}[htbp]
\centering
\subfigure[]{
\includegraphics[width=4.05cm,height=3.5cm]{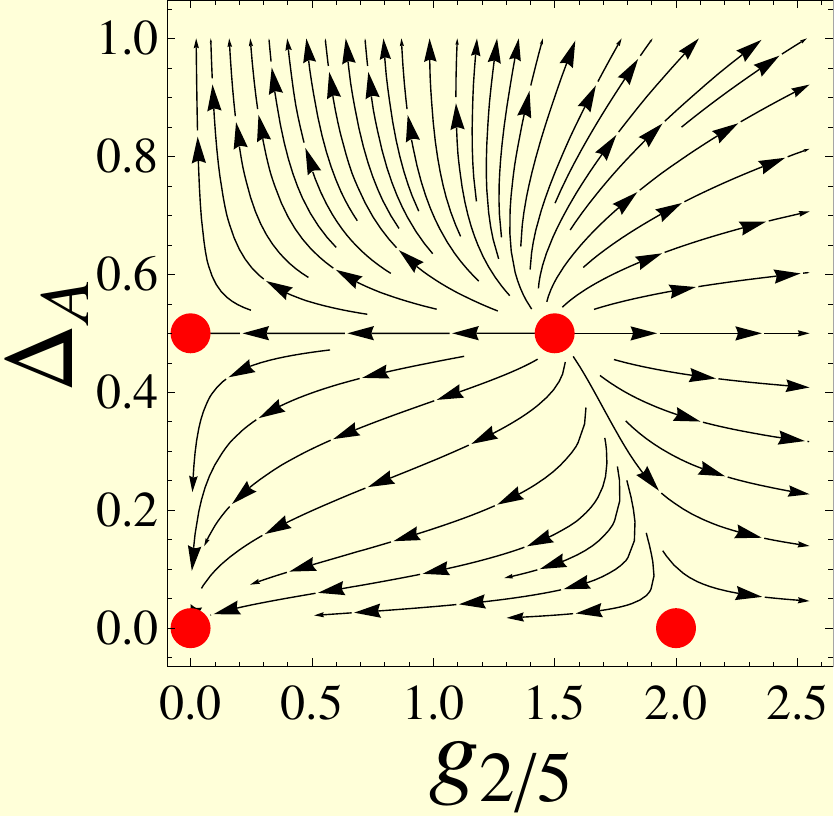}
\label{OddP-int-a}
}
\subfigure[]{
\includegraphics[width=4.05cm,height=3.5cm]{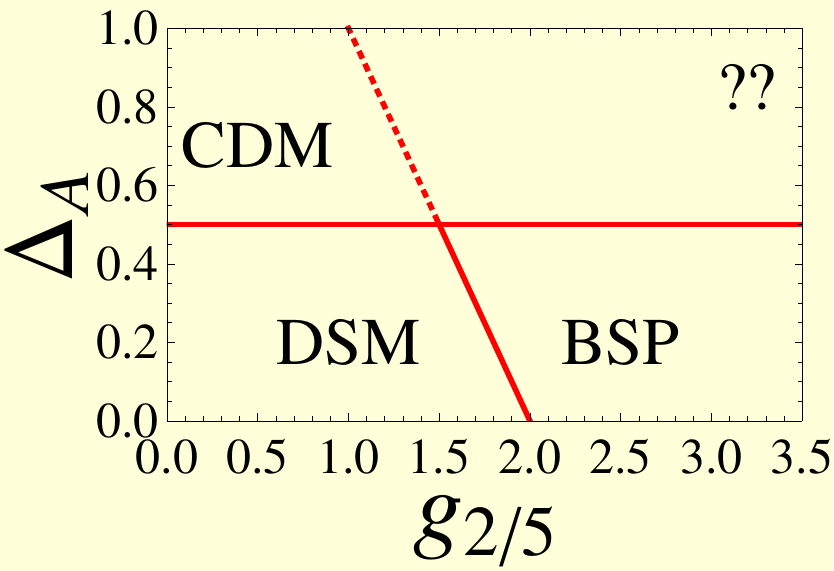}
\label{OddP-int-b}
}
\label{OddP-int}\caption[]{ (a) RG flow and (b) phase diagram in $g_{2/5}-\Delta_{A}$ plane. When the transition is driven by interaction $g_2$ or $g_5$ the order parameters in the strong coupling phases are $\langle \bar{\Psi} \Psi \rangle$ or $\langle \bar{\Psi} i \gamma_5 \Psi \rangle$, respectively. The dotted line represents a crossover boundary, right of which interaction and disorder immediately flow to a strongly coupled phase (strong  interaction and disorder), the nature of which is unknown to us (hence, the question marks in the right hand corner of the phase diagram). Figures are generated upon substituting $\epsilon_1=2$ and $\epsilon_2=1$.  }\label{oddP-disorder}
\end{figure}

The full problem of the nature of the interacting DSM in the presence of disorder is, of course, a formidable challenge for which we can only provide partial and incomplete resolution. The vanishing DSM DOS in the clean noninteracting limit enables certain simplifications allowing some progress, which we now are ready to discuss in this section. So far we have established that in clean DSMs, when the strength of short-range interactions exceeds a threshold, the system can find itself in various BSPs. The continuous QPTs to the BSPs take place through QCPs, which are mean-field or Gaussian in nature, and are characterized by the exponents $\nu=\epsilon^{-1}_1$ and $z=1$, and various physical quantities (e.g., the mass gap in the ordered phase) exhibit logarithmic violation of scaling, since the system lives at the upper critical dimension ($d_u=3$).

The non-interacting DSM remains stable against weak but generic quenched disorder, since weak disorder is an irrelevant perturbation. However, strong enough disorder can drive DSM through a QPT into the CDM phase. In the presence of only CSP disorder (potential and axial), critical exponents at the disorder controlled itinerant QCPs or the line of QCPs are $\nu=\epsilon^{-1}_2$ and $z=1+\frac{\epsilon_2}{2}$ to the leading-order in $\epsilon$-expansion, which may not be too accurate in this context since $\epsilon_2=1$ in $d=3$. Although it has been argued recently that $z=3/2$ is exact in this problem~\cite{pixley}. By contrast, a CSB disorder (spin-orbit), drives the DSM into a CDM phase through a QCP that is characterized by the exponents $\nu=\epsilon^{-1}_2$ and $z=1+\frac{9}{2} \epsilon_2$. The CLEs near two distinct dirty QCPs being equal is likely to be an artifact of the one-loop calculation, but expected to be different in general.

With the weak-coupling RG analyses in place for the clean interacting and the non-interacting dirty DSM separately, we are now in a position to investigate their interplay, treating both interaction and disorder on an equal footing. To understand the interplay of interaction and disorder in three dimensional DSMs, we perform RG calculations to the quadratic order in both interaction and disorder couplings. The relevant Feynman diagrams are shown in Fig.~\ref{Feynman-Diag} $(xi)-(xv)$ [in addition to the diagrams $(ii)-(v)$ and $(vii)-(x)$ in Fig.~\ref{Feynman-Diag}].

We here address the competition between interaction and disorder within the framework of a double $\epsilon$-expansion, a detailed analysis of which is presented in Appendix~\ref{double-epsilon}. In addition, we show explicit computation of all diagrams from Fig.~\ref{Feynman-Diag} for a simpler model with one interaction and one disorder couplings, namely $g_5$ and $\Delta_A$. The resulting flow and phase diagrams are discussed in subsection~\ref{intdis-simplest}, see also Fig.~\ref{oddP-disorder}. The prescription laid out in Appendix~\ref{double-epsilon} can easily be taken over to arrive at the coupled flow equations for generic interaction and disorder, as shown in Eqs.~(\ref{RG-Full}) and (\ref{CSB-disorder-RG-int}). It should be noted that the double $\epsilon$-expansion, we implement here is different than the one introduced in Refs.~\cite{dorogovtsev, cardy, lawrie}, in the context of disordered bosonic systems. In Refs.~\cite{dorogovtsev, cardy, lawrie}, one of the $\epsilon$s captures the deviation from the upper critical dimensions $d=4$ where four-boson interaction coupling is merginal, while the second $\epsilon$ (namely $\epsilon_\tau$) is introduced in the imaginary time co-ordinates. Thus during this procedure of double $\epsilon$-expansion the quenchness of random impurities is sacrificed in order to capture the ultraviolet divergences. By contrast, the double $\epsilon$-expansion scheme we introduce here leaves the disorder vertices infinitely correlated in time (quench disorder) and the ultraviolet divergences of various diagrams [$(xi)-(xv)$ in Fig.~\ref{Feynman-Diag}] are captured by performing the shell integral about appropriate merginal dimensions [namely about $d_c=1(2)$ to capture interaction (disorder) driven corrections to disorder (interaction)], as shown in Appendix~\ref{double-epsilon}. Given that in clean interaction and dirty noninteracting systems, $\epsilon$-expansions about one and two spatial dimensions gives quantitatively correct results, we believe that when these two perturbations are present simultaneously, the double $\epsilon$-expansion possibly yields qualitatively correct picture, at least when they are not too strong. Our theory thus provides the stability of the disordered phase (i.e., along the ordinate in Fig.~\ref{oddP-disorder}) in the presence of weak interaction as well as the stability of the interacting phase (i.e., along the abscissa of Fig.~\ref{oddP-disorder}) in the presence of weak disorder, but unable to assess the actual nature of the strong coupling phases [denoted by the question marks in the upper right hand quadrant of Fig.~\ref{OddP-int-b}].

\subsection{Chiral symmetric disorder and interaction}

We first consider the competition between electron-electron interactions and CSP disorder. Thus, we start by taking into accunt only potential and axial disorders. Interestingly the flow equations for the CSP disorder do not receive any perturbative correction from the short-range interactions at the one loop level. The RG flow equations for various couplings are given by
\begin{widetext}
\begin{eqnarray}\label{RG-Full}
\frac{d v}{d l} &=& v\big(z-1-\Delta_V-\Delta_{A} \big), \:
\frac{d \Delta_V}{d l} = \Delta_V \big(-\epsilon_2+2 \Delta_V+2\Delta_{A} \big), \:
\frac{d \Delta_{A}}{d l} = \Delta_{A} \big(-\epsilon_2+2 \Delta_V+2\Delta_{A} \big),
\nonumber \\
\frac{d g_1}{d l} &=& - \epsilon_1 \; g_1-\frac{1}{3} \big(g_1 g_2+ g_1 g_5+2 g_2 g_5 \big) + g_1 \big(\Delta_V + \Delta_{A} \big)-\frac{4}{3} \Delta_V \big( g_2 +g_5 \big), \nonumber \\
\frac{d g_2}{d l} &=& -\epsilon_1 \; g_2 +g^2_2 -\frac{2}{3} \big( g_1 g_2 -g_2 g_5 +g_1 g_5 \big) + g_4 \big(g_2-g_5 \big) + g_2 \big(\Delta_{A}-\frac{5}{3} \Delta_V \big)-\frac{8}{3} \; g_5 \Delta_V, \nonumber \\
\frac{d g_4}{d l} &=&  -\epsilon_1 \; g_4 + \frac{1}{3} \big( g_1 g_2 + g_1 g_5 -4 g_2 g_5 \big) +g_4 \big( \Delta_V +\Delta_{A}\big)+ \frac{4}{3} \big(g_2+g_5 \big) \Delta_V, \nonumber \\
\frac{d g_5}{d l} &=& -\epsilon_1 \; g_5 +g^2_5 -\frac{2}{3} \big(g_1 g_5-g_2 g_5+g_1 g_2 \big)+ g_4 \big(g_5-g_2 \big) + g_5 \big( \Delta_{A}-\frac{5}{3} \Delta_{V} \big) -\frac{8}{3}\; g_2 \Delta_V.
\end{eqnarray}
\end{widetext}
Analysis of these coupled flow equations is an involved task. Also, in the presence of potential disorder we cannot find any subset of coupling constants, which remains closed under coarse grainning. Therefore, we are compelled to analyze the full set of coupled flow equations for five coupling constants ($g_1,g_2,g_4,g_5,\Delta_{V}$). But, one can gain valuable insight by considering a simpler model.

\subsubsection{Simple model with $g_5$ and axial disorder ($\Delta_{A}$)}\label{intdis-simplest}

Let us first consider a model with only two coupling constants $g_5$ and $\Delta_{A}$, which remains closed under the RG procedure, and no additional coupling gets generated through the loop corrections. The flow equations of $v, \Delta_{A}$ and $g_5$ can be readily obtained from Eq.~(\ref{RG-Full}) by setting $g_1=g_2=g_4=\Delta_V=0$. The corresponding flow diagram in the $(g_5, \Delta_{A})$ plane is shown in Fig.~\ref{OddP-int-a}. The coupled flow equations support \emph{four} fixed points: $(i)$ a fully stable fixed point at $(g_5, \Delta_{A})=(0,0)$ representing a robust clean, noninteracting DSM, $(ii)$ an interacting Gaussian QCP at $(g_5, \Delta_{A})=(\epsilon_1,0)$ that governs that transition out of DSM into a ${\mathcal P}$ and ${\mathcal T}$ symmetry breaking insulator (pseudoscalar mass, characterized by a constant axion angle $\theta_{ax}=\mbox{sgn}(m_2) \frac{\pi}{2}$) when the coupling constant $g_5$ is strong enough, $(iii)$ a non-interacting dirty itinerant QCP at $(g_5, \Delta_{A})=(0,\frac{\epsilon_2}{2})$, which, on the other hand, describes a disorder-controlled QPT toward the formation of a CDM, and $(iv)$ a fully unstable MCP at $(g_5, \Delta_{A})=(\epsilon_1-\frac{\epsilon_2}{2}, \frac{\epsilon_2}{2})$, resulting from the interplay of disorder and interaction. At the MCP, three distinct phases, namely, the DSM, the CDM and a ${\mathcal P}$, ${\mathcal T}$-odd insulator, meet. Since $\Delta_{A}$ does not receive any correction from interaction ($g_5$), the phase boundaries between the disorder controlled CDM phase, and the interaction driven BSP and the DSM, are parallel to the $g_5$-axis (see Fig.~\ref{OddP-int-b}). Such an outcome is possibly an artifact of one loop calculation, but the topology of the phase diagram shown in Fig.~\ref{oddP-disorder} should remain valid qualitatively.

We realize that similar RG flow equations and phase diagram can be obtained if we replace the interaction coupling $g_5$ by $g_2$ (due to the underlying chiral symmetry of DSM). When $g_2>g^\ast_2=\epsilon_1$, the DSM becomes susceptible toward the formation of a BSP that lacks only a continuous chiral $U_c(1)$ symmetry, but preserves ${\mathcal C}$, ${\mathcal P}$ and ${\mathcal T}$ symmetries (scalar mass). The discussion below addresses both possibilities (either $g_5$ or $g_2$ nonzero).

In the presence of sufficiently weak disorder ($\Delta_{A}<\Delta^{\ast}_{A}=\frac{\epsilon_2}{2}$), the boundary between the DSM and the BSP (a ${\mathcal P}$, ${\mathcal T}$-odd insulator) shifts toward weaker interactions as one enhances the strength of the axial disorder. Such a behavior can be appreciated by comparing the strength of interactions at the MCP ($g^{M}_5=\epsilon_1-\frac{\epsilon_2}{2}$) and at the clean interacting QCP ($g^\ast_5=\epsilon_1$). Since $g^{M}_5<g^\ast_5$ (for $\epsilon_2>0$), sufficiently weak axial disorder, somewhat surprisingly, enhances the ordering tendency in the DSM. This is one of the main results of our analysis in interacting and dirty DSM, which can be justified in the following way. From Eq.~(\ref{anomalousdim}), we note that anomalous dimensions for Dirac mass operators (both scalar and pseudo-scalar, and thus axionic) is increased by axial disorder. As a result axial disorder boosts the formation of all mass orders in DSMs, as we show in next subsubsection. However, we fail to provide such intuitive justification on the role of regular potential and spin-orbit disorder on ordering tendencies in DSM, as the RG flow equations gets terribly coupled.

By contrast, when $\Delta_{A}>\epsilon_2/2$, but $g_5 \ll g^{M}_5$, the DSM gives away to the CDM phase, and sufficiently weak interaction is irrelevant in the extreme close vicinity of the diffusive QCP, located at $(g_5,\Delta_{A})=(0,\epsilon_2/2)$. Hence, our weak coupling RG analysis suggests that both interaction and disorder controlled QCPs are stable against sufficiently weak disorder and interaction, respectively, as one approaches the QCPs from the DSM side of the transitions. This stability is also an important finding of our theory.

However, inside the broken symmetry phase, even sufficiently weak disorder can generate random mass or bond disorder for the order parameter field. Even though such mass disorder is absent in the bare theory, it can be generated in the ordered phase, since the correlation length $\xi \sim \Lambda^{-1} (g_5-g^\ast_5)^{-\nu}$ provides the infrared cutoff for the flow of disorder coupling. Notice that at the clean QCP $\nu =1/2<\frac{2}{3}$ (an exact result), and the \emph{Harris criterion} is satisfied~\cite{harris}. Therefore, the interacting QCP is unstable against the coupling of the order parameter field with the mass or bond disorder, toward a new disorder and interaction controlled QCP with $\nu \geq 2/3$~\cite{harris, chayes}. Extracting the influence of the mass disorder at the clean interacting QCP is beyond the scope of the present weak coupling RG analysis about the lower critical dimension ($d_{l}=z=1$), and remains a future problem of interest.

Furthermore, in the metallic phase the average DOS near the Dirac point increases, which can enhance the effect of interactions. The flow diagram in Fig.~\ref{OddP-int-a} suggests that even sufficiently weak interaction ultimately grows under RG when $\Delta_{A}>\epsilon_2/2$, and eventually the system runs to an unknown strong disorder and interaction controlled phase, which is inaccessible by perturbative RG appraoch. Our weak coupling analysis is inadequate to address the exact nature of such strong disorder and interaction controlled phase. Nevertheless, in a sufficiently clean or weakly interacting system, two QCPs we find from the weak coupling RG calculation, can still describe the \emph{crossover} behavior of various physical quantities over a sufficiently large crossover length scale.

\subsubsection{Generic interaction and axial disorder ($\Delta_A$)}

Having developed some intuition about the possible quantum phases and their stability, by keeping just one each of interaction and disorder terms in the RG analysis, we now discuss the generic situation in the presence of axial disorder. We have already emphasized the subtlety and shortcomings of the weak coupling RG calculation when the system enters into a strong coupling phase. Without delving into the fate of our analysis in the strong coupling limit (which definitely goes well beyond the scope of the double-$\epsilon$ expansion scheme), one can still arrive at some limited, but valuable conclusions regarding the instability of Dirac quasiparticles (driven by either interaction or disorder) and approximately guess the qualitative structure of the phase diagram of interacting DSM in a random environment, as discussed below.

\begin{widetext}
\begin{center}
\begin{table}[h]
  \begin{tabular}{|c||c|c|}
     \hline
QCP & Critical points with disorder & Multi-critical points (with two unstable directions) with axial disorder \\
     \hline \hline
C1 & $(-0.125 \epsilon_1, 0.5 \epsilon_1,-0.375 \epsilon_1, 0.5 \epsilon_1, 0)$ & $(-0.345 \epsilon_1 + 0.5\epsilon_2, 0.625 \epsilon_1-0.5 \epsilon_2, -0.53 \epsilon_1 + 0.5 \epsilon_2, 0.625 \epsilon_1- 0.5 \epsilon_2, 0.5\epsilon_2)$ \\
     \hline
C3 & $(0.185 \epsilon_1, 0.645 \epsilon_1, 0.405 \epsilon_1, -0.455 \epsilon_1, 0) $ & $(0.39 \epsilon_1- 0.5 \epsilon_2, 0.735 \epsilon_1 - 0.5 \epsilon_2, 0.555 \epsilon_1 - 0.5 \epsilon_2, -0.59 \epsilon_1 + 0.5 \epsilon_2, 0.5 \epsilon_2)$ \\
     \hline
C2 & $(0.185 \epsilon_1, -0.455 \epsilon_1, 0.405 \epsilon_1, 0.645 \epsilon_1, 0) $ & $(0.39 \epsilon_1- 0.5 \epsilon_2, -0.59 \epsilon_1 + 0.5 \epsilon_2, 0.555 \epsilon_1 - 0.5 \epsilon_2, 0.735 \epsilon_1 - 0.5 \epsilon_2, 0.5 \epsilon_2)$ \\
     \hline
C4 & $(-2 \epsilon_1, -\epsilon_1, 0, -\epsilon_1, 0) $ & $(-1.75 \epsilon_1+0.5 \epsilon_2, -\epsilon_1+0.5 \epsilon_2, 0,-\epsilon_1+0.5 \epsilon_2, 0.5 \epsilon_2)$ \\
     \hline
CA & $(0,0,0,0,0.5 \epsilon_2) $ & -------------------- \\
     \hline
 \end{tabular}
\caption{The first column shows the symbols for various QCPs in either clean interacting (C1, C2, C3, C4) or dirty noninteracting (CA) DSM. Second column displays locations of various QCPs in the presence of only axial disorder. The first four QCPs correspond to the ones in clean interacting system, describing transitions to various BSPs, and the fifth one represents disorder-driven DSM-CDM QPT. The third column shows the locations of various MCPs associated with each interacting QCP. Coupling constants at various fixed points are quoted in the following order $(g_1,g_2,g_4,g_5,\Delta_{A})$.}\label{table-2}
\end{table}

\begin{table}[h]
  \begin{tabular}{|c||c|c|}
     \hline
QCP & Critical points with disorder & Multi-critical points (with two unstable directions) with potential disorder \\
     \hline \hline
C1 & $(-0.125 \epsilon_1, 0.5 \epsilon_1,-0.375 \epsilon_1, 0.5 \epsilon_1, 0)$ & $(-0.415 \epsilon_1 - 0.5\epsilon_2, 0.465 \epsilon_1+0.5 \epsilon_2, -0.445 \epsilon_1 - 0.5 \epsilon_2, 0.465 \epsilon_1+0.5 \epsilon_2, 0.5\epsilon_2)$ \\
     \hline
C3 & $(0.185 \epsilon_1, 0.645 \epsilon_1, 0.405 \epsilon_1, -0.455 \epsilon_1, 0) $ & $(0.385 \epsilon_1- 0.5 \epsilon_2, 0.705 \epsilon_1 - 0.5 \epsilon_2, 0.595 \epsilon_1 - 0.5 \epsilon_2, -0.645 \epsilon_1 + 0.5 \epsilon_2, 0.5 \epsilon_2)$ \\
     \hline
C2 & $(0.185 \epsilon_1, -0.455 \epsilon_1, 0.405 \epsilon_1, 0.645 \epsilon_1, 0) $ & $(0.385 \epsilon_1- 0.5 \epsilon_2, -0.645 \epsilon_1 + 0.5 \epsilon_2, 0.595 \epsilon_1 - 0.5 \epsilon_2, 0.705 \epsilon_1 - 0.5 \epsilon_2, 0.5 \epsilon_2)$ \\
     \hline
C4 & $(-2 \epsilon_1, -\epsilon_1, 0, -\epsilon_1, 0) $ & $(-2.675 \epsilon_1-0.5 \epsilon_2, -0.84 \epsilon_1-  0.5 \epsilon_2, 0.005 \epsilon_1-0.5 \epsilon_2,-0.84 \epsilon_1 - 0.5 \epsilon_2, 0.5 \epsilon_2)$ \\
     \hline
CV & $(0,0,0,0,0.5 \epsilon_2) $ & -------------------- \\
     \hline
  \end{tabular}
\caption{Same as Table~\ref{table-2}, but in the presence of only potential disorder. Coupling constants at various fixed points are quoted in the following order $(g_1, g_2, g_4, g_5, \Delta_V)$.}\label{table-3}
\end{table}
\end{center}
\end{widetext}

We now take into account all short-range interactions ($g_1, g_2, g_4, g_5$) and the axial disorder. The coupled flow equations all together support \emph{five} QCPs. Four of them correspond to the ones in the clean interacting system, summarized in Sec.~III, describing continuous QPTs to various BSPs. The remaining one is solely controlled by axial disorder in a noninteracting DSM, capturing DSM-CDM QPT. These critical points are tabulated in the second column of Table~\ref{table-2}.

In addition, we also find \emph{four} MCPs (with \emph{two} unstable directions), summarized in the third column of Table~\ref{table-2}. These MCPs play the same role as the one at $(g_5,\Delta_{A})=(\epsilon_1-\frac{\epsilon_2}{2},\frac{\epsilon_2}{2})$ in the $(g_5,\Delta_A)$ plane, as shown in Fig.~\ref{oddP-disorder}, and together with the QCPs determine the phase boundaries between DSM and various BSPs, as discussed in the previous subsection. The relative strength of interactions at a given QCP (say $g^\ast_i$) and the corresponding MCP (say $g^M_i$) determines the role of sufficiently weak disorder on the ordering tendencies in DSM. Notice that $\Delta_{A}=\epsilon_2/2$ at all MCPs, and the sign of all interaction couplings at a given critical point and the corresponding MCP is same (upon setting $\epsilon_1=2$ and $\epsilon_2=1$). From Table~\ref{table-2}, we find that $g^M_i<g^\ast_i$ for each QCP. Therefore, axial disorder enhances the propensity of interaction driven BSPs in three-dimensional DSMs, and a representative phase diagram in the $g_2-g_5$ plane is shown in Fig.~\ref{pd-Diracaxial} for $\Delta_A=0.3$.

\subsubsection{Generic interaction and potential disorder ($\Delta_V$)}

Next we attempt to understand the role of potential disorder in three-dimensional interacting DSMs. The RG flow equations for $v, \Delta_V$ and $g_j$'s (for $j=1,2,4,5$) can be obtained from Eq.~(\ref{RG-Full}), upon neglecting the contribution from $\beta_{\Delta_{A}}$ and setting $\Delta_{A}=0$ in the rest of the flow equations.

Once again we obtain five QCPs, which are summarized in the second column of Table~\ref{table-3}. Four critical points correspond to the ones in clean interacting system. The dirty (noninteracting) QCP is located at $\Delta_V=\frac{\epsilon_2}{2}$ and $g_j=0$ for $j=1,2,4,5$. In addition, we find four MCPs (with two unstable directions). The third column of Table~\ref{table-3} displays the location of the MCPs associated with each interacting QCPs. Comparing the strength of interactions at various QCP and at the corresponding MCP (for $\epsilon_1=2, \epsilon_2=1$), we conclude that potential disorder also enhances the ordering tendency toward the insulating states through the formation of pseudo-scalar (QCP C2) and scalar (QCP C3) masses, where $\langle \bar{\Psi} i \gamma_5 \Psi \rangle \neq 0$ and $\langle \bar{\Psi} \Psi \rangle \neq 0$, respectively.

On the other hand, potential disorder appears to oppose the formation of an axionic insulator (takes place through QCP C1), since $g^\ast_i> g^M_i$ for $i=1,2,4,5$ near C1.  A representative phase diagram of interacting Dirac femrions in the $g_2-g_5$ plane is shown in Fig.~\ref{pd-Diracpotential} for $\Delta_V=0.3$. In addition, potential disorder also suppresses the pairing instability in the $s$-wave channel (through QCP C4) of massless Dirac fermions.

\subsection{Chiral symmetry breaking disorder and interaction}

Finally, we address the interplay between electron-electron interaction and CSB disorder (spin-orbit). As shown in Sec.~\ref{CSB-disorder-dirty}, the RG analysis does not close only with spin-orbit disorder as it generates axial disorder through loop corrections. Thus we need to account for both spin-orbit and axial disorder, even if the bare model contain no axial impurity. The RG flow equations for various coupling constants to one-loop order are 
\begin{widetext}
\begin{eqnarray}\label{CSB-disorder-RG-int}
\frac{dv}{dl} &=& v(z-1-\Delta_{A}-3\Delta_{SO}), \quad \frac{d \Delta_{A}}{dl} = -\epsilon_2 \Delta_{A} +2 \Delta^2_{A}-6 \Delta_{A} \Delta_{SO} + 4 \Delta^2_{SO}, \nonumber \\
\frac{d \Delta_{SO}}{dl} &=& -\epsilon_2 \Delta_{SO}-\frac{2}{3} \Delta^2_{SO} + 2 \Delta_{SO} \Delta_{A}
+ \frac{\Delta_{SO}}{3} \left(-g_1+g_2+g_4-g_5 \right), \nonumber \\
\frac{dg_1}{dl} &=& -\epsilon_1 g_1-\frac{1}{3} \left(g_1 g_2+ g_1 g_5+2 g_2 g_5 \right) 
+ g_1 (3 \Delta_{SO}+ \Delta_A) -\frac{4}{3} \left( -3 g_2 + 2 g_5 +2 g_4 \right) \Delta_{SO}, \nonumber \\ 
\frac{dg_2}{dl} &=& -\epsilon_1 g_2 +g^2_2 -\frac{2}{3} \left( g_1 g_2 -g_2 g_5 +g_1 g_5 \right)
 +   g_4 (g_2-g_5) + g_2 \Delta_A + \frac{4}{3} \left( 3 g_1 + \frac{9}{4} g_2  + 2 g_4 + 2 g_5 \right) \Delta_{SO}, \nonumber \\
\frac{dg_4}{dl} &=& -\epsilon_1 g_4 + \frac{1}{3} \left( g_1 g_2 + g_1 g_5 -4 g_2 g_5\right) + g_4 (\Delta_A- 9 \Delta_{SO}) + \frac{4}{3} (2 g_4- 4 g_5) \Delta_{SO} \nonumber \\
\frac{dg_5}{dl} &=& -\epsilon_1 g_5 +g^2_5 -\frac{2}{3} \left(g_1 g_5-g_2 g_5+g_1 g_2 \right)
+ g_4 (g_5-g_2) + g_5 \Delta_A -\frac{4}{3} \left( 4 g_4 + \frac{19}{4} g_5 \right) \Delta_{SO}.
\end{eqnarray} 
\end{widetext}
The above set of flow equations supports only \emph{four} QCPs, which were previously found in the clean interacting systems, namely C1, C2, C3 and C4 (see Sec.~\ref{clean-interacting}). However, the CSB disorder driven dirty QCP at $\left( \Delta_A, \Delta_{SO} \right)= (9/10,6/5) \epsilon_2$ (in the noninteracting system) becomes a MCP with two unstable directions in the presence of electronic interactions. This result can be substantiated from the following observation. The DCE at the dirty QCP is $z=1+9 \epsilon_2 /2$. Therefore, the scaling dimension of short-ranged interaction at this QCP is $[g]=z-d=-2 + 9 \epsilon_2/2=5/2$ (upon setting $\epsilon_2=1$ for three dimensional DSM). Hence, sufficiently weak short-ranged interaction is a relevant perturbation at the dirty QCP driven by the CSB disorder. Enhancement of electronic interaction near the CSB disorder driven QCP can also be understood from the fact that near this QCP DOS diverges according to $\varrho(E)\sim |E|^{-5/11}$. Consequently, such a QCP becomes unstable against infinitesimal interactions, and turns into a MCP with two unstable directions.

A representative phase diagram of interacting DSM for sufficiently weak spin-orbit disorder in the $(g_2,g_5)$-plane is shown in Fig.~\ref{pd-Diracspinorbit}. Notice that due to the lack of continuous chiral symmetry in the presence of spin-orbit disorder, the phase diagram in Fig.~\ref{pd-Diracspinorbit} lacks the symmetry $g_2 \leftrightarrow g_5$. By contrast, the phase diagrams in the clean interacting DSM (see Fig.~\ref{pd-DiracClean}), and also the ones in the presence of axial (see Fig.~\ref{pd-Diracaxial}) or potential (see Fig.~\ref{pd-Diracpotential}) disorders are symmetric under $g_2 \leftrightarrow g_5$. Such symmetry stems from the underlying chiral symmetry of massless Dirac femrions, which remains preserved even when the DSM is subject to CSP disorder, but gets broken in the presence of CSB disorder. Otherwise, weak spin-orbit disorder, although reduces the pairing tendency of Dirac fermions in the pseudo-scalar and axionic mass channels, but substantially increases the phase phase available for scalar mass generation, as shown in Fig.~\ref{pd-Diracspinorbit}.

\section{Discussion and conclusion}\label{discussion}

To summarize, in this work we have addressed the effects of (i) short-range repulsive electron-electron interaction, (ii) random quenched disorder (time-reversal symmetric), and (iii) the interplay between interaction and disorder, in the three dimensional DSM. 

In clean system, we show that when finite range interactions are sufficiently strong DSM becomes unstable toward the formation of various BSPs, among which (a) regular insulator that only lacks the continuous chiral $U_c(1)$ symmetry, (b) microscopic parity and time-reversal symmetry breaking insulator, and (c) an axionic insulator (also ${\mathcal P}, {\mathcal T}$ odd) (see Fig.~\ref{pd-DiracClean}). When the ordered phase lacks microscopic ${\mathcal P}$ and ${\mathcal T}$ symmetries, it supports a magneto-electric effect captured by the axionic term~\cite{{topo-fdth}}        
\begin{equation}
S_{em}=-\frac{e^2}{32 \pi^2} \int d^4 \epsilon^{\mu \nu \rho \lambda} \theta_{ax} \; F_{\mu \nu} F_{\rho \lambda},
\end{equation}
where the \emph{axion angle} $\theta_{ax}$ is a constant/dynamic variable in pseudo-scalar/axionic insulating phase, and $F_{\mu \nu}$ is the electromagnetic field strength tensor. The QPTs into the BSPs are mean-field in nature and the QCPs are characterized by the exponents $\nu=1/2$ and $z=1$, which we capture here performing an $\epsilon$-expansion about the lower critical dimension $d_l=1$. At the QCPs a pseudo Lorentz symmetry gets restored, since $z=1$, and the Fermi velocity remains non-critical across the DSM-BSP QPT. The long range tail of the Coulomb interaction is shown to enhance the ordering tendencies for weaker interaction stregths. Thus, by tuning the effective dielectric constant in Dirac materials, so as to increase the long range Coulomb interaction, one can drive the system through DSM-BSP QPTs, see Figs.~\ref{pd-DiracClean} and \ref{pd-DiracCoulomb}.

The non-interacting dirty DSM is shown to be stable against sufficiently weak, but generic (time-reversal symmetric) randomness. Nonetheless, with increasing disorder strength beyond a threshold, the DSM can undergo a QPT and enter into the CDM phase. We here focused on two different types of disorder (a) CSP and (b) CSB. The DSM-CDM QPT driven by CSP disorder is characterized by the exponents $\nu=1$ and $z=3/2$ (to one loop order). On the other hand, when the DSM-CDM QCP is tuned by CSB disorder, the exponents take the values $\nu=1$ and $z=11/2$ (to one loop order). Here, we extract these exponents by performing an $\epsilon$-expansion around the lower critical dimension for DSM-CDM QPT, which is $d_l=2$. Scaling of various physical quantities at DSM-CDM QCPs are shown in Table~\ref{table-intro}. We emphasize that a great deal of theoretical work has already been done on the disorder driven QPT in noninteracting DSMs and our work complements the existing literature~\cite{fradkin, shindou, goswami-chakravarty, herbut-Imura, radziovsky, roy-dassarma, ominato, pixley, dassarma-hwang-transport}.

We also study the interplay of interaction and disorder in DSM. In the presence of both interaction and disorder it is not possible to find a unique lower critical dimension of the theory about which one can perform an $\epsilon$-expansion. To circumvent this technical barrier, we implement a double $\epsilon$-expansion to address this challenging question. The fact that DSM remains stable against weak interaction and disorder, which one can reconcile from the double-$\epsilon$ expansion, gives us some confidence that our results are at least qualitatively correct, in the weak-coupling regime. Direct numerical work will, however, be necessary in the future to check the quantitative validity of the double $\epsilon$-expansion technique. We find that in the presence of chiral symmetric disorder, both clean interacting and dirty diffusive QCPs are stable, as one approaches them from the DSM side of the transitions. Our analysis suggests that weak axial disorder enhances the propensity of any ordering in DSM [see Fig.~\ref{pd-Diracaxial}], while potential disorder is beneficiary to the formation of regular and ${\mathcal P}$, ${\mathcal T}$ breaking insulator only. By contrast, the presence of potential disorder supresses the condensation of massless Dirac fermions into an axionic insulator [see Fig.~\ref{pd-Diracpotential}]. The presence of CSB spin-orbit disorder ruins the symmetry of interacting DSM under chiral rotation. The spin-orbit disorder, however, increases the propensity of scalar mass generation substantially, see Fig.~\ref{pd-Diracspinorbit}.

Nevertheless, the ultimate fate of these outcomes in the strong coupling limit remains unknown at this stage. We realize that the clean interacting QCPs satisfy the Harris criterion~\cite{harris, chayes}. Hence, it is natural to anticipate that mass disorder (giving rise to random $T_c$) is a relevant perturbation at these QCPs. Therefore, the ultimate long wavelength behavior of any BSP even in sufficiently weakly disordered DSM will be governed by a new QCP where both disorder and interaction are finite and $\nu \geq 2/d$~\cite{harris, chayes}. If the randomness is sufficiently weak, the clean interacting QCPs can dictate crossover benhavior over a sufficiently large scale. Although the weak coupling analysis suggests that CSP disorder controlled QCP between DSM and CDM is stable against sufficiently weak interaction, the importance of interaction inside the diffusive metallic phase is also beyond the scope of our analysis (We mention as an aside that the corresponding problem of the fate of a three-dimensional metallic Fermi liquid in the presence of strong disorder and interaction is still an open question even after 40 years of intensive research). In contrast, we find that CSB breaking disorder driven DSM-CDM QCP becomes unstable against infinitesimally weak interactions. Our work deals with the stability of interacting, but dirty DSMs, and shed light on the influence of sufficiently weak disorder on the instability of DSM toward the formation of various BSPs, see Figs.~\ref{pd-Diracaxial}, \ref{pd-Diracpotential}, \ref{pd-Diracspinorbit}.

Even though we focus on single copy of four-component massless Dirac fermions with a genuine chiral symmetry, our analysis can be generalized for a wide variety of systems, supporting linearly dispersing quasiparticle excitations around few isolated points in the Brillouin zone, such as the topological DSM, which has recently been realized in Cd$_2$As$_3$~\cite{cdas} and Na$_3$Bi~\cite{nabi}, and Weyl semimetals, discovered in inversion asymmetric TaAs~\cite{taas-1, tasas-2, taas-3}, NbAs~\cite{nbas-1}, TaP~\cite{tap-1}, and time-reversal symmetry breaking YbMnBi$_2$~\cite{borisenko}, Sr$_{1-y}$MnSb$_2$~\cite{chiorescu}. The topological DSM accommodates two copies of four-component massless Dirac femrions that possesses a genuine chiral symmetry. Two Dirac points in these materials are protected by time-reversal, inversion and four-fold rotational symmetry in the tetragonal environment~\cite{shovkovy, furusaki}. Weyl semimetals enjoy a bonafide chiral symmetry, as it is tied with the \emph{translational symmetry} in the continuum limit~\cite{roy-sau}. In the former system there are additional inter-valley scattering processes due to the presence of multiple nodes, while in Weyl semimetals the lack of time-reversal and/or inversion symmetries, makes the system susceptible to generic disorder (time-reversal-symmetry breaking disorder, for example). Therefore, to properly address the role of electronic interaction and/or disorder in such systems one needs to account for several additional coupling constants that although turns the problem into a rich harbor of a plethora of phases, associated phase transitions and quantum critical phenomena, the analysis gets rapidly lengthy in the absence of symmetries (in Weyl semimetal) or with increasing number of nodes (in topological DSM). However, the qualitative structure of the phase diagram and the stability of various critical points and phases should remain unchanged from the results presented in the current work, with some quantitative differences depending on the system. For example, (a) the semimetallic phase (in topological DSM and Weyl semimetals) should remain stable in these materials against sufficiently weak, but generic interaction and disorder, (b) interaction driven QPTs are mean-field in nature~\cite{aji, nandkishore}, (c) disorder can drive these system through non-Gaussian itinerant QCP and place them into a diffusive metallic phase~\cite{roy-bera, brouwer, altland, chen-Song, ohtsuki, hughes, nandkishore-disorder}.

Before concluding, we remark on several open (and potentially important) questions in the context of interacting dirty DSM, which are, however, beyond the scope of present perturbative analysis. In particular, we want to qualitatively discuss below three topics of possible relevance: the strong-coupling situation, disorder-induced rare regions, and the experimental observability of the quantum phases and QPTs discussed in our work.

Throughout this article we have emphasized that our weak-coupling RG analysis can only ascertain the stability of interaction-driven BSPs against various types of randomness to certain extend, but cannot demonstrate the ultimate ground state in Dirac systems when both interaction and disorder flow to strong coupling. The stability of the BSPs to weak enough disorder stems from the fact that the ordered phases are fully gapped, providing some immunity to disorder. We also believe that the results obtained from weak-coupling analysis, should at least describe crossover behavior, due to the stable nature of the various fixed points. At the end, however, the nature of strongly interacting disordered phase in Dirac materials still remains an open question. We know that without any interaction, disorder, by itself, drives DSM into a CDM phase at intermediate coupling, and ultimately into an Anderson insulator at stronger disorder~\cite{pixley}. We can then ask the following question: What would happen if one turns on interaction in the strongly disordered phase? Motivated by the corresponding problem in ordinary three dimensional metals (Fermi liquid), we can speculate that the ultimate fate of the strongly disordered and strongly interacting three dimensional Dirac systems would be an `\emph{axionic glass}' phase, where the interaction-driven ${\mathcal P}, {\mathcal T}$ symmetry breaking order will possess only short-range correlations, with the whole system being Anderson localized globally. Such an axionic glass, where the gap is likely to display random spatial fluctuation due to the Anderson localization, would not possess any long range ordering, but depending on the relative strength between disorder and interaction, it might exhibit considerable short-range order~\cite{galss-comment}. At this stage we can only speculate that the unknown strong-coupling phase deep in the upper right hand corner of the quantum phase diagram in Fig.~\ref{OddP-int-b} is some type of \emph{Dirac glass}. Much more work is necessary to definitively establish the existence (or not) of such an glassy phase (axionic or more generally, Dirac) in a strong-coupling situation.

Second, it has recently been argued that the noninteracting DSM can actually develop exponentially small DOS even at infinitesimal disorder arising from resonances associated with the disorder-induced rare region phenomena (the so-called Griffiths physics)~\cite{nandkishore-disorder, pixley-unpublished}. Such rare region effects (if generic) invalidate the basic scaling argument in the noninteracting dirty system that disorder is irrelevant in three dimensional Dirac and Weyl semimetals, and therefore raising question about the quantum criticality associated with the formation of CDM phase at finite disorder discussed in Sec.~\ref{dirty-DSM} and Refs.\cite{fradkin, shindou, goswami-chakravarty, herbut-Imura, radziovsky, roy-dassarma, ominato, pixley, roy-bera, brouwer, altland, chen-Song, ohtsuki, hughes}. If such rare region with finite DOS are indeed present, then the noninteracting DSM-CDM critical point becomes `hidden' or `avoided'. Since rare region effects is nonperturbative, and thus inaccessible by perturbative RG analysis, we can only comment on very recent numerical work, shows that their effect to be quantitatively miniscule~\cite{pixley-unpublished}. Thus, the quantum critical physics of the noninteracting dirty system should continue to be operationally effective except at the lowest (largest) energy (length) scale. Although the existence of the rare regions in the \emph{noninetracting} dirty DSM is a matter of considerable fundamental interest, the question of interest to the current work is whether rare region physics affects our RG considerations for the \emph{interacting} DSM. It is well-known that repulsive interaction should strongly suppress rare region effects, and we believe that our conclusions for the interacting DSM remain unaffected by any exponentially weak rare region effects.

Finally, we comment on a practical question regarding the observability of the various predicted BSPs and CDM phase in experiments. Identifying these phases can be a challenging task in actual experiments. First, if the gap is small in a BSP, which might the situation in a weakly correlated material, it may become difficult to detect it in an experiment at finite temperature. Second, accessing a QCP (interaction or disorder driven) in Dirac materials requires the chemical potential to be fine-tuned at the Dirac point, so that the clean noninteracting system is indeed a true semimetal, which can be quite challenging to achieve in an experiment. In addition, even if the nominal `average' chemical potential is tuned to the Dirac point by producing overall charge neutrality, random fluctuations in the local dopant density may produce electron and hole `puddles', where the chemical potential randomly fluctuates spatially compared with energy of the Dirac point. This effect is well-established in a prototypical two dimensional Dirac system, graphene~\cite{Adam-PNAS, Rossi-DasSarma, Das-Sarma-RMP}, where these puddles dominate experiments around the Dirac point. However, every QCP is associated with a quantum critical regime, which extends over finite energy, temperature, frequency~\cite{sachdev}. It is quite often the situation that a QCP gets masked by some other phases at lowest energy scale. Nonetheless, the existence of QCP manifests through critical scaling of various physical quantities (thermodynamic and transport), inside the \emph{quantum critical fan}. In addition, in strongly interacting Dirac semimetals, the renormalized chemical potential can get pinned close to the Dirac point (for example, if the DSM arises from hybridization between $d$ and $f$ electrons). Hence, even if rare regions or puddles may set natural infrared cutoffs for the critical regime associated with the disorder-driven DSM-CDM QCP in an ideal noninteracting systems, their effects should be substantially suppressed near interacting QCP toward the formation of BSPs. Therefore, our proposed critical scaling behavior of specific heat, DOS, conductivity (both optical and dc) should manifest the existence of underlying QCPs, even if their existence at the lowest energy scale gets masked by various effects.

\acknowledgements

This work was supported by  NSF-JQI-PFC and LPS-MPO-CMTC. We thank P. Goswami, I. F. Herbut, C. Honerkamp, V. Juri\v ci\'c, D.  V. Khveshchenko, J. D. Sau for stimulating discussions. B. R. is thankful to Aspen Center of Physics for hospitality during the Winter Conference (2015), where part of this work was finalized.

\appendix

\section{Long-range Coulomb interaction}~\label{coulomb-long-range}

\begin{figure}[htb]
\includegraphics[width=8.2cm,height=4.5cm]{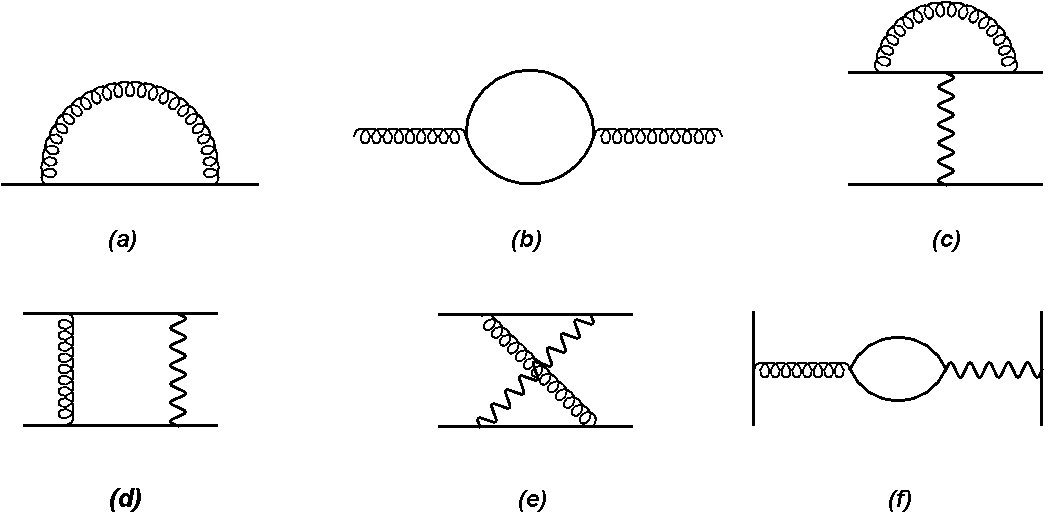}
\caption{One-loop diagrams capturing the compeition between long range and short range components of the Coulomb interaction. The solid and spiral lines respectively represent fermion and gauge field. All diagrams produce logarithmically divergent contributions. }\label{Coulomb-Fig}
\end{figure}

In this appendix we address the role of long range tail of the Coulomb interaction, as well as its interplay with the short-range (local) interactions in three dimensional DSM. The imaginary time action in the presence of both long range and short range components of Coulomb interaction reads as 
\begin{eqnarray}\label{action-LR}
S &=& \int d^3 x d \tau \bigg\{ \bar{\Psi} \left[ \gamma_0 \left( \partial_\tau + i g \phi \right) + v \gamma_j \partial_j \right] \Psi \nonumber \\
&+& \frac{1}{2} \left( \partial_j \phi \right)^2 - L_{int} \bigg\},
\end{eqnarray}
where $g=\sqrt{4 \pi v \alpha}$ and $\alpha=e^2/(4 \pi \varepsilon v)$ is the fine structure constant and $\varepsilon$ is the dielectric constant of the medium. Short range parts of the Coulomb interaction is captured by $L_{int}$, defined in Eq.~(\ref{int-reduced}). To obtain the low energy behavior of the model in Eq.~(\ref{action-LR}), we perform a RG calculation to one loop order and compute the flow of various coupling constant to the quadratic order. The pertinent Feynman diagrams are shown in Fig.~\ref{Coulomb-Fig} [in addition to $(ii)-(v)$ in Fig.~\ref{Feynman-Diag}].

The RG flow equations of various coupling constants are given by 
\begin{eqnarray}
\frac{d v}{dl} &=& \frac{2 \alpha}{3 \pi} v, \: \: \frac{d \alpha}{d l}=-\frac{4 \alpha^2}{3 \pi}, \nonumber \\
\frac{dg_1}{dl} &=& -\epsilon_1 g_1-\frac{1}{3} \left(g_1 g_2+ g_1 g_5+2 g_2 g_5 \right) \nonumber \\
&-& \frac{4 \alpha}{3 \pi} g_1 + \frac{2 \alpha}{3 \pi} \left(g_2 +g_5 \right), \nonumber \\
\frac{dg_2}{dl} &=& -\epsilon_1 g_2 +g^2_2 -\frac{2}{3} \left( g_1 g_2 -g_2 g_5 +g_1 g_5 \right) \nonumber \\
& + & g_4 (g_2-g_5) + \frac{4 \alpha}{3 \pi} \left(g_2 +g_5 \right), \nonumber \\
\frac{dg_4}{dl} &=& -\epsilon_1 g_4 + \frac{1}{3} \left( g_1 g_2 + g_1 g_5 -4 g_2 g_5\right) - \frac{2 \alpha}{3 \pi} \left(g_2 +g_5 \right), \nonumber \\
\frac{dg_5}{dl} &=& -\epsilon_1 g_5 +g^2_5 -\frac{2}{3} \left(g_1 g_5-g_2 g_5+g_1 g_2 \right) \nonumber \\
&+& g_4 (g_5-g_2) + \frac{4 \alpha}{3 \pi} \left(g_2 +g_5 \right).
\end{eqnarray}
Notice that even in the presence of long range Coulomb interaction, the flow equations continue to enjoy the symmetry under $g_2 \leftrightarrow g_5$. Such symmetry stems from the underlying chiral symmetry of low energy Dirac Hamiltonian, which remains unaffected upon incorporating long range instantaneous density-density interaction.

\begin{figure}[htb]
\includegraphics[width=7.0cm,height=5.0cm]{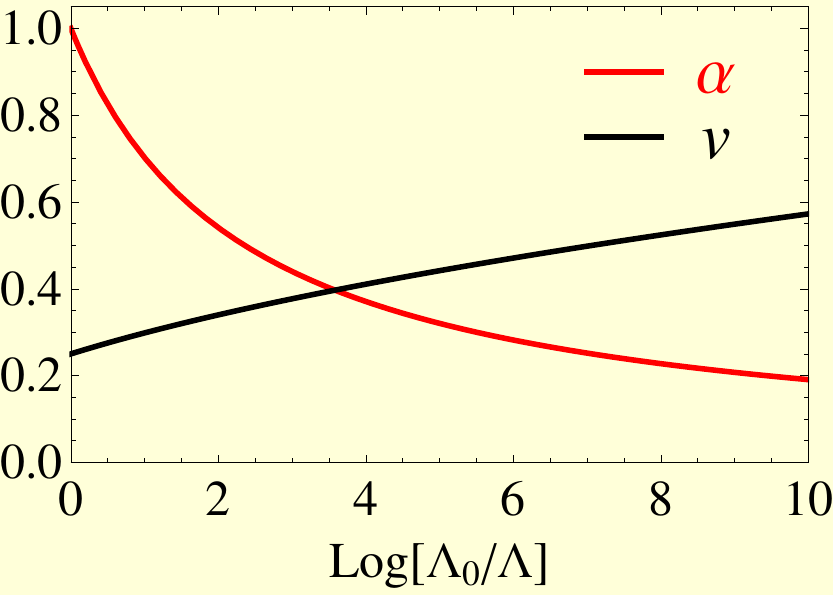}
\caption{(Color online) RG flows for the fine structure constant ($\alpha$) and the Fermi velocity ($v$) due to long range Coulomb interaction. Here, we set $\alpha_0=1.0$ and $v_{0}=0.25$, where the quantities with subscript ``$0$" correspond to their bare values. $\Lambda$ represents the running infrared cut-off, and $\Lambda_0$ to the ultraviolet cut-off. Thus, as we approach the infrared side of the theory, $\Lambda$ decreases monotonically.}\label{Coulomb-flow}
\end{figure}

If we only focus on the long range tail of the Coulomb interaction and neglects its short range peices, we find that as the system approaches deep infrared regime the fine structure constant decreases monotonically. However, the Fermi velocity increases logarithmically~\cite{goswami-chakravarty, isobe-nagaosa, hosur, throckmorton-3d, roy-migdal}. This situation is depicted in Fig.~\ref{Coulomb-flow}.

In the presence of short-range pieces of the Coulomb interaction, we find the there are all together four QCPs, reported in Sec.~\ref{clean-interacting}. Therefore, long range tail of the Coulomb interaction does not change the quantum critical behavior in three dimensinal DSMs. However, due to corrections to the flow equations of short range Coulomb interactions, arising from its long range tail, the phase boundary between DSM and various BSPs changes, but in a \emph{non-universal} fashion, and a representative phase diagram in $g_2-g_5$ plane is shown in Fig.~\ref{pd-DiracCoulomb}. Comparing the phase boundaries in Figs.~\ref{pd-DiracClean} and \ref{pd-DiracCoulomb}, we find that long range Coulomb interaction enhances the propensity of various ordering (insulation) in DSMs. This outcome stems from the fact that long range Coulomb interaction increases the anomalous dimension of all Dirac mass operators.

\section{Fierz identity}\label{Append-fierz}

We devote this Appendix to demosntrate how one can reduce the number of linearly independent couplings using, so called the \emph{Fierz identity}~\cite{HJR}. Let us define a eight component vector as
\begin{eqnarray}
X^\top &=& \big[ \left( \bar{\Psi} \gamma_0 \Psi \right)^2, \left( \bar{\Psi} \Psi \right)^2, \left( \bar{\Psi} \gamma_0 \gamma_j \Psi \right)^2,\left( \bar{\Psi} \gamma_0 \gamma_5 \Psi \right)^2, \nonumber \\
&& \left( \bar{\Psi} i\gamma_5 \Psi \right)^2, \left( \bar{\Psi} \gamma_l \gamma_k \Psi \right)^2, \left( \bar{\Psi} \gamma_5 \gamma_j \Psi \right)^2, \left( \bar{\Psi} i\gamma_j \Psi \right)^2 \big]. \nonumber \\
\end{eqnarray}
The Fierz transformation allows one to write each quartic term as linear combination of the remaining, which follows from the following relation
\begin{eqnarray}
\left[\bar{\Psi} (x) M \Psi(x) \right]\left[\bar{\Psi} (y) N \Psi(y) \right]=-\frac{1}{16} \mbox{Tr} \left[M \Gamma_a N \Gamma_b \right] \nonumber \\
\left[\bar{\Psi}(x) \Gamma_a \Psi(y) \right] \left[\bar{\Psi}(y) \Gamma_b \Psi(x) \right],
\end{eqnarray}
and for contact interactions, as considered in Eq.~(\ref{Lint3D}), $x=y$. The \emph{minus} sign in the right hand side of the above equation comes from the Grasmann nature of the fermionic fields, $\bar{\Psi}$ and $\Psi$. The space of four-dimensional matrices is spanned by the basis $\Gamma_a$, $a=1,2, \cdots, 16$. The set of eight equations of constraint can be written compactly as $F X=0$, where
\begin{eqnarray}
F=\left( \begin{array}{c c c c c c c c}
5 & 1 & 1 & 1 & 1 & 1 & 1 & 1 \\
1 & 5 & -1 & -1 & -1 & 1 & 1 & -1 \\
3 & -3 & 3 & -3 & 3 & 1 & -1 & 1 \\
1 & -1 & -1 & 5 & -1 & -1 & 1 & 1 \\
1 & -1 & 1 & -1 & 5 & -1 & 1 & -1 \\
3 & 3 & 1 & -3 & -3 & 3 & -1 & 1 \\
3 & 3 & -1 & 3 & 3 & -1 & 3 & -1 \\
3 & -3 & 1 & 3 & -3 & 1 & -1 & 3
\end{array}
\right).
\end{eqnarray}

The rank of the above matrix ($F$) is \emph{four}. Hence, out of eight contact interaction terms in Eq.~(\ref{Lint3D}) only $8-4=4$ are linearly independent. For convenience, we chose $g_1$, $g_2$, $g_4$ and $g_5$ as independent couplings. Then rest of the quartic terms can be expressed as linear combinations of these four independent couplings according to
\begin{eqnarray}\label{constraintcouplings}
g_3=-g_1+g_2+g_4-2 g_5,  g_6=-g_1-2 g_2 +g_4+g_5, \nonumber \\
g_7=-2 g_1-g_2 -g_4-g_5, g_8=-g_1 +g_2 -2 g_4+g_5.
\end{eqnarray}
\\

\begin{figure}[htb]
\includegraphics[width=8.2cm,height=4.5cm]{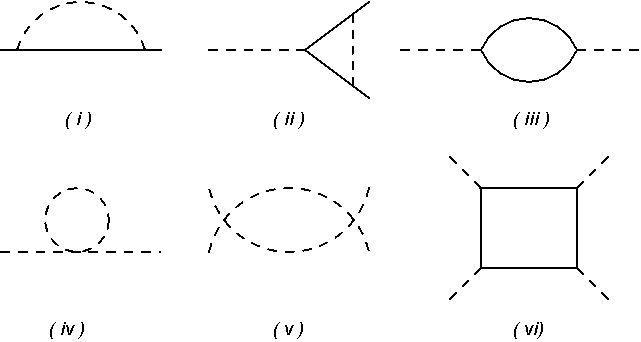}
\caption{Diagrams contributing to the renormalization of $\lambda$, $g$ and $m^2$ [see Eq.~(\ref{GNY})] to the leading order in $\epsilon$. The dotted and solid lines respectively represent boson and fermion.}\label{GNY-Fig}
\end{figure}

\section{Order parameter theory: Gross-Neveu-Yukawa formalism in $d=3$}\label{Append-GNY}

In this Appendix, we display the order parameter description of the QPT, out of DSM into the ${\mathcal P}$, ${\mathcal T}$ odd insulator (pseudo-scalar mass). The corresponding imaginary action reads as
\begin{equation}\label{GNY}
S=\int d^dx \: {\mathcal L}_{total},
\end{equation}
where ${\mathcal L}_{total}={\mathcal L}_{F} + {\mathcal L}_{B} + {\mathcal L}_{B-F}$. Various components of ${\mathcal L}_{total}$ are given by
\begin{eqnarray}
{\mathcal L}_{F} &=&\bar{\Psi} \gamma_\mu \partial_\mu \Psi, \nonumber \\
{\mathcal L}_{B-F} &=& g \; \Phi \; \bar{\Psi} i \gamma_5 \Psi, \nonumber \\
{\mathcal L}_{B} &=& \frac{1}{2} \left( \partial_\mu \Phi \right)^2 +\frac{1}{2} m^2 \; \Phi^2 +\frac{\lambda}{4!} \Phi^4.
\end{eqnarray}
The above formalism is also known as \emph{Gross-Neveu-Yukawa} theory~\cite{zinn-justin, moshe-moshe}. We now perform an $\epsilon$-expansion around the upper critical dimension (space-imaginary time) $d_u=4$, where $\epsilon=4-d$. The diagrams that give rise to corrections to various coupling constants appearing in the theory to the leading order in $\epsilon$ are shown in Fig.~\ref{GNY-Fig}.

\begin{table}[h]
  \begin{tabular}{|c||c|c|c|c|c|}
     \hline
  Source & Order parameter & C1 & C2 & C3 & C4 \\
     \hline \hline
$\Delta_1$ & chemical potential & 0 & 0 & 0 & 0 \\
\hline
$\Delta_2$ & scalar mass & 1.75 & -0.51 & 1.69 & -2\\
\hline
$\Delta_3$ & spin-orbit coupling & 0.25 & 0.13 & 0.13 & 0 \\
\hline
$\Delta_4$ & axial chemical potential & 0 & 0 & 0 & 0 \\
\hline
$\Delta_5$ & pseudo-scalar mass & 1.75 & 1.69 & -0.51 & -2\\
\hline
$\Delta_6$ & magnetization & 0.083 & -0.44 & 0.29 & -0.67\\
\hline
$\Delta_7$ & axial magnetization & -0.33 & -0.52 & -0.52 & 2.67\\
\hline
$\Delta_8$ & current & 1 & -0.27 & -0.27 & 0\\
\hline \hline
$\Delta_S$ & s-wave pairing & -0.5 & -0.77 & -0.77 & 4\\
\hline
$\Delta_{op}$ & odd-parity pairing & 1.5 & -0.41 & 0.41 & 0\\
\hline
$\Delta_{V,j}$ & vector (nodal) pairing & -1.67 & -0.59 & 0.88 & 1.33\\
\hline
$\Delta_{V,0}$ & time-like vector pairing & 0 & 0 & 0 & 0 \\
\hline
  \end{tabular}
\caption{Physical meaning and anomalous dimensions of various fermion bilinear at different QCPs (C1, C2, C3, and C4, see also Sec.~\ref{clean-interacting}). Numbers are in units of $\epsilon_1/2$. }\label{table-anomalous}
\end{table}

The RG flow equations (infrared) are given by 
\begin{eqnarray}
\beta_{g^2} &=&\epsilon g^2 -(2 N +3) g^4, \nonumber \\
\beta_{\lambda} &=& \epsilon \lambda -\frac{3 \lambda^2}{2} -4 N \lambda g^2 + 24 N g^4,
\end{eqnarray}
after taking $g^2 N_d \to g^2$ and $\lambda N_d \to \lambda$, where $N_d= S_d/(2 \pi)^d$, $S_d$ is the surface area of $d$-dimensional unit sphere, and $N$ is the number of $4$-component Dirac fermions (thus $N=1$ in our problem). These two coupled flow equations support only trivial solution $(g^2_\ast,\lambda_\ast)=(0,0)$ in $d=4$ ($\epsilon=0$).

On the other hand, the flow equation of mass ($m$) of the order parameter field that tunes the transition out of symmetric DSM phase to a ${\mathcal P}$, ${\mathcal T}$ odd insulator, is
\begin{equation}
\beta_{m^2}= m^2 \left( 1+ 2 N g^2 + \frac{\lambda}{2} \right),
\end{equation}
from which we determine the CLE ($\nu$)
\begin{equation}
\nu^{-1}=2 - (2 N g^2_\ast +\frac{\lambda_\ast}{2})=2.
\end{equation}
Therefore, the CLE is $\nu=\frac{1}{2}$ at the Gaussian QCP.

\section{Susceptibility of order parameters}\label{Append-susceptibility}

\begin{figure}[htb]
\includegraphics[width=8.2cm,height=2.0cm]{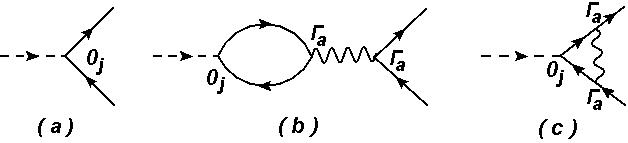} \\
\includegraphics[width=6.2cm,height=2.3cm]{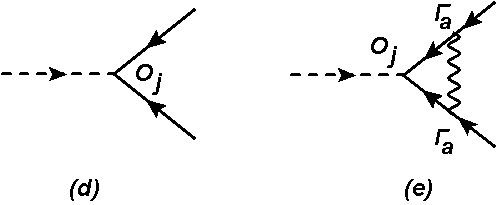}\\
\includegraphics[width=8.2cm,height=2.3cm]{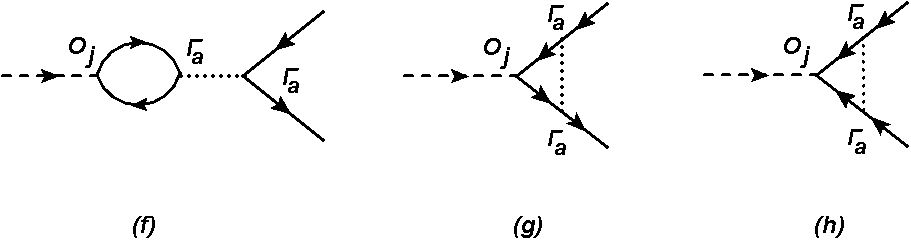}
\caption{Order parameter vertex is represented by the matrix $O_j$. Bare order parameter vertex in the particle-hole and particle-particle channels are shown in diagrams (a) and (d), respectively. Renormalization of bare vertex by four-fermion interactions $\left(\bar{\Psi} \Gamma_a \Psi \right)^2$ arises from diagrams (b) and (c) (particle-hole channel), and (e) (pairing channels). Disorder induced renormalization of fermion bilinears in particle-hole [(f) and (g)] and particle-particle [(h)] channels.  }\label{susceptibility-diag}
\end{figure}

In this appendix we compute the anomalous dimensions and RG flow equations of various order parameters (in particle-hole and particle-particle channels). Analysis of flows of various fermion bilinears allows us to pin the nature of BSPs across various QCPs unambiguously. To proceed with the calculation, we consider all the symmetry allowed order parameters and define an action $S_{s}=-\int d^3x d\tau {\mathcal L}_s$, where the Lagrangian density for the source terms is
\begin{eqnarray}
{\mathcal L}_{S} &=& \Delta_1 \bar{\Psi}\gamma_0 \Psi + \Delta_2 \bar{\Psi} \Psi + \Delta_3 \bar{\Psi}\gamma_0 \gamma_j \Psi + \Delta_4 \bar{\Psi}\gamma_0 \gamma_5 \Psi \nonumber \\
&+& \Delta_5 \bar{\Psi} i \gamma_5 \Psi + \Delta_6 \bar{\Psi} \gamma_j \gamma_k \Psi + \Delta_7 \bar{\Psi}\gamma_5 \gamma_j \Psi  \nonumber \\
&+& \Delta_8\bar{\Psi}i \gamma_j \Psi + \Delta_{S} \Psi^\dagger i\gamma_0 \gamma_5 \gamma_2 \Psi^\ast+ \Delta_{op}\Psi^\dagger i\gamma_0  \gamma_2 \Psi^\ast  \nonumber \\
&+& \Delta_{V,1}\Psi^\dagger \gamma_3  \Psi^\ast + \Delta_{V,2}\Psi^\dagger i\gamma_0 \gamma_5  \Psi^\ast+\Delta_{V,3}\Psi^\dagger \gamma_1  \Psi^\ast \nonumber \\
&+& \Delta_{V,0}\Psi^\dagger i\gamma_0\gamma_1 \gamma_3 \Psi^\ast.
\end{eqnarray}
RG flow equations for various order parameters ($\Delta_j$s) due to the four-fermions interactions (see Fig.~\ref{susceptibility-diag}) and disorders are given by
\begin{eqnarray}
 \bar{\beta}_{\Delta_1} &=& \Delta_V+\Delta_A+3 \Delta_{SO},\nonumber \\
\bar{\beta}_{\Delta_2}  &=& \frac{-g_1+3 g_2 + g_4 + g_5}{2}- \Delta_V+ \Delta_A+3 \Delta_{SO}, \nonumber \\
\bar{\beta}_{\Delta_3}  &=& \frac{-g_1+g_2+g_4-g_5}{6} + \frac{\Delta_V-\Delta_{A}-\Delta_{SO}}{3}, \nonumber \\
\bar{\beta}_{\Delta_4}  &=& \Delta_V+ \Delta_A -3\Delta_{SO}\nonumber  
\end{eqnarray}
\begin{eqnarray} 
\bar{\beta}_{\Delta_5}  &=& \frac{1}{2} \left(-g_1+g_2+g_4 +3 g_5 \right) - \Delta_V +\Delta_A+3 \Delta_{SO}, \nonumber \\
\bar{\beta}_{\Delta_6}  &=& \frac{1}{6} \left(g_1+g_2-g_4-g_5 \right) + \frac{\Delta_V}{3} -\frac{\Delta_A}{3}+ \frac{\Delta_{SO}}{3}, \nonumber \\
\bar{\beta}_{\Delta_7} &=& -\frac{1}{3} \left( g_1+g_2+g_4 + g_5\right) - \frac{\Delta_V}{3}-\frac{\Delta_A}{3}+ \frac{\Delta_{SO}}{3}, \nonumber \\
\bar{\beta}_{\Delta_8} &=& -\frac{1}{3} \left( g_1-g_2+ g_4 -g_5 \right) - \frac{\Delta_V}{3} -\frac{\Delta_A}{3}-\frac{\Delta_{SO}}{3} , \nonumber \\
\bar{\beta}_{\Delta_S} &=& -\frac{1}{2} \left( g_1+g_2+ g_4 +g_5 \right) -\Delta_V+\Delta_A -3 \Delta_{SO}, \nonumber\\
\bar{\beta}_{\Delta_{op}} &=& -\frac{1}{2} \left( g_1-g_2+ g_4 -g_5 \right)+ \Delta_V -\Delta_A-3 \Delta_{SO} \nonumber \\
\bar{\beta}_{\Delta_{V,j}} &=& \frac{1}{3} \left(- g_1+g_2+ g_4 -g_5 \right)+ \frac{\Delta_V}{3}+\frac{\Delta_A}{3}+\frac{\Delta_{SO}}{3}, \nonumber \\
\bar{\beta}_{\Delta_{V,0}} &=&\Delta_V +\Delta_A-3 \Delta_{SO},
\end{eqnarray}
where 
\begin{equation}
\bar{\beta}_{X} = \frac{d \log \Delta_X}{dl}-1
\end{equation}
When we substitute the values for the coupling constants at various QCPs, quantities in the right hand side of each equation yields to the \emph{anomalous dimension} of associated fermion bilinear. The anomalous dimension of each fermion bilinear near various interacting QCPs and their physical meanings are quoted in Table~\ref{table-anomalous}. To determine the nature of BSPs, we simultaneously run the flow equations of quartic coupling constants ($g_{1,2,4,5}$), disorder (only for dirty DSM) and fermionic susceptibilities. The leading divergent channel as the coupling constants accede threshold values (critical couplings) determines the actual BSP. The resulting phase diagrams are shown in Fig.~\ref{Dirac-PD}.

\section{Details of $\epsilon$ and double $\epsilon$-expansion}\label{double-epsilon}   

In the main part of the paper, we have presented the RG calculation in clean interacting (dirty noninteracting) system based on an $\epsilon$-expansion, performed about one (two) spatial dimensions in Sec.~\ref{clean-interacting} (\ref{dirty-DSM}). In the presence of both electronic interaction and disorder we carried the RG calculation by invoking the notion of a double $\epsilon$-expansion, see Sec~\ref{interaction-disorder}. Here we present some essential details of the diagramatic calculation. We layout all the key steps for evaluating different classes of diagram in the presence of generic interaction and disorder. Furthermore, we display explicit evaluation of each and every diagrams shown in Fig.~\ref{Feynman-Diag} for a particular set of interaction and disorder couplings, namely $g_5$ which when strong gives rise to pseudo scalar mass in clean system and $\Delta_A$ (axial disorder), which when strong supports a diffusive metal in a dirty noninteracting system. In order to keep the discussion on technical details coherent, we discuss three cases separately (i) the $\epsilon_1$-expansion around $d=1$ for clean interacting model, (ii) the $\epsilon_2$ expansion around $d=2$ for dirty noninteracting system, and (iii) the double $\epsilon$-expansion in the presence of both interaction and disorder. The fermion Greens function as a function of Matsubara frequency reads as 
\begin{equation}
G(i\omega, \mathbf{k})=-\; \frac{i\omega \gamma_0 +i \gamma_j v k_j}{\omega^2+v^2 k^2},
\end{equation}
where summation over repeated spatial indices ($j=1,2,3$) is assumed. 
\\

\begin{widetext}

First we discuss the clean interacting model. We here schematically denote the coupling constant as $g_a$ when the interaction vertex is accompanied by a $4 \times 4$ matrix $\Gamma_a$. For different interaction channel (accompanied by distinct $4 \times 4$ matrices) one can read off appropriate coupling constant from Eq.~(\ref{int-reduced}). The contribution from diagram $(ii)$ in Fig.~\ref{Feynman-Diag} goes as 
\begin{eqnarray}
&& (2,ii)=  -2 \; g^2_a \; \left( \bar{\Psi} \Gamma_a \Psi \right)^2 \:\: {\mathbf{Tr}} \int^{\prime} \frac{d^d \mathbf k}{(2\pi)^d} \int^{\infty}_{-\infty} \frac{d\omega}{2 \pi} \left[\Gamma_a G(i \omega, \mathbf k) \Gamma_a G(i \omega, \mathbf k) \right] 
= 4 g^2_5 \left( \bar{\Psi} i \gamma_5 \Psi \right)^2 \;\int^{\prime} \frac{d^d \mathbf k}{(2\pi)^d} \frac{1}{vk} \nonumber \\
&& = g^2_5 \; \left( \bar{\Psi} i \gamma_5 \Psi \right)^2 \: \left[ \frac{8 S_d}{v (2 \pi)^d} \right] \; \int^{\Lambda}_{\Lambda e^{-l}} k^{d-2} dk  = (2 g^2_5) \; \left( \bar{\Psi} i \gamma_5 \Psi \right)^2 \: \left[\frac{2 S_d \Lambda^{\epsilon_1}}{v (2\pi)^d} \right] \: l + {\mathcal O}(l^2),
\end{eqnarray} 
for $\Gamma_a=i\gamma_5$, where $d=1+\epsilon_1$ and ${\mathbf{Tr}}$ is taken over four component $\gamma$ matrices. Throughout this Appendix all femrionic fields are slow variables, obtained after integrating over the fast Wilsonian shell. Contribution from diagram $(iii)$ in Fig.~\ref{Feynman-Diag} is given by
\begin{eqnarray}
 (2,iii) = 4 g_a g_b \left( \bar{\Psi} \Gamma_a \Psi \right) \bar{\Psi} \int^{\prime} \frac{d^d \mathbf k}{(2\pi)^d} \int^{\infty}_{-\infty} \frac{d\omega}{2 \pi} \left[ \Gamma_b G(i \omega, \mathbf k) \Gamma_a G(i \omega, \mathbf k) \Gamma_b \right] \Psi  =-g^2_5 \left( \bar{\Psi} i \gamma_5 \Psi \right)^2 \left[\frac{2 S_d \Lambda^{\epsilon_1}}{v (2\pi)^d} \right] l  +  {\mathcal O}(l^2),
\end{eqnarray} 
for $\Gamma_a=\Gamma_b=i \gamma_5$. Contributions from diagrams $(iv)$ and $(v)$ goes as  
\begin{eqnarray}
&& (2, iv)+(2,v)= 4  g_a g_b \int^{\prime} \frac{d^d \mathbf k}{(2\pi)^d} \int^{\infty}_{-\infty} \frac{d\omega}{2 \pi} \left( \bar{\Psi} \Gamma_a G(i\omega,\mathbf k) \Gamma_b \Psi\right) \; \bar{\Psi} \left[ \Gamma_b G(i\omega,\mathbf k) \Gamma_a + \Gamma_a G(-i\omega,-\mathbf k) \Gamma_b \right] \Psi \nonumber \\
&&= -\frac{4}{3} g^2_5 (1-1) \left( \bar{\Psi} \gamma_j \Psi \right)^2 \int^{\prime} \frac{d^d \mathbf k}{(2\pi)^d} \int^{\infty}_{-\infty} \frac{d\omega}{2 \pi} \frac{v^2 k^2}{(\omega^2+ v^2 k^2)^2} -4 g^2_5 (1-1) \left( \bar{\Psi} \gamma_0 \Psi \right)^2 \; \int^{\prime} \frac{d^d \mathbf k}{(2\pi)^d} \int^{\infty}_{-\infty} \frac{d\omega}{2 \pi} \frac{\omega^2}{(\omega^2+ v^2 k^2)^2} \nonumber \\
&&=\left[-\frac{1}{3}(1-1) \right] g^2_5 \left( \bar{\Psi} \gamma_j \Psi \right)^2 \left[\frac{2 S_d \Lambda^{\epsilon_1}}{v (2\pi)^d} \right] l + (1-1) g^2_5 \left( \bar{\Psi} \gamma_0 \Psi \right)^2 \left[\frac{2 S_d \Lambda^{\epsilon_1}}{v (2\pi)^d} \right] l + {\mathcal O} (l^2),
\end{eqnarray}
for $\Gamma_a=\Gamma_b=i \gamma_5$. Due to a special property of the Greens function, namely $G(-i \omega, -\mathbf k)=-G(i \omega, \mathbf k)$, contributions from diagrams $(iv)$ and $(v)$ mutually cancel each other whenever $\Gamma_a=\Gamma_b$. Thus after the evaluating diagrams $(ii)$ and $(iii)$ in Fig.~\ref{Feynman-Diag} we arrive at the flow equation for $g_5$ displayed in Eq.~(\ref{g5RG}) in terms of dimensionless coupling constant, defined as $g_5 \left[\frac{2 S_d \Lambda^{\epsilon_1}}{v (2\pi)^d} \right] \to g_5$. Thus following the above prescription, contributions from these four Feynman diagrams [$(ii)-(v)$ in Fig.~\ref{Feynman-Diag}] can be evaluated in the presence of generic four-fermion interaction [with various combinations of $4 \times 4$ matrices $\Gamma_a$ and $\Gamma_b$ as shown in Eq.~(\ref{int-reduced})], which ultimately leads to the flow equations announced in Eq.~(\ref{RGintcoup}). Notice that diagrams $(iv)$ and $(v)$ can potentially generate new four-femrion interaction [such as $g_{3,6,7,8}$ from Eq.~(\ref{Lint3D})] that are not contained in $L_{int}$, shown in Eq.~(\ref{int-reduced}). Whenever such new four-fermion interactions are generated we rewrite them in term of $g_{1,2,4,5}$ by using the Fierz constraints, shown in Eq.~(\ref{constraintcouplings}).
\\

Next we display the details of diagramatic calculation in a dirty noninteracting system. For notational convenience, we denote the disorder coupling as $\Delta_a$ when the disorder vertex is accompanied by a $4 \times 4$ matrix $\Gamma_a$. Then we present the calculation explicitly for the axial disorder, for which $\Gamma_a=\gamma_0 \gamma_5$ and $\Delta_a=\Delta_A$. For various other choices of $\Gamma_a$ and corresponding coupling constants readers should see Eq.~(\ref{disorder-action}). 

The correction to the fermionic self-energy arising from diagram $(vii)$ in Fig.~\ref{Feynman-Diag} is given by 
\begin{eqnarray}
(2, vii) &=&  \Delta_a \bar{\Psi}_\alpha \left( \int^{\prime} \frac{d^d \mathbf k}{(2\pi)^d} \: \Gamma_a G(i \omega, \mathbf k) \Gamma_a \right) \Psi_\alpha = \Delta_A \: \left(\bar{\Psi} i\omega \gamma_0 \Psi\right) \: \frac{S_d}{(2 \pi)^d} \; \int^{\Lambda}_{\Lambda e^{-l}} \; k^{d-3} dk \nonumber \\
&=& \Delta_A \: \left( \bar{\Psi}_\alpha i\omega \gamma_0 \Psi_\alpha \right) \: \frac{S_d \Lambda^{\epsilon_2}}{(2 \pi)^d} l + \mathcal{O}(l^2) 
\rightarrow \Delta_A \: \left(\bar{\Psi}_\alpha i\omega \gamma_0 \Psi_\alpha \right) \; l + \mathcal{O}(l^2),
\end{eqnarray}
for $d=2+\epsilon_2$ and $\Gamma_a =\gamma_0 \gamma_5$, after introducing dimensionless disorder coupling, defined as $\Delta_A \left[ \frac{S_d \Lambda^{\epsilon_2}}{v^2 (2 \pi)^d} \right] \to \Delta_A$. From the self energy correction we find the field renormalization coefficient, defined as $\Psi \to Z^{-1/2}_\Psi \Psi$, to be $Z_\Psi=e^{d l} [1-\Delta_A]$. The renormalization coefficient for Fermi velocity ($v$), defined as $v \to Z^{-1}_v v$, is given by $Z_v=e^{(z-1)l} [1-\Delta_A]$. From $Z_v$ we obtain the flow equation of $v$ announced in Eq.~(\ref{dirtyflow}) (for $\Delta_V=0$). Here $\alpha$ is the replica index.

Contribution from diagram $(viii)$ in Fig.~\ref{Feynman-Diag} goes as 
\begin{eqnarray}\label{disorder-RG-vert}
&&(2, viii)= 4 \Delta_a \Delta_b \; \left( \bar{\Psi}_\alpha \Gamma_a \Psi_\alpha \right) \bar{\Psi}_\beta \left( \int^{\prime}  \frac{d^d \mathbf k}{(2\pi)^d} \left[ \Gamma_b G(0, \mathbf k) \Gamma_a G(0, \mathbf k) \Gamma_b \right] \right) \Psi_\beta \nonumber \\
&&= (2 \Delta^2_A) \; \left( \bar{\Psi}_\alpha \gamma_0 \gamma_5 \Psi_\alpha \right) \left( \bar{\Psi}_\beta \gamma_0 \gamma_5 \Psi_\beta \right) \;\int^{\prime} \frac{d^d \mathbf k}{(2\pi)^d} \frac{1}{v^2k^2} = (2 \Delta^2_A) \; \left( \bar{\Psi}_\alpha \gamma_0 \gamma_5 \Psi_\alpha \right) \left( \bar{\Psi}_\beta \gamma_0 \gamma_5 \Psi_\beta \right) \; \left[ \frac{S_d}{v^2 (2 \pi)^d} \right] \: \int^{\Lambda}_{\Lambda e^{-l}} k^{d-3} dk   
\nonumber \\
&&= (2 \Delta^2_A) \; \left( \bar{\Psi}_\alpha \gamma_0 \gamma_5 \Psi_\alpha \right) \left( \bar{\Psi}_\beta \gamma_0 \gamma_5 \Psi_\beta \right) \; \left[ \frac{S_d \Lambda^{\epsilon_2}}{v^2 (2 \pi)^d} \right] \; l + {\mathcal O}(l^2),
\end{eqnarray}
where $d=2+\epsilon_2$ and for $\Gamma_a=\Gamma_b=\gamma_0 \gamma_5$. Together the contribution from diagrams $(ix)$ and $(x)$ reads as 
\begin{eqnarray}
&& (2, ix)+(2,x)= 4  \Delta_a \Delta_b \int^{\prime} \frac{d^d \mathbf k}{(2\pi)^d} \left( \bar{\Psi}_\alpha \Gamma_a G(0,\mathbf k) \Gamma_b  \Psi_\alpha \right) \; \bar{\Psi}_\beta \left[ \Gamma_b G(0,\mathbf k) \Gamma_a + \Gamma_a G(0,-\mathbf k) \Gamma_b \right] \Psi_b \nonumber \\
&&=4 \Delta^2_A (1-1) \left( \bar{\Psi}_\alpha \gamma_j \Psi_\alpha \right) \left( \bar{\Psi}_\beta \gamma_j \Psi_\beta \right) \int^{\prime} \frac{d^d \mathbf k}{(2\pi)^d} \frac{1}{v^2 k^2}
=4 \Delta^2_A (1-1) \left( \bar{\Psi}_\alpha \gamma_j \Psi_\alpha \right) \left( \bar{\Psi}_\beta \gamma_j \Psi_\beta \right) \left[ \frac{S_d \Lambda^{\epsilon_2}}{v^2 (2 \pi)^2} \right] \; l + \mathcal O (l^2). \nonumber \\
\end{eqnarray}
when $\Gamma_a=\Gamma_b=\gamma_0 \gamma_5$. Thus due to the special property of Greens finction (odd function of frequency and momentum) contribution of these two diagrams mutually cancel each other. However, for generic disorder (for exmaple spin-orbit disorder) these two diagrams do not cancel each other and their net contribution can be evaluated from the first line of the above expression. Hence, from Eq.~(\ref{disorder-RG-vert}), we immediately arrive at the flow equation of disorder coupling $\Delta_A$, as shown in the last equation of Eq.~(\ref{dirtyflow}) in terms of dimensionless coupling (after setting $\Delta_V=0$). 
\\

Finally we expose the details of the diagramatic calculation in the presence of both electronic interaction and disorder. When these two perturbations are simultaneously present one needs to account for additional Feynman diagrams, namely $(xi)-(xv)$ in Fig.~\ref{Feynman-Diag}. Calculations in the presence of generic interaction and disorder can be carried out following the prescription, highlighted below, as we have given the expression for each such mixed diagrams for arbitrary interaction and disorder vertices. For further illustration, we here also present evaluation of these diagrams in the presence of interaction in the $g_5$ channel and axial disorder ($\Delta_A$).

Notice that diagram $(xi)$ in Fig.~\ref{Feynman-Diag} renormalizes disorder vertex. Correction to disorder vertex $\Delta_a$ (associated with matrix $\Gamma_a$) due to electronic interaction $g_b$(accompanied by matrix $\Gamma_b$), arising from the diagram $(xi)$, is given by 
\begin{eqnarray}
&&(2,xi)= g_b \Delta_a \left( \bar{\Psi}_\alpha \Gamma_a \Psi_\alpha \right) \bar{\Psi}_\beta \left[ \int^{\prime} \frac{d^d \mathbf k}{(2\pi)^d} \int^{\infty}_{-\infty} \frac{d\omega}{2 \pi} \left[ \Gamma_b G(i\omega, \mathbf k) \Gamma_a G(i\omega, \mathbf k) \Gamma_b \right] \right] \Psi_\beta
= g_5 \Delta_A \left( \bar{\Psi}_a \gamma_0 \gamma_5 \Psi_a \right)  \nonumber \\ 
&&\times \left( \bar{\Psi}_b \gamma_0 \gamma_5 \Psi_b \right) \frac{S_d}{v(2 \pi)^d} \; (1-1) \; \int^{\Lambda}_{\Lambda e^{-l}} k^{d-2} dk  =  (g_5 \Delta_A) \left( \bar{\Psi}_a \gamma_0 \gamma_5 \Psi_a \right) \left( \bar{\Psi}_b \gamma_0 \gamma_5 \Psi_b \right) \; \frac{S_d \Lambda^{\epsilon_1}}{v(2 \pi)^d} \; (1-1) \; l + \mathcal{O} (l^2),
\end{eqnarray}
for $d=1+\epsilon_1$, $\Gamma_a=\gamma_0 \gamma_5$ and $\Gamma_b=i \gamma_5$. Although the explicit contribution of such diagram is trivial, this exercise sets the stage to carry out the perturbative analysis to capture the interplay of generic interaction and disorder.

On the other hand, contribution for diagram $(xii)$ that renoramlizes interaction coupling constant $g_a$ goes as 
\begin{eqnarray}
(2, xii)&=& 2 g_a \Delta_b \left( \bar{\Psi}_\alpha \Gamma_a \Psi_\alpha \right) \bar{\Psi}_\beta \int^{\prime} \frac{d^d \mathbf k}{(2\pi)^d} \left[ \Gamma_b G(0, \mathbf k) \Gamma_a G(0, \mathbf k) \Gamma_b \right] \Psi_\beta 
= (2 g_5 \Delta_A) \; \left( \bar{\Psi}_\alpha i \gamma_5 \Psi_\alpha \right)^2 \frac{S_d}{(2 \pi)^d} \int^{\Lambda}_{\Lambda e^{-l}} k^{d-3} dk \nonumber \\
&=& (2 g_5 \Delta_A) \; \left( \bar{\Psi}_\alpha i \gamma_5 \Psi_\alpha \right)^2  \: \: \left[ \frac{S_d \Lambda^{\epsilon_2}}{(2 \pi)^d} \right] \; l + \mathcal{O} (l^2) 
\end{eqnarray}
for $d=2+\epsilon_2$, $\Gamma_a=i \gamma_5$ and $\gamma_b=\gamma_0 \gamma_5$. Diagrams $(xiii)$ and $(xiv)$ also renormalize the interaction vertex to the one-loop order. The contribution from these two diagrams together reads as
\begin{eqnarray}
&& (2, xiii)+ (2, xiv)= 4 g_a \Delta_b \int^{\prime} \frac{d^d \mathbf k}{(2 \pi)^d} \left( \bar{\Psi}_\alpha \Gamma_b G(0, \mathbf k) \Gamma_a \Psi_\alpha \right) \; \bar{\Psi}_\alpha \left[\Gamma_b G(0, \mathbf k) \Gamma_a + \Gamma_a G(0, -\mathbf k) \Gamma_b \right] \Psi_\alpha \nonumber \\
 && = \frac{4}{3} g_5 \Delta_A \left( \bar{\Psi}_\alpha \gamma_0 \gamma_j \Psi_\alpha \right)^2 (1-1) \frac{S_d}{v^2 (2 \pi)^d} \int^{\Lambda}_{\Lambda e^{-l}} k^{d-3} dk 
= \frac{4}{3} g_5 \Delta_A \left( \bar{\Psi}_\alpha \gamma_0 \gamma_j \Psi_\alpha \right)^2 (1-1) \: \left[\frac{S_d \Lambda^{\epsilon_2}}{v^2 (2 \pi)^d} \right] \; l + \mathcal O (l^2),
\end{eqnarray}
for $d=2+\epsilon_2$, and $\Gamma_a=i \gamma_5$, $\Gamma_b=\gamma_0 \gamma_5$. Although contribution from these two diagrams vanishes for $\Gamma_a=i\gamma_5$ and $\Gamma_b=\gamma_0 \gamma_5$, in general yields renormalization to four femrion interaction. Note that set of these two diagrams can in principle generate new four fermion terms proportional to $g_j$ with $j=3, 6,7,8$. When such four-fermion terms are generated, we rewrite them in terms of $g_k$ where $k=1,2,4,5$ by using the Fierz constraints from Eq.~(\ref{constraintcouplings}), so that the interacting Lagrangian $L_{int}$ from Eq.~(\ref{int-reduced}) remains closed under renormalization group procedure. The last Feynman diagram $(xv)$ from Fig.~\ref{Feynman-Diag} also renormalizes interaction vertex and its contribution reads as 
\begin{eqnarray}
&& (2, xv)=-4 g_b \Delta_a \left( \bar{\Psi} \Gamma_a \Gamma_b \Psi \right)^2\; \mathbf{Tr} \; \int^{\prime} \frac{d^d \mathbf k}{(2 \pi)^d} \; \left[ \Gamma_a G(0, \mathbf k) \Gamma_b G(0, \mathbf k) \right]
= \mp 4 g_b \Delta_a \left( \bar{\Psi} \Gamma_a \Gamma_b \Psi \right)^2\; \mathbf{Tr} (\Gamma_a \Gamma_b) \; \left[ \frac{S_d}{v^2 (2 \pi)^2}\right] \nonumber \\
&& \times \int^{\Lambda}_{\Lambda e^{-l}} k^{d-3} dk
=\mp 4 g_b \Delta_a \left( \bar{\Psi} \Gamma_a \Gamma_b \Psi \right)^2\; \mathbf{Tr} (\Gamma_a \Gamma_b) \; \left[ \frac{S_d \Lambda^{\epsilon_2}}{v^2 (2 \pi)^2}\right] \; l + \mathcal O (l^2), 
\end{eqnarray}
where the $\mp$ sign depends on weather $\Gamma_b$ anti-commute or commutes with $G(0, \mathbf k)$. Notice that due to $\mathbf{Tr}$ this diagram can contributes only when $\Gamma_a=\Gamma_b$. Hence, in our calculation such diagram contributes only when we seek to understand the interplay of potential disorder $\Delta_V$ and interaction in the channel $g_1$. In a model for dirty interacting DSM, defined in terms of two coupling constants $g_5$ and $\Delta_A$, upon collecting the contributions from all these diagrams, we arrive at the flow equations 
\begin{eqnarray}
\frac{d g_5}{dl}&=& -\epsilon_1 g_5 + g^2_5 \left[ \frac{2 S_d \Lambda^{\epsilon_1}}{v (2 \pi)^d}\right] + g_5 (2\Delta_A-\Delta_A) \left[ \frac{ S_d \Lambda^{\epsilon_2}}{v^2 (2 \pi)^d}\right], \nonumber \\
\frac{d \Delta_A}{dl}&=&-\epsilon_2 \Delta_A + 2 \Delta^2_A \left[ \frac{S_d \Lambda^{\epsilon_2}}{v^2 (2 \pi)^d}\right] + a g_5 \Delta_A \left[ \frac{2 S_d \Lambda^{\epsilon_1}}{v (2 \pi)^d}\right], \: \:
\frac{dv}{dl}= v\left(z-1-\Delta_A \left[ \frac{S_d \Lambda^{\epsilon_2}}{v^2 (2 \pi)^d}\right] \right).
\end{eqnarray}
For these choices of the coupling constants $a=0$. Multiplying the first equation by $\left[ \frac{2 S_d \Lambda^{\epsilon_1}}{v (2 \pi)^d}\right]$ and the second one by $\left[ \frac{S_d \Lambda^{\epsilon_2}}{v^2 (2 \pi)^d}\right]$, and introducing the dimensionless coupling constants defined above, we arrive at the following flow equations    
\begin{eqnarray}
\frac{d g_5}{dl}=-\epsilon_1 g_5 + g^2_5 + g_5 \Delta_A, \: 
\frac{d\Delta_A}{dl}=-\epsilon_2 \Delta_A + 2 \Delta^2_A + a \Delta_A g_5, \: 
\frac{dv}{dl}=v(z-1-\Delta_A),
\end{eqnarray}
which can readily be obtained from Eq.~(\ref{RG-Full}) upon setting $g_1=g_2=g_4=0$ and $\Delta_V=0$. The model for interacting dirty DSM with only these two couplings constants has been discussed in details in subsection~\ref{intdis-simplest}.  
\\

\emph{Hence, to evaluate the mixed diagrams [$(xi)-(xv)$ in Fig.~\ref{Feynman-Diag}], one first need to identify the coupling constant (in the interaction or disorder channel) that gets renormalized from a given diagram. When disorder (interaction) coupling gets renormalized by interaction (disorder), the shell integration needs to be evaluated about one (two) spatial dimension(s).} This is the key feature of double $\epsilon$-expansion to address the interplay of interaction and disorder in a three dimensional DSM.  We emphasize again that this technique fails completely for the strongly interacting regime where both interaction and disorder and strong as indicated in the upper right hand quadrant of Fig.~\ref{OddP-int-b} with question marks. \\

 We here sketched all the crucial steps for evaluating each and every diagram shown in Fig.~\ref{Feynman-Diag} in terms of arbitrary $4 \times 4$ matrices $\Gamma_a$ and $\Gamma_b$ (or the interaction and disorder coupling constants). The announced steps can now be readily taken over to compute the perturbative corrections in the presence of generic interaction and disorder. The results are quoted in Eq.~(\ref{RGintcoup}) for clean interacting model, in Eqs.~(\ref{dirtyflow}) and (\ref{CSB-disorder}) for dirty noninteracting system, and Eqs.~(\ref{RG-Full}) and (\ref{CSB-disorder-RG-int}) in the presence of interaction and disorder.

\end{widetext}


\begin{thebibliography}{}

\bibitem{belitz} D. Belitz and T. R. Kirkpatrick, Rev. Mod. Phys. {\bf 66}, 261 (1994), and references therein.

\bibitem{belitz-vojta} D. Belitz, T. R. Kirkpatrick, and T. Vojta, Rev. Mod. Phys. {\bf 77}, 579 (2005), and references therein. 

\bibitem{mott-davis} N. F. Mott, and E. A. Davis, \emph{Electronic Processes in Non-Crystalline Materials} ( Oxford University Press, 2nd ed., 1979)

\bibitem{pollak} M. Pollak, M. Ortuno, and A. Frydman, \emph{The Electron Glass}, (Cambridge University Press, 1st ed., 2012). 

\bibitem{filkenstein} A. M. Finkelstein, Zh. Eksp. Teor. Fiz. {\bf 84}, 168 (1983), Sov. Phys. JETP {\bf 57}, 97 (1983).

\bibitem{dashwang-1}  S. Das Sarma, E. H. Hwang, Scientific Reports {\bf 5}, 16655 (2015), and references therein.

\bibitem{abrikosov} A. A. Abrikosov, L. P. Gorkov, and I. E. Dzyaloshinski, \emph{Methods of Quantum Field Theory in Statistical Physics} (Dover Publications, 1975).

\bibitem{pines} D. Pines, and P. Nozieres, \emph{Theory Of Quantum Liquids: Normal Fermi Liquids} (Westview Press, 1994)  

\bibitem{shankar} R. Shankar, Rev. Mod. Phys. {\bf 66}, 129 (1994).

\bibitem{sachdev} S. Sachdev, \emph{Quantum Phase Transitions} (Cambridge University Press, 2nd ed., 2007).

\bibitem{abrahams} \emph{50 Years of Anderson Localization}, edited by by E. Abrahams (World Scientific Publishing Company, 1st ed., 2010).


\bibitem{balatsky} For pedagogical introduction to Dirac materials see  T. O. Wehling, A. M. Black-Schaffer, A. V. Balatsky, Adv. Phys. {\bf 76}, 1 (2014).


\bibitem{fu-kane} L. Fu, and C. L Kane, Phys. Rev. B {\bf 76}, 045302 (2007).

\bibitem{model-TI} C-X. Liu, X-L. Qi, H. Zhang, X. Dai, Z. Fang, and S-C. Zhang, Phys. Rev. B {\bf 82}, 045122 (2010).

\bibitem{review-TI-1} M. Z. Hasan, and C. L. Kane, Rev. Mod. Phys. {\bf 82}, 3045 (2010).

\bibitem{review-TI-2} X-L. Qi, and S-C. Zhang, Rev. Mod. Phys. {\bf 83}, 1057 (2011).

\bibitem{hassan-cava} S.-Y. Xu, Y. Xia, L. A. Wray, S. Jia, F. Meier, J. H. Dil, J. Osterwalder, B. Slomski, A. Bansil, H. Lin, R. J. Cava, and M. Z. Hasan, Science {\bf 332}, 560 (2011).

\bibitem{ando} T. Sato, K. Segawa, K. Kosaka, S. Souma, K. Nakayama, K. Eto, T. Minami, Y. Ando, and T. Takahashi, Nat. Phys. {\bf 7}, 840 (2011).

\bibitem{hassan-neupane} M. Brahlek, N. Bansal, N. Koirala, S.-Y. Xu, M. Neupane, C. Liu, M. Z. Hasan, and S. Oh, Phys. Rev. Lett. {\bf 109}, 186403 (2012). 

\bibitem{armitage} L. Wu, M. Brahlek, R. V. Aguilar, A. V. Stier, C. M. Morris, Y. Lubashevsky, L. S. Bilbro, N. Bansal, S. Oh, N. P. Armitage, Nat. Phys. {\bf 9}, 410 (2013). 

\bibitem{TPT-BiT} X. Xi, C. Ma, Z. Liu, Z. Chen, W. Ku, H. Berger, C. Martin, D. B. Tanner, G. L. Carr, Phys. Rev. Lett. {\bf 111}, 155701 (2013).

\bibitem{TKI-exp-1}  S. Wolgast, C. Kurdak, K. Sun, J. W. Allen, Dae-Jeong Kim, and Zachary Fisk, Phys. Rev. B {\bf 88}, 180405(R) (2013).

\bibitem{TKI-exp-2} M. Neupane,	N. Alidoust, S-Y. Xu,	T. Kondo,	Y. Ishida,	D. J. Kim,	C. Liu,	I. Belopolski,	Y. J. Jo,	T-R. Chang,	H-T. Jeng,	T. Durakiewicz,	L. Balicas,	H. Lin,	A. Bansil, S. Shin,	Z. Fisk, and M. Z. Hasan, Nature Communication {\bf 4}, 2991 (2013).

\bibitem{TKI-exp-3} N. Xu,	P. K. Biswas,	J. H. Dil,	R. S. Dhaka,	G. Landolt,	S. Muff,	C. E. Matt,	X. Shi,	N. C. Plumb,	M. Radovi\'{c},	E. Pomjakushina,	K. Conder,	A. Amato,	S. V. Borisenko,	R. Yu,	H.-M. Weng,	Z. Fang,	X. Dai,	J. Mesot,	H. Ding	and M. Shi, Nature Communication {\bf 5}, 4566 (2014).

\bibitem{TKI-exp-4} M. Xia, J. Jiang, Z. R. Ye, Y. H. Wang, Y. Zhang, S. D. Chen, X. H. Niu, D. F. Xu, F. Chen, X. H. Chen, B. P. Xie, T. Zhang, and D. L. Feng, Sci. Rep. 4, 5999 (2014).

\bibitem{TKI-exp-5}  M. Neupane, S-Y. Xu, N. Alidoust, G. Bian, D. J. Kim, C. Liu, I. Belopolski, T.-R. Chang, H.-T. Jeng, T. Durakiewicz, H. Lin, A. Bansil, Z. Fisk, and M. Z. Hasan , Phys. Rev. Lett. {\bf 114}, 016403 (2015).

\bibitem{roy-TKI}  B. Roy, J. D. Sau, M. Dzero, and V. Galitski, Phys. Rev. B {\bf 90},155314 (2014).

\bibitem{coleman-review}  M. Dzero, J. Xia, V. Galitski, P. Coleman, arxiv:1506.05635. 

\bibitem{fisk-pressure}  Y. Zhou, D-J. Kim, P. F. S. Rosa, Q. Wu, J. Guo, S. Zhang, Z. Wang, D. Kang, C. Zhang, W. Yi, Y. Li, X. Li, J. Liu, P. Duan, M. Zi, X. Wei, Z. Jiang, Y. Huang, Y-F. Yang, Z. Fisk, L. Sun, Z. Zhao, arxiv:1501.03901.

\bibitem{dornhaus} R. Dornhaus, G. Nimtz, and B. Schlicht, \emph{Narrow-Gap Semicounductors}, (Springer-Verlag, 1983).

\bibitem{cdas} S. Borisenko, Q. Gibson, D. Evtushinsky, V. Zabolotnyy, B. Buechner, and R. J. Cava, Phys. Rev. Lett. {\bf 113}, 027603 (2014).

\bibitem{nabi} Z. K. Liu, B. Zhou, Z. J. Wang, H. M. Weng, D. Prabhakaran, S.-K. Mo, Y. Zhang, Z. X. Shen, Z. Fang, X. Dai, Z. Hussain, and Y. L. Chen, Science, {\bf 343}, 864 (2014).


\bibitem{goswami-chakravarty} P. Goswami, and S. Chakravarty, Phys. Rev. Lett. {\bf 107}, 196803 (2011).

\bibitem{hosur} P. Hosur, S. Parameswar, A. Vishwanath, Phys. Rev. Lett. {\bf 108}, 046602 (2012).

\bibitem{isobe-nagaosa} H. Isobe, N. Nagaosa, Phys. Rev. B {\bf 86}, 165127 (2012); \emph{ibid} {\bf 87}, 205138 (2013).

\bibitem{nomura} A. Sekine, and K. Nomura, Phys. Rev. B {\bf 90}, 075137 (2014).

\bibitem{gonzalez} J. Gonzalez, Phys. Rev. B {\bf 92}, 125115 (2015).

\bibitem{kim-moon} E-G. Moon, Y. B. Kim, arxiv:1409.0573. 

\bibitem{throckmorton-3d}  R. E. Throckmorton, J. Hofmann, E. Barnes, S. Das Sarma, Phys. Rev. B {\bf 92}, 115101 (2015). 

\bibitem{roy-goswami-sau} B. Roy, P. Goswami, J. D. Sau, Phys. Rev. B {\bf 93}, 041101 (2016).


 
\bibitem{fradkin} E. Fradkin, Phys. Rev. B {\bf 33}, 3263 (2986)

\bibitem{shindou} R. Shindou, and S. Murakami, Phys. Rev. B {\bf 79}, 045321 (2009).

\bibitem{herbut-Imura} K. Kobayashi, T. Ohtsuki, K.-I. Imura, and I. F. Herbut, Phys. Rev. Lett. {\bf 112}, 016402 (2014).

\bibitem{radziovsky} S. V. Syzranov, L. Radzihovsky, and V, Gurarie, Phys. Rev. Lett. {\bf 114}, 166601 (2015);  S. V. Syzranov, V. Gurarie, L. Radzihovsky, Phys. Rev. B {\bf 91}, 035133 (2015).

\bibitem{roy-dassarma} B. Roy and S. Das Sarma, Phys. Rev. B {\bf 90}, 241112 (2014).

\bibitem{ominato} Y. Ominato and M. Koshino, Phys. Rev. B {\bf 89}, 054202 (2014).

\bibitem{pixley} J. Pixley, P. Goswami, and S. Das Sarma, Phys. Rev. Lett. {\bf 115}, 076601 (2015), arxiv:1505.07938.

\bibitem{dassarma-hwang-transport}  S. Das Sarma, E. H. Hwang, Phys. Rev. B {\bf 91}, 195104 (2015).

\bibitem{fiete}  R. Lundgren, P. Laurell, G. A. Fiete, Phys. Rev. B {\bf 90}, 165115 (2014).

\bibitem{herbut-book} I. F. Herbut, \emph{A Modern Approach to Critical Phenomena} (Cambridge University Press, UK, Cambridge, 2007).


\bibitem{coulomb-ref} Although in this paper we do not address the effects of long range Coulomb interaction in details, we devote Appendix~\ref{coulomb-long-range} to discuss its effects in clean interacting DSM. A representative phase diagram is shown in Fig.~\ref{pd-DiracCoulomb}. The long range Coulomb interaction does not alter the critical behavior in DSM, only shifts the phase boundary between DSM and BSPs in a non-universal fashion, but to weaker interaction strength, see Figs.~\ref{pd-DiracClean} and \ref{pd-DiracCoulomb}. 

\bibitem{boundary} When the difference among the susceptibilities (dimensionless) associated with scalar and pseudoscalar mass channels is less than $10^{-2}$, we identify the broken symmetry phase as axionic insulator, which is a linear superposition of two masses. The phase diagrams presented here should, however, correspond to a generic situation in the presence of generic repulsive interaction in DSM.    

\bibitem{peskin} M. E. Peskin and D. V. Schroeder, \emph{An Introduction to Quantum Field Theory} (Addison-Wesley, Reading, MA, 1995).

\bibitem{khveshchenko}  D. V. Khveshchenko, Phys. Rev. Lett. {\bf 87}, 246802 (2001). 

\bibitem{herbut-original} I. F. Herbut, Phys. Rev. Lett. {\bf 97}, 146401 (2006).

\bibitem{HJR} I. F. Herbut, V. Juri\v{c}i\'{c}, B. Roy, Phys. Rev. B {\bf 79}, 085116 (2009).

\bibitem{aji} H. Wei, S-P. Chao, and V. Aji, Phys. Rev. Lett. {\bf 109}, 196403 (2012).

\bibitem{sczhang} Z. Wang and S-C. Zhang, Phys. Rev. B {\bf 87}, 161107(R) (2013).

\bibitem{nandkishore} J. Maciejko, and R. Nandkishore, Phys. Rev. B {\bf 90} 035126 (2014).

\bibitem{nomura-Weyl}  A. Sekine, and K. Nomura, J. Phys. Soc. Jpn. {\bf 82}, 033702 (2013).



\bibitem{brouwer} B. Sbierski, G. Pohl, E. J. Bergholtz, and P. W. Brouwer, Phys. Rev. Lett. {\bf 113}, 026602 (2014); B. Sbierski, E. J. Bergholtz, P. W. Brouwer, Phys. Rev. B {\bf 92}, 115145 (2015).

\bibitem{altland} A. Altland, and D. Bagrets, Phys. Rev. Lett. {\bf 114}, 257201 (2015).

\bibitem{chen-Song} C-Z. Chen, J. Song, H. Jiang, Q-F. Sun, Z. Wang, X. C. Xie, Phys. Rev. Lett. {\bf 115}, 246603 (2015).

\bibitem{ohtsuki} S. Liu, T. Ohtsuki, R. Shindou, Phys. Rev. Lett. {\bf 116}, 066401 (2016).

\bibitem{roy-bera} S. Bera, J. D. Sau, B. Roy, Phys. Rev. B {\bf 93}, 201302 (2016).

\bibitem{hughes} H. Shapourian, T. L. Hughes, Phys. Rev. B {\bf 93}, 075108 (2016).

\bibitem{nandkishore-disorder} For possible rare region effects in Weyl semimetals, see R. Nandkishore, D. A. Huse, S. L. Sondhi, Phys. Rev. B {\bf 89}, 245110 (2014).

\bibitem{harris} A. B. Harris, J. Phys. C {\bf 7}, 1671 (1974).

\bibitem{chayes} J. T. Chayes, L. Chayes, D. S. Fisher, and T. Spencer, Phys. Rev. Lett. 57, 2999 (1986).



\bibitem{long-range-explanation-1} Notice that we here neglect the long range tail of the Coulomb interaction and focus on its short-range components. In the presence of underlying lattice, long range Coulomb interaction is always accompanied by the short-range ones. Thus phase diagram of interacting three dimensional DSM can be studied by tuning the strength of the long range Coulomb interaction in lattice based numerical simulations. For numerical simulation in monolayer graphene (a prototypical two-dimensional DSM) with long-range Coulomb interaction see J. Drut, and T. A. L\"ahde, Phys. Rev. Lett. {\bf 102}, 026802 (2009); Phys. Rev. B {\bf 79}, 165425 (2009). 

\bibitem{long-range-explanation-2} The critical exponents extracted near the semimetal-insulator transition in Ref.~\cite{long-range-explanation-1} match reasonably well with the ones computed near a QCP in graphene for CSB ordering. See B. Rosenstein, H-L. Yu, and A. Kovner, Phys. Lett. B {\bf 314}, 381 (1993); I. F. Herbut, V. Juri\v{c}i\'{c}, O. Vafek, Phys. Rev. B {\bf 80}, 075432 (2009).  

\bibitem{semenoff} For field theoretic analysis on the role of long range Coulomb interaction near semimetal-insulator QPT in single-layer graphene, see V. Juri\v{c}i\'{c}, I. F. Herbut, G. W. Semenoff, Phys. Rev. B {\bf 80}, 081405 (2009). 

\bibitem{sankar-twoloop} See also  E. Barnes, E. H. Hwang, R. E. Throckmorton, S. Das Sarma, Phys. Rev. B {\bf 89},  235431 (2014). 

\bibitem{peccei-quinn} R. D. Peccei and H. R. Quinn, Phys. Rec. Lett. {\bf 38}, 1440 (1977).

\bibitem{weinberg} S. Weinberg, Phys. Rev. Lett. {\bf 40}, 223 (1978).

\bibitem{wilczek} F. Wilczek, Phys. Rev. Lett. {\bf 40}, 279 (1978).

\bibitem{vanderbilt-axion} A. M. Essin, J. E. Moore, D. Vanderbilt, Phys. Rev. Lett. {\bf 102}, 146805 (2009).

\bibitem{zhang-axion} R. Li, J. Wang, X-L. Qi, and S-C. Zhang, Nature Phys. {\bf 6}, 284 (2010).

\bibitem{callan}  C. G. Callan, and J. A. Harvey, Nucl. Phys. B {\bf 250}, 427 (1985).

\bibitem{ohsaku} T. Ohsaku, Phys. Rev. B {\bf 65}, 024512 (2002).

\bibitem{fuberg} L. Fu, and E. Berg, Phys. Rev. Lett. {\bf 105}, 097001 (2010).

\bibitem{GR-fieldtheory} P. Goswami, and B. Roy, arxiv:1211.4023.   

\bibitem{axion-SC} P. Goswami, and B. Roy, Phys. Rev. B {\bf 90},  041301(R) (2014).

\bibitem{NJL} Y. Nambu, and G. Jona-Lasinio, Phys. Rev. {\bf 122}, 345 (1961).

\bibitem{gross-neveu} D. J. Gross, A. Neveu, Phys. Rev. D {\bf 10}, 3235 (1974).

\bibitem{zinn-justin} J. Zinn-Justin, \emph{Quantum Field Theory and Critical Phenomena} (Oxford Science Publications, Oxford, 2002).

\bibitem{moshe-moshe} M. Moshe, and J. Zinn-Justin, Phys. Rept. {\bf 385}, 69 (2003).

\bibitem{miransky} V. A. Miransky, \emph{Dynamical Symmetry Breaking In Quantum Field Theories}, (World Scientific Publishing Company, 1993).

\bibitem{multicritical-roy} For symmetry restoration near a QCP in $(2+1)$-dimensions, see B. Roy, Phys. Rev. B {\bf 84}, 113404 (2011); B. Roy, V. Juri\v{c}i\'{c} Phys. Rev. B {\bf 90}, 041413(R) (2014). 

\bibitem{roy-lorentz} B. Roy, V. Juri\v{c}i\'{c}, I. F. Herbut, J. High Energy Phys. {\bf 04}, 018 (2016) .

\bibitem{catalysis-3D} V. P. Gusysnin, V. A. Miransky, and I. A. Shovkovy, Nucl. Phys. B {\bf 462}, 249 (1996).

\bibitem{roy-sau} B. Roy, J. D. Sau, Phys. Rev. B {\bf 92}, 125141 (2015).

\bibitem{roy-catalysis-scaling} I. F. Herbut, and B. Roy, Phys. Rev. B {\bf 77}, 245438 (2008); B. Roy, M. P. Kennett, S. Das Sarma, Phys. Rev. B {\bf 90},  201409(R) (2014).


\bibitem{comment-ladder-crossing} Potentially new disorder vertices can be generated from Figs.~\ref{Feynman-Diag} $(ix)$ and $(x)$. However, Dirac kernel is an \emph{odd} function of frequency (Matsubara) and momentum, and two $\gamma$ matrices satisfy the algebra $\left\{\gamma_0,\gamma_5 \right\}=0$. Consequently, contribution from these two diagrams cancel out when $\Gamma_a=\Gamma_b=\gamma_0$ or $\gamma_0 \gamma_5$, as well as when $\Gamma_a=\gamma_0$ and $\Gamma_b=\gamma_0 \gamma_5$ or vice-versa. This conclusion remains unchaged order by order in perturbation theory. For detailed discussion on such explicit cancellation at two-loop order, see Ref.~\cite{roy-dassarma}.  

\bibitem{juricic} B. Roy, V. Juri\v{c}i\'{c}, and S. Das Sarma, arXiv:1603.00017 

\bibitem{wegner} F. Wegner, Z. Phys. B {\bf 25}, 327 (1976).

\bibitem{dorogovtsev} S. N. Dorogovtsev, Phys. Lett. {\bf 76A}, 169 (1980).

\bibitem{cardy} D. Boyanovsky and J. L. Cardy, Phys. Rev. B {\bf 26}, 154 (1982).

\bibitem{lawrie} I. D. Lawrie and V. V. Prudnikov, J. Phys. C {\bf 17}, 1655 (1984). 

\bibitem{topo-fdth} X-L. Qi, T. L. Hughes, and S-C. Zhang, Phys. Rev. B {\bf 78}, 195424 (2010).  



\bibitem{taas-1} C. Zhang, Z. Yuan, S. Xu, Z. Lin, B. Tong, M. Z. Hasan, J. Wang, C. Zhang, S. Jia, arxiv:1502.00251. 

\bibitem{tasas-2} S-Y. Xu, I. Belopolski, N. Alidoust, M. Neupane, C. Zhang, R. Sankar, S-M. Huang, C-C. Lee, G. Chang, B. Wang, G. Bian, H. Zheng, D. S. Sanchez, F. Chou, H. Lin, S. Jia, M. Z. Hasan, Science {\bf 347}, 294 (2015).

\bibitem{taas-3} B. Q. Lv, H. M. Weng, B. B. Fu, X. P. Wang, H. Miao, J. Ma, P. Richard, X. C. Huang, L. X. Zhao, G. F. Chen, Z. Fang, X. Dai, T. Qian, H. Ding, Phys. Rev. X {\bf 5}, 031013 (2015).

\bibitem{nbas-1} S-Y. Xu, N. Alidoust, I. Belopolski, C. Zhang, G. Bian, T-R. Chang, H. Zheng, V. Strokov, D. S. Sanchez, G. Chang, Z. Yuan, D. Mou, Y. Wu, L. Huang, C-C. Lee, S-M. Huang, B. K. Wang, A. Bansil, H-T. Jeng, T. Neupert, A. Kaminski, H. Lin, S. Jia, M. Z. Hasan, arxiv:1504.01350.

\bibitem{tap-1} N. Xu, H. M. Weng, B. Q. Lv, C. Matt, J. Park, F. Bisti, V. N. Strocov, D. gawryluk, E. Pomjakushina, K. Conder, N. C. Plumb, M. Radovic, G. Autes, O. V. Yazyev, Z. Fang, X. Dai, G. Aeppli, T. Qian, J. Mesot, H. Ding, M. Shi, arxiv:1507.03983.

\bibitem{borisenko} S. Borisenko, D. Evtushinsky, Q. Gibson, A. Yaresko, T. Kim, M. N. Ali, B. Buechner, M. Hoesch, R. J. Cava, arxiv:1507.04847.

\bibitem{chiorescu} J. Y. Liu, J. Hu, D. Graf, S.M.A. Radmanesh, D.J. Adams, Y.L. Zhu, G.F. Chen, X. Liu, J. Wei, I. Chiorescu, L. Spinu, Z.Q. Mao, arxiv:1507.07978.



\bibitem{shovkovy} E. V. Gorbar, V. A. Miransky, and I. A. Shovkovy, Phys. Rev. B {\bf 88}, 165105 (2013). 

\bibitem{furusaki} T. Morimoto, A. Furusaki, Phys. Rev. B {\bf 89}, 235127 (2014).


\bibitem{galss-comment} Since the ${\mathcal P}, {\mathcal T}$-odd insulator, characterized by constant axion angle $\theta_{ax}=\mbox{sgn} (m_2) \pi/2$, breaks an Ising symmetry, the jump in the axion angle across two domains is $\Delta \theta_{ax}= \pm \pi$. Thus, the boundary between two domains of ${\mathcal P}, {\mathcal T}$-odd insulator accommodates gapless mode.  

\bibitem{pixley-unpublished} J. Pixley, D. A. Huse, S. Das Sarma, Phys.Rev.X {\bf 6}, 021042 (2016).

\bibitem{Adam-PNAS} S. Adam, E. H. Hwang, V. M. Galitski, S. Das Sarma, PNAS {\bf 104}, 18392 (2007).
 
\bibitem{Rossi-DasSarma} E. Rossi and S. Das Sarma, Phys. Rev. Lett. {\bf 101}, 166803 (2008).

\bibitem{Das-Sarma-RMP} S. Das Sarma, S. Adam, E. H. Hwang, and E. Rossi, Rev. Mod. Phys. {\bf 83}, 407 (2011).

\bibitem{roy-migdal} B. Roy, J. D. Sau, S. Das Sarma, Phys. Rev. B {\bf 89}, 165119 (2014).



\end{thebibliography}
\end{document}